\documentclass[12pt,a4paper]{report}
\usepackage[cp1250]{inputenc}
\usepackage{wrapfig}

\usepackage{amsmath}
\usepackage{graphicx}
\usepackage{amssymb}
\usepackage{epsfig}
\usepackage{upgreek}
\usepackage{bm}
\usepackage{siunitx}
\usepackage{docmute} 
\usepackage{hyperref}
\usepackage{bookmark}
\usepackage{xcolor}
\hypersetup{
    linktocpage=true,
    hidelinks,
    colorlinks=false,
    citebordercolor=black,
    linkbordercolor=black,
    urlbordercolor=black,
}

\usepackage{tikz}
\usetikzlibrary{arrows.meta}
\usepackage[normalem]{ulem}
\usepackage{adjustbox}
\usepackage{authblk}
\usepackage{titlesec}

\newcounter{count}
\makeatletter
\@addtoreset{figure}{count}
\@addtoreset{equation}{count}
\@addtoreset{table}{count}
\makeatother

\newcommand{\titl}[1]{{\centering\Large\bf #1\par}\bigskip}
\newcommand{\name}[1]{{\centering\rm\normalsize #1\par}\bigskip}

\newcommand{\adr}[1]{{\it \normalsize #1\par}\medskip}

\titleformat{\section}{\normalfont\small\bfseries}{\thesection}{1em}{}

\begin{document}
\begin{titlepage}
\begin{center}

\text{\large Mini-Proceedings}\\[0.2cm]
{\bf
\text{\Large Fourth International Workshop}\\
\text{\Large on the Extension Project}\\
\text{\Large for the J-PARC Hadron Experimental Facility}\\
\text{\Large (HEF-ex 2024)}\\[1.5cm]
}

\includegraphics[width=1\textwidth]{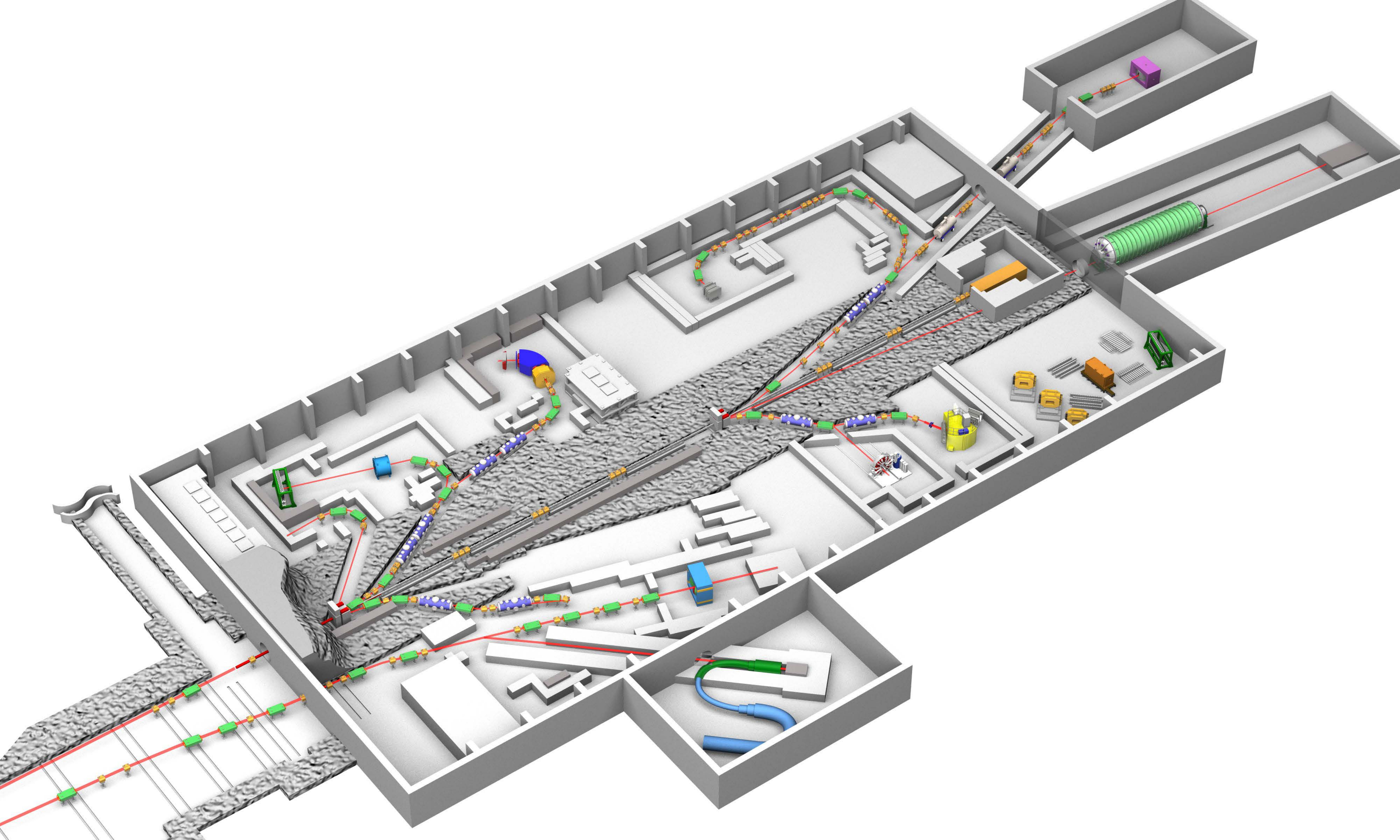}\\[2cm]

\def\thefootnote{\fnsymbol{footnote}}

\newcommand{\Jlab}{1}
\newcommand{\IPNS}{2}
\newcommand{\Yukawa}{3}
\newcommand{\LNF}{4} 
\newcommand{\Giessen}{5}
\newcommand{\iTHEMS}{6}
\newcommand{\KEKTheory}{7}
\newcommand{\SOKENDAI}{8}
\newcommand{\ASRC}{9}
\newcommand{\OsakaElectro}{10}
\newcommand{\Genova}{11}
\newcommand{\Beihanga}{12}
\newcommand{\Beihangc}{13}
\newcommand{\Beihangd}{14}
\newcommand{\SCNT}{15}
\newcommand{\CNS}{16}
\newcommand{\Julich}{17}
\newcommand{\KMI}{18}
\newcommand{\Nagoya}{19}
\newcommand{\Tohoku}{20}
\newcommand{\Nanjing}{21}
\newcommand{\Hosei}{22}
\newcommand{\TITECH}{23}
\newcommand{\Kyoto}{24}
\newcommand{\Bonn}{25}
\newcommand{\RCNP}{26}
\newcommand{\KEKACC}{27}
\newcommand{\TMU}{28}
\newcommand{\Tokyo}{29}
\newcommand{\Nishina}{30}
\newcommand{\Osaka}{31}
\newcommand{\Akita}{32}
\newcommand{\JlabT}{33}
\newcommand{\William}{34}
\newcommand{\Osakas}{35}
\newcommand{\KyotoP}{36}
\newcommand{\Czech}{37}
\newcommand{\Oxford}{38}
\newcommand{\Washington}{39}
\newcommand{\Showa}{40}
\newcommand{\Nippon}{41}
\newcommand{\KyotoS}{42}
\newcommand{\Beihangb}{43}
\newcommand{\CASa}{44}
\newcommand{\CASb}{45}
\newcommand{\Peking}{46}

\name{
P.~Achenbach$^{\Jlab}$,
K.~Aoki$^{\IPNS*}$,
S.~Aoki$^{\Yukawa}$,
C.~Curceanu$^{\LNF}$,
S.~Diehl$^{\Giessen}$,
T.~Doi$^{\iTHEMS}$,
M.~Endo$^{\KEKTheory,\SOKENDAI}$,
M.~Fujita$^{\ASRC}$, 
T.~Fukuda$^{\OsakaElectro}$,
H.~Garcia-Tecocoatzi$^{\Genova}$,
L.S.~Geng$^{\Beihanga,\Beihangc,\Beihangd,\SCNT}$,
T.~Gunji$^{\CNS}$,
C.~Hanhart$^{\Julich}$,
M.~Harada$^{\KMI,\Nagoya,\ASRC}$,
T.~Harada$^{\OsakaElectro}$,
S.~Hayakawa$^{\Tohoku}$,
B.R.~He$^{\Nanjing}$,
E.~Hiyama$^{\Tohoku}$,
R.~Honda$^{\IPNS}$,
Y.~Ichikawa$^{\Tohoku,\ASRC}$,
M.~Isaka$^{\Hosei}$,
D.~Jido$^{\TITECH}$,
A~.Jinno$^{\Kyoto}$,
K.~Kamada$^{\Tohoku}$, 
Y.~Kamiya$^{\Bonn}$,
Y.K.~Kong$^{\Nagoya}$,
Z.W.~Liu$^{\Beihanga}$,
J.X.~Lu$^{\Beihangd}$,
K.~Miwa$^{\Tohoku}$,
K.~Mizutani$^{\RCNP}$,
K.~Murakami$^{\TITECH,\iTHEMS}$,
R.~Muto$^{\KEKACC}$,
K.~Murase$^{\TMU}$,
S.~Nagao$^{\Tokyo}$,
T. Nanamura$^{\Nishina}$,
H.~Nanjo$^{\Osaka}$
Y.~Nara$^{\Akita}$,
A.~Ohnishi$^{\Yukawa,\dagger}$,
J.W.~Qiu$^{\JlabT,\William}$,
A.~Sakaguchi$^{\Osakas}$,
F.~Sakuma$^{\Nishina*}$,
E.~Santopinto$^{\Genova}$,
A.~Scordo$^{\LNF}$,
T.~Sekihara$^{\KyotoP}$,
C.~Seong$^{\Tohoku}$, 
N.V.~Shevchenko$^{\Czech}$,
J.R.~Stone$^{\Oxford}$,
I.~Strakovsky$^{\Washington}$,
K.~Suzuki$^{\RCNP}$,
H.~Takahashi$^{\IPNS}$,
M.~Takizawa$^{\Showa}$,
H.~Tamura$^{\Tohoku,\ASRC}$,
N.~Tomida$^{\Kyoto}$,
A.~Umeya$^{\Nippon}$,
O.~V\'azquez Doce$^{\LNF}$,
J.~Yamagata-Sekihara$^{\KyotoS}$,
T.O.~Yamamoto$^{\ASRC}$,
Y.~Xiao$^{\Beihangb,\Beihangc}$,
Z.~Yin$^{\TITECH}$,
Z.~Yu$^{\JlabT}$, and
B.S.~Zou$^{\CASa,\CASb,\Peking}$
}

\footnote[0]{$^*$editors: sakuma@ribf.riken.jp, kazuya.aoki@kek.jp}

\newpage

\adr{\footnotesize
$^{\Jlab}$ Thomas Jefferson National Accelerator Facility, Newport News, VA, 23606, USA\\
$^{\IPNS}$ Institute of Particle and Nuclear Studies, High Energy Accelerator Research Organization, Tsukuba 305-0801, Japan\\
$^{\Yukawa}$ Yukawa Institute for Theoretical Physics, Kyoto University, Kyoto 606-8502, Japan\\
$^{\LNF}$ Laboratori Nazionali di Frascati, INFN, Via E. Fermi 54, I-00044 Frascati(RM), Frascati, Italy\\
$^{\Giessen}$ Justus Liebig University Giessen, 35390 Giessen, Germany\\
$^{\iTHEMS}$ Interdisciplinary Theoretical and Mathematical Sciences Program (iTHEMS), RIKEN, Wako 351-0198, Japan\\
$^{\KEKTheory}$ KEK Theory Center, Tsukuba, Ibaraki 305--0801, Japan\\ 
$^{\SOKENDAI}$ Graduate Institute for Advanced Studies, SOKENDAI, Tsukuba, Ibaraki 305--0801, Japan\\
$^{\ASRC}$ Advanced Science Research Center, Japan Atomic Energy Agency, Tokai 319-1195, Japan\\
$^{\OsakaElectro}$ Research Center for Physics and Mathematics, Osaka Electro-Communication University, Neyagawa, Osaka, 572-8530, Japan\\
$^{\Genova}$ INFN, Sezione di Genova, Via Dodecaneso 33, 16146 Genova, Italy\\
$^{\Beihanga}$ School of Physics, Beihang University, Beijing 102206, China\\
$^{\Beihangc}$ Peng Huanwu Collaborative Center for Research and Education, Beihang University, Beijing 100191, China\\
$^{\Beihangd}$ Beijing Key Laboratory of Advanced Nuclear Materials and Physics, Beihang University, Beijing 102206, China\\
$^{\SCNT}$  Southern Center for Nuclear-Science Theory (SCNT), Institute of Modern Physics, Chinese Academy of Sciences, Huizhou 516000, China\\
$^{\CNS}$ Center for Nuclear Study, Graduate School of Science, the University of Tokyo, Hongo
Campus. 7-3-1, Hongo, Bunkyo-ku, Tokyo, 113-0033, Japan\\
$^{\Julich}$ IAS-4, Forschungszentrum J\"ulich, J\"ulich, Germany\\
$^{\KMI}$ Kobayashi-Maskawa Institute for the Origin of Particles and the Universe, Nagoya University, Nagoya, 464-8602, Japan\\
$^{\Nagoya}$ Department of Physics, Nagoya University, Nagoya, 464-8602, Japan \\
$^{\Tohoku}$ Department of Physics, Tohoku University, Sendai, 980-8578, Japan\\
$^{\Nanjing}$ Department of Physics, Nanjing Normal University, Nanjing 210023, PR China\\
$^{\Hosei}$ Hosei University, 2-17-1 Fujimi, Chiyoda, Tokyo 102-8160, Japan\\
$^{\TITECH}$ Department of Physics, Tokyo Institute of Technology, Megro, Tokyo 152-8551, Japan\\
$^{\Kyoto}$ Department of Physics, Faculty of Science, Kyoto University, Kyoto 606-8502, Japan\\
$^{\Bonn}$ Helmholtz Institut f\"ur Strahlen- und Kernphysik and Bethe Center for Theoretical Physics, Universit\"at Bonn, D-53115 Bonn, Germany\\
$^{\RCNP}$ Research Center for Nuclear Physics, Osaka University, Japan \\
$^{\KEKACC}$ Accelerator Laboratory, High Energy Accelerator Research Organization, Tsukuba 305-0801, Japan \\
$^{\TMU}$ Department of Physics, Tokyo Metropolitan University, Hachioji 192-0397, Japan\\
$^{\Tokyo}$ The University of Tokyo, 7-3-1 Hongo Bunkyo, Tokyo, 113-0033, Japan\\
$^{\Nishina}$ RIKEN Nishina Center for Accelerator-Based Scienc, RIKEN, Wako 351-0198, Japan\\
$^{\Osaka}$ Department of Physics, Osaka University, Toyonaka, Osaka 560-0043, Japan\\
$^{\Akita}$ Akita International University, Yuwa, Akita-city 010-1292, Japan\\
$^{\JlabT}$ Theory Center, Jefferson Lab, Newport News, Virginia 23606, USA\\
$^{\William}$ Department of Physics, William \& Mary, Williamsburg, Virginia 23187, USA\\
$^{\Osakas}$ Student Life Cycle Support Center, Osaka University, Toyonaka, Osaka 560-0043, Japan\\
$^{\KyotoP}$ Graduate School of Life and Environmental Sciences, Kyoto Prefectural University, Sakyo-ku, Kyoto 606-8522, Japan\\
$^{\Czech}$ Nuclear Physics Institute, 25068 \v{R}e\v{z}, Czech Republic \\
$^{\Oxford}$ University of Oxford, Department of Physics (Astrophysics), Oxford, United Kingdom\\
$^{\Washington}$ Institute for Nuclear Studies, Department of Physics, The George Washington University, Washington, DC 20052, USA\\
$^{\Showa}$ Showa Pharmaceutical University, Machida, Tokyo, 194-8543, Japan\\
$^{\Nippon}$ Liberal Arts and Sciences, Nippon Institute of Technology, Saitama 345-8501, Japan\\
$^{\KyotoS}$ Department of Physics, Kyoto Sangyo University, Kyoto 603-8555, Japan\\
$^{\Beihangb}$ School of Space and Environment, Beihang University, Beijing 102206, China\\
$^{\CASa}$ CAS Key Laboratory of Theoretical Physics, Institute of Theoretical Physics, Chinese Academy of Sciences, Beijing 100190, China\\
$^{\CASb}$ School of Physical Sciences, University of Chinese Academy of Sciences, Beijing 100049, China\\
$^{\Peking}$ School of Physics, Peking University, Beijing 100871, China\\
}

\end{center}
\end{titlepage}
 
\clearpage

\tableofcontents
\clearpage

\phantomsection
\addcontentsline{toc}{section}{
{\bf Introduction} \\
F.~Sakuma on behalf of the HEF-ex 2024 Organizing Committee}

\titl{
Introduction
}

\name{
F.~Sakuma~$^{1}$ on behalf of the HEF-ex 2024 Organizing Committee
}

\adr{
$^1$ RIKEN, Wako 351-0198, Japan
}


The Fourth International Workshop on the Extension Project for the J-PARC Hadron Experimental Facility (HEF-ex 2024) was held on February 19 -- 21, 2024.
The workshop was organized by Hadron Hall Users' Association (HUA) and supported by J-PARC Center, KEK Theory Center, and RCNP.

We have extensively discussed the facility extension project with the nuclear and high-energy physics communities worldwide, through several HEF-ex workshops, to refine and facilitate the project; their outputs were summarized as a third white paper~[1].
The project was finally selected as a top-priority project to be budgeted in the KEK mid-term plan (FY2022-26) in KEK-PIP2022 (Project Implementation Plan) and is about to be realized.
Detailed information on the extension project can be found on HUA's website~[2].

In the fourth workshop, discussions were continuously devoted to the physics case that connected both the ``present'' and the ``future'' Hadron Experimental Facility at J-PARC.
To this end, this workshop covered a wide range of topics in flavor, hadron, and nuclear physics related to both experimental and theoretical activities being conducted at the facility as follows:
\begin{itemize}
    \setlength{\parskip}{0cm} 
    \setlength{\itemsep}{0cm} 
    \item S=-1 and -2 hypernuclei
    \item baryon-baryon and meson-baryon interactions
    \item meson in nuclei
    \item baryon spectroscopy
    \item kaon rare decays
    \item $\mu - e$ conversion
    \item future facilities and instrumentation
\end{itemize}

The workshop was held at J-PARC in a hybrid format.
The workshop included 177 participants from 15 countries (35 participants from abroad), 99 of whom were on-site participants (12 participants from abroad).
The scientific program comprised plenary sessions by invited speakers and parallel sessions by invited and contributed speakers. 
There were 16 plenary and 44 parallel talks.
The entire program and presentation files are available online at the workshop website~[3].

\vfill  

\noindent{\bf References }
\begin{description}
\setlength\itemsep{-3pt}
\item{[1]} \href{https://doi.org/10.48550/arXiv.2110.04462}{Taskforce on the extension of the Hadron Experimental Facility, arXiv:2110.04462 [nucl-ex], (2021).}
\item{[2]} \url{http://www.rcnp.osaka-u.ac.jp/~jparchua/en/hefextension.html}
\item{[3]} \url{https://kds.kek.jp/event/46965}
\end{description}

\stepcounter{count}
\clearpage

\phantomsection
\addcontentsline{toc}{section}{
{\bf Hadron and Strangeness Physics with CLAS12 -- Complementarity to Experiments at J-PARC HEF} \\
P.~Achenbach}


\titl{Hadron and Strangeness Physics with CLAS12 \\ {\small Complementarity to Experiments at J-PARC HEF}}
\name{Patrick Achenbach$^{1}$}
\adr{$^{1}$Thomas Jefferson National Accelerator Facility, Newport News, VA, 23606, USA}

\paragraph{Introduction.}
%
The Thomas Jefferson National Accelerator Facility (JLab) experimental research program is addressing pressing science questions, among them the determination of the spectrum and structure of hadrons. Research at JLab with the CLAS12 spectrometer in Hall~B is having a profound influence on this field by using the precision of the electromagnetic interaction and its ability to ``tune'' the distance scale being probed~[1]. Experimental science with CLAS12 is complementary to two core lines of research at the Hadron Experimental Facility (HEF) at the Japan Proton Accelerator Research Complex (J-PARC): (i) strangeness nuclear physics and (ii) hadron physics, with flavor physics being its third core line not addressed here~[2,3]. The combined scientific output is dramatically improving our understanding of QCD in the ``strong'' (non-perturbative) regime, where it is still poorly understood, and it elucidates nuclear matter containing strange quarks, from the laboratory to deep inside neutron stars.

\paragraph{\boldmath Hadrons with Strangeness $S = -1$.} 
%
Studies of the spectrum of excited nucleon states and their structure in terms of the $Q^2$-evolution of the electrocouplings represent a source of information on many facets of the emergence of strong QCD underlying the generation of nucleon states of different quantum numbers~[4]. 

A key aspect of the former CLAS and the new CLAS12 spectrometers is their large acceptance that spans the full center-of-mass energy ($W$) and kaon angular range where new nucleon states could be and still can be discovered. The highest impact on spectroscopy of nucleon resonance states in the mass range above 1.7\,GeV have been studies of $KY$ final states with polarization observables. In terms of $N^*$ structure studies, using $KY$ final states in electroproduction experiments will eventually be necessary to cross-check the extracted electro-couplings from the $N\pi\pi$ channels. This will be important to understand model-dependent effects and to reduce the associated systematic uncertainties through channels with very different resonant and non-resonant contributions. As the couplings of higher-mass $N^*$ to $N\pi$, $N\pi\pi$, and $KY$ are not understood, $KY$ studies may allow unique access to states of non-standard configurations such as molecular states or gluonic hybrids as well as to conventional $qqq$ states.

The CLAS Collaboration measured the differential photoproduction cross sections of the $\Sigma^0$(1385), the $\Lambda$(1405), and the $\Lambda$(1520) in the $\gamma p \to K^+Y$ reactions for photon beam energies from near production threshold to $W=$ 2.85\,GeV with high precision~[5]. The line shape of the $\Lambda$(1405) has not only been studied extensively using real photoproduction with this data from CLAS, but also using electroproduction~[6], although with limited statistics. As the $\Lambda$(1405) is considered to be dynamically generated by the kaon--nucleon interaction, kaon beam experiments provide a crucial complementary approach to study this strange resonance.  

With CLAS12, $K^+Y$ electroproduction cross sections are measured from an unpolarized proton target with a focus on the $K^+\Lambda$ and $K^+\Sigma^0$ final states to study high-lying nucleon excited states~[7,8]. From exclusive final states, the electrocouplings for the most prominent $N^*$ and $\Delta^*$ states are extracted~[9]. Higher statistics electroproduction analyses would clearly be useful to study the two-pole model of the $\Lambda$(1405). 

Nucleon resonances in the non-strangeness sector will be investigated with the E45 experiment at the J-PARC K1.8
beamline with partial wave analyses of the $\pi N \to \pi\pi N$ reactions~[10]. In the past, most resonance properties were determined primarily from partial-wave analysis of the $\pi N \to \pi N$ data. However, many of these resonances have strong decay branching ratios to the $\pi\pi N$ final state. The J-PARC E31 experiment has been performed to investigate the $\Lambda$(1405) spectrum shape through a measurement of the $K^- d \to n\Sigma^0\pi^0$ reaction~[11]. The result clearly shows the interference between the $I = 0$ and $I = 1$ amplitudes in the $\pi^\pm\Sigma^\mp$ spectra, as theoretically expected. The J-PARC E72 experiment was proposed to provide high-precision differential cross sections for the $K^- p \to \eta\Lambda$ reaction to elucidate the resonance contributions around 1670\,MeV and to confirm the existence of the $\Lambda^*$(1665) with $J = 3/2$~[12]. 

A more complete understanding of nucleon resonances requires reaction data for both hadronic and electromagnetic probes implementing coupled-channels effects. 

\paragraph{\boldmath Hadrons with Strangeness $S = -2$ and $S = -3$.}
%
Baryons with multiple strange quarks play an important role in the quark model and our understanding of hadrons. In particular, the $\Omega$ baryon contains only a single flavor of constituent quarks, $s$, and its spectrum is expected to provide unique information on the study of the internal motion of quarks. However, only two $\Omega$ and six $\Xi$ states are considered to be well-established and more information about excited states is scarce~[13]. 

The production mechanisms of multi-strange baryons are studied in exclusive reactions with CLAS12 through the associated production of kaons, which allows tagging of the production of baryons with net strangeness, or through their weak decays, which are identified by a displaced vertex. One aim is to collect a large statistics sample of $\Xi$ baryons that will be used to search for new and missing excited states with the possibility to measure their quantum numbers, as well as to determine the mass splitting of ground and excited state isospin doublets. These data samples will also provide the statistics necessary for measuring the beam polarization transfer and induced polarization of the ground state $\Xi$ in the reaction $\gamma^{(*)} p \to \Xi^-K^+K^-$. Clear signals of the $\Xi$(1320) and $\Xi$(1530) are seen in the $(e'K^+K^+)$ missing mass spectra that will allow to study the $Q^2$-evolution of the cross section to explore the production mechanism. CLAS12 is expected to observe the $\Omega$ baryon for the first time in electroproduction~[14]. A promising search strategy is the reconstruction of the $(p\pi^-K^-)$ invariant mass spectrum using detached vertex cuts. 

Planned as one of the new beamlines in the J-PARC HEF extension project, K10 will provide a highly purified, high-intensity kaon beam with momenta up to 10\,GeV$\!/c$. With this beam large numbers of multi-strange baryons such as $\Xi$ and $\Omega$ could be produced. Spectroscopy of strange baryons at K10 will provide crucial ingredients of the structure of hadrons as composite systems of quarks and gluons~[2,3].

\paragraph{\boldmath $Y\!N$ Interactions.} 
%
Experimental input on hyperon--nucleon ($Y\!N$) interactions is needed to establish realistic models of nuclear forces, where the $\Sigma N$ channel is closely related to the $\Lambda N$ channel because of the strong $\Lambda N$--$\Sigma N$ coupling in the $S = -1$ sector. Modern $Y\!N$ interactions have been tested with hypernuclear data because $\Lambda$ binding energies and hypernuclear energy levels reflect the underlying two-body interactions~[15]. Both, at JLab and at J-PARC, several past and future hypernuclear experiments follow this approach. 

Studies of $KN$ interactions in either of the initial or final state can be used for better understanding of final-state interaction contributions that will be essential to develop and tune both single-channel and coupled-channels reaction models. However, data on two-body scattering reactions are scarce. 

At JLab, the CLAS detector and a photon beam impinging on an extended hydrogen target was utilized to study the $\Lambda p \to \Lambda p$ elastic scattering cross section in the $\Lambda$ momentum range of $0.9-2.0$\,GeV$\!/c$~[16]. These are the first data on this reaction since the 1970s. The technique is also applied to measurements from CLAS using a deuteron target to access $\Lambda n$ scattering. 

In the J-PARC E40 experiment at the K1.8 beamline, $\Sigma^- p$ and $\Sigma^+ p$ elastic scatterings and the $\Sigma^- p \to \Lambda n$ inelastic scattering reactions were systematically studied~[17--19].  New data on the $\Lambda N$ interactions in free space will be obtained from $\Lambda p$ scattering experiments planned at the new K1.1 and high-p beamlines at J-PARC. For the high momentum of the incident $\pi^-$ beam, $\Lambda$ hyperons in a wide momentum range ($0.4 < p$ (GeV$/c$) $< 2.0$) can be utilized with a large acceptance forward spectrometer. In addition, the systematic high-precision $(\pi, K^+)$ spectroscopy of $\Lambda$-hypernuclei at the High Intensity High Resolution (HIHR) beamline will clarify the density dependence of the $\Lambda N$ interaction in medium with unprecedented precision~[2,3]. 

The extensive data on low-energy $\Lambda N$ and $\Sigma N$ scattering from CLAS and E40 are already providing precise constraints to $\chi$EFT calculations. In future, the $YN$ interactions will be further investigated systematically by different reaction and isospin channels.

\paragraph{Technical Synergies.}
%
Ongoing developments in computing and networking technology aim at replacing standard hardware-based triggers for the data acquisition system in current and future experiments with a streaming or triggerless readout scheme, both at JLab~[20] and J-PARC~[21]. This allows to integrate the whole detector information for efficient real-time data tagging and$/$or selection. In contrast to hardware-triggered readout, the streaming readout concept relies on modern digital data processing. It can provide large factors of data reduction, removes nearly all deadtime, introduces less restrictions for filter criteria and potentially less bias, and opens unique opportunities for the application of AI/ML. 

At JLab, CLAS12 is at the forefront of this ambitious program with a data rate for the full spectrometer in streaming mode estimated to be on the level of 50\,GB$/$s and expecting a data reduction factor of 10 or higher. At J-PARC, Experiments E16 and E50 require systems with high data and event rates, for which a prototype of streaming readout software was developed. Both laboratories are internationally cooperating to overcome the present limitations of the readout systems and the implementation of AI/ML is expected at all stages from design, simulation, filtering, calibration, and reconstruction to analysis.

\paragraph{Concluding Remarks.}
%
The J-PARC HEF was constructed to explore low-energy QCD dynamics through the use of high-intensity secondary beams as well as 30-GeV primary proton beams. Since the first beam was delivered to the J-PARC HEF in February 2009, several hadron and strangeness physics experiments have been successfully performed and many more are proposed. Since Spring 2018, experiments with the CLAS12 detector and electrons beams of up to 11\,GeV at JLab provide data with a large kinematic coverage for the exclusive electroproduction of ground and excited, non-strange and strange hadrons. This spectrometer also allows for studies of the production and decays of multi-strange hadrons such as $\Xi$ and $\Omega$ hyperons. CLAS12 has acquired high statistics in several separate running periods and will take much more data over the next years. Many parts of the experimental programs at the J-PARC HEF and its future extension are complementary with CLAS12 at JLab. 

\paragraph{Acknowledgment.}
This material is based upon work supported by the U.S.\ Department of Energy, Office of Science, Office of Nuclear Physics under contract DE-AC05-06OR23177.\\

-----------

%
\noindent{\bf References }
\begin{description}
\setlength\itemsep{-3pt}

\item{[1]} 
J. Arrington {\em et al.}, Prog. Part. Nucl. Phys. {\bf 127} (2022) 103985.

\item{[2]} 
K. Aoki {\em et al.} (Taskforce on the Extension of the Hadron Experimental Facility), \href{https://doi.org/10.48550/arXiv.2110.04462}{arXiv:2110.04462 [nucl-ex], (2021)}; EPJ Web Conf. {\bf 271} (2022) 11001.

\item{[3]} 
H. Ohnishi, F. Sakuma, and T. Takahashi, Prog. Part. Nucl. Phys. {\bf 113} (2020) 103773.

\item{[4]} 
M. Ding, C. D. Roberts, and S. M. Schmidt, Particles {\bf 6} (2023) 57.

\item{[5]} 
K. Moriya {\em et al.} (CLAS Collaboration), Phys. Rev. C {\bf 87} (2013) 035206. 

\item{[6]} 
H. Y. Lu {\em et al.} (CLAS Collaboration), Phys. Rev. C {\bf 88} (2013) 045202. 

\item{[7]} 
D. S. Carman {\em et al.}, Experiment E12-06-108A,
\href{https://www.jlab.org/exp_prog/proposals/14/E12-06-108A.pdf}{Proposal to JLab PAC 42 (2014)}.
 
\item{[8]} 
D. S. Carman {\em et al.}, Experiment E12-16-010A, 
\href{https://www.jlab.org/exp_prog/proposals/16/PR12-16-010A.pdf}{Proposal to JLab PAC 44 (2016)}. 

\item{[9]} 
R. Gothe {\em  et al.}, Experiment E12-09-003, \href{https://www.jlab.org/exp_prog/proposals/09/PR12-09-003.pdf}{Proposal to JLab PAC 34 (2009)}.

\item{[10]} 
K. H. Hicks {\em  et al.}, Experiment E45,  
\href{http://j-parc.jp/researcher/Hadron/en/pac_1207/pdf/P45_2012-3.pdf}{Proposal to J-PARC PAC 15 (2012)}.

\item{[11]} 
S. Aikawa (J-PARC E31 Collaboration), Phys. Lett. B {\bf 837} (2023) 137637. 

\item{[12]}
K. Tanida, {\em  et al.}, Experiment E72, \href{http://j-parc.jp/researcher/Hadron/en/pac_1801/pdf/P72_2018-9.pdf}{Proposal to J-PARC PAC 25 (2018)}.

\item{[13]}
R. L. Workman {\em et al.} (Particle Data Group), Prog. Theor. Exp. Phys. {\bf 2022} (2022) 083C01. 

\item{[14]} 
L. Guo {\em et al.}, Experiment E12-12-008, \href{https://www.jlab.org/exp_prog/proposals/12/PR12-12-008.pdf}{Proposal to JLab PAC 39 (2012)}. 

\item{[15]} 
A. Gal, E. V. Hungerford, and D. J. Millener, Rev. Mod. Phys. {\bf 88} (2016) 035004.

\item{[16]} 
J. Rowley {\em et al.} (CLAS Collaboration), Phys. Rev. Lett. {\bf 127} (2021) 272303.

\item{[17]} 
K. Miwa {\em et al.} (J-PARC E40 Collaboration), Phys. Rev. Lett. {\bf 128} (2022) 072501.

\item{[18]} 
K. Miwa {\em et al.} (J-PARC E40 Collaboration), Phys. Rev. C {\bf 104} (2021) 045204.

\item{[19]} 
T. Nanamura {\em et al.}, Prog. Theor. Exp. Phys. {\bf 2022} (2022) 093D01.

\item{[20]}
F. Ameli {\em et al.}, Eur. Phys. J. Plus {\bf 137} (2022) 958. 

\item{[21]}
T. Takahashi, R. Honda, Y. Igarashi and H. Sendai, IEEE Trans. Nucl. Sci. {\bf 70} (2023) 922.

\end{description}

\stepcounter{count}
\clearpage

\phantomsection
\addcontentsline{toc}{section}{
{\bf Kaonic atoms at the DA$\Phi$NE Collider in Italy: a strangeness Odyssey} \\
C.~Curceanu on behalf of the SIDDHARTA-2 collaboration}

\titl{Kaonic atoms at the DA$\Phi$NE Collider in Italy:
a strangeness Odyssey}

\name{
Catalina Curceanu on behalf of the SIDDHARTA-2 collaboration$^{1}$
}

\adr{
$^1$ Laboratori Nazionali di Frascati, INFN, Via E. Fermi 54, I-00044 Frascati(RM), Frascati, Italy \\
}

Kaonic atoms represent an ideal tool to study the low-energy regime of Quantum Chromodynamics (QCD) in
the strangeness sector, which cannot be described with a perturbative approach. They enable to directly access the $K^{-}$N interaction at threshold, without the need of an extrapolation as in the case of scattering experiments, since the relative energy between the kaon and the nucleus is already at the level of few keV. 

\par Starting in the late 1990s, a new era of kaonic atoms studies was initiated by the KpX experiment at KEK [1] in Japan, followed by the DEAR [2] and SIDDHARTA [3] experiments at DA$\Phi$NE.
The SIDDHARTA collaboration performed the most precise measurement of kaonic hydrogen and the first exploratory study of kaonic deuterium [4]. Moreover, the kaonic helium 4 and 3 transitions
to the 2p level were measured, for the first time in gas in He$^{4}$ [5,6] and for the first time ever in He$^{3}$ [7].
To achieve the challenging kaonic deuterium measurement goal, the SIDDHARTA-2 experiment was designed to perform high precision X-ray spectroscopy in the high radiation environment of a particle collider. For this reason, the SIDDHARTA-2 collaboration developed innovative fast and very precise Silicon Drift Detectors (SDD) [8] for X-ray spectroscopy and a series of other detectors, such as a kaon trigger and three veto systems to reduce the background [9].

The initial optimization of the experimental apparatus was accomplished through the measurement of kaonic helium 4 transitions to the 2p level [10] which have yields about 100 times higher than the expected transitions on the 1s level in kaonic deuterium. The result turned out to be the most precise measurement in a gas target, providing a new experimental input for the theoretical models. Using the gaseous target allowed to observe and measure, also, for the first time, M-lines in kaonic helium, in particular the M$_\beta$, M$_\gamma$, and M$_\delta$ transitions [11].

In 2023, before starting the kaonic deuterium measurement, a refined optimization and performance check of the experimental apparatus was realized measuring, for the first time, high-n X-ray transition in kaonic neon with the unprecedented precision below 1 eV. This result not only provide new data for the kaonic atoms database, setting a precedent for future high-precision kaonic atomic measurements, but also confirm the use of these type of measurements for extracting the charged kaon mass, a real puzzle presently, since two very precise measurements already exist, but they are not compatible with each other.

The kaonic helium-4 and the kaonic neon results demonstrated the excellent performance of the SIDDHARTA-2 setup in terms of X-ray detection accuracy and background suppression capability, paving the way for the kaonic deuterium measurement. The kaonic deuterium data taking campaign began in May 2023, aiming to collect data for a total integrated luminosity of about 800 pb$^{-1}$. The first two phases (Run1 and Run2) have been completed in 2023, by collecting an integrated luminosity of about 500 pb$^{-1}$. The preliminary analysis looks very promising. The third phase  was initiated in February 2024 and is ongoing. 
This measurement, long awaited by the scientific community, is fundamental for a better and more precise understanding of strong interaction. 
Moreover, the SIDDHARTA-2 collaboration put forward a proposal for future precision measurements of kaonic atoms charting the periodic table: EXKALIBUR proposal [12], by using a series of innovative technologies and methods.

The experiments at the DA$\Phi$NE collider represent a unique opportunity in the world to unlock the secrets of the QCD in the strangeness sector and contribute to a better understanding of the role of strangeness in the Universe, from nuclei to the stars.

\vfill  

\noindent{\bf References }
\begin{description}
\setlength\itemsep{-3pt}
\item{[1]} M. Iwasaki, et al. Phys. Rev. Lett.{\bf 78}(1997) 3067.
\item{[2]} DEAR Collaboration, Phys. Rev. Lett.  {\bf 94} (2005) 212302 .
\item{[3]} M. Bazzi et al., Phys.
Lett. B {\bf 704} (2011)  113.

\item{[4]}  M. Bazzi et al.,  Nuclear Physics A {\bf 907} (2013) 69.
\item{[5]} M. Bazzi et al, Phys. Lett. B  {\bf  681} (2009) 310.
\item{[6]} M. Bazzi et al, Phys. Lett. B {\bf 714}(2012) 40.
\item{[7]} M. Bazzi et al, Phys. Lett. B {\bf 697 }(2011) 199 .

\item{[8]} M. Miliucci, M. Iliescu, F. Sgaramella  et al., Meas. Sci. Technol. {\bf 33} (9)  (2022) 95502.

\item{[9]} C. Curceanu et al., Rev. Mod. Phys. {\bf 91} (2019) 025006.

\item{[10]}  D. Sirghi, F. Sirghi, F. Sgaramella et al., J. Phys. G Nucl. Part. Phys. {\bf 49} (5) (2022)  55106.

\item{[11]} F. Sgaramella et al.,  J. Phys. G: Nucl. Part. Phys. {\bf 51} (2024)  055103.

\item{[12]} C. Curceanu et al.,  Front. Phys. 11:1240250 (2023).

\end{description}

\stepcounter{count}
\clearpage

\phantomsection
\addcontentsline{toc}{section}{
{\bf Hadron physics with antiproton-induced reactions at PANDA} \\
S.~Diehl for the PANDA collaboration}

\titl{Hadron physics with antiproton-induced reactions at PANDA}


\name{Stefan Diehl$^{1}$ for the PANDA collaboration}


\adr{$^1$ Justus Liebig University Giessen, 35390 Giessen, Germany}


Antiproton-induced reactions in the GeV energy regime offer a lot of unique opportunities to study QCD in the confinement region and the underlying fundamental symmetries. The PANDA experiment at FAIR [1] will use stored anti-protons from the High Energy Storage Ring (HESR) with a momentum range between 1.5 GeV/c and 15 GeV/c, interacting with a cluster jet and pellet targets ($\bar{p}p$) and foils ($\bar{p}A$). The luminosity at the peak intensity will be $2\cdot10^{32} cm^{-2}s^{-1}$ with a resolution of $\delta p/p = 2\cdot10^{-4}$ and an interaction rate of $2\cdot 10^{-7} s^{-1}$. In high-resolution mode, a beam resolution of $\delta p/p = 4 \cdot 10^{-5}$ at a luminosity of $2\cdot10^{31} cm^{-2}s^{-1}$ can be achieved.
PANDA is designed as a modular multi-purpose device with excellent forward acceptance and resolution and a full coverage up to backward angles, a wide dynamic range of particle momenta from 0.1 GeV/c to 8 GeV/c, as well as an excellent momentum measurement in a magnetic field ($\Delta p/p \approx$ 1\%). It will provide particle ID with a wide momentum range, electromagnetic calorimetry, and high-resolution vertex detection.

The PANDA physics program (see also [2], [3]) reaches from spectroscopy, over strangeness physics and nucleon structure, up to hadrons in nulei and hypernuclei.
In the field of hadron spectroscopy, the search for exotic hadrons like glueballs and hybrids is a key part of the program. Due to the high resolution of the HESR beam, precise energy scans can be performed to measure the line-shape of the hadronic states, which is directly related to their origin and sheds light on their nature.
In the field of strangeness physics, PANDA can produce hyperons in strong interactions with a high production rate of several thousand to million events per day, depending on the final state. It can therefore be seen as a hyperon factory and will provide precise studies of hyperons and access to potential CP violations.
In the field of nucleon structure, PANDA can access electromagnetic form factors in the time-like region with higher precision and in an extended kinematic range, compared to previous and ongoing experiments like BES3. It is also expected to provide the first measurements of these form factors in the unphysical region.
Furthermore, Transition Distribution Amplitudes (TDA) as well as Generalized Distribution Amplitudes (GDAs) can be accessed in annihilation reactions. Based on the Drell-Yan process, also Transverse Parton Distribution functions (TMDs) become accessible.

Altogether, PANDA is a unique facility with a broad physics program employing anti-protons. It puts precision spectroscopy alongside a high discovery potential with the application of modern detector technologies.
\\
\\
*The work is supported by BMBF.

\vfill  

\noindent{\bf References }
\begin{description}
\setlength\itemsep{-3pt}

\item{[1]} PANDA web page, \url{https://panda.gsi.de/}

\item{[2]} PANDA collaboration, Physics Performance Report for: PANDA, \url{https://panda.gsi.de/oldwww/archive/public/panda_pb.pdf}

\item{[3]} Barucca, G., Davi, F., Lancioni, G. et al. PANDA Phase One. Eur. Phys. J. A 57, 184 (2021), \url{https://doi.org/10.1140/epja/s10050-021-00475-y}

\end{description}

\stepcounter{count}
\clearpage

\phantomsection
\addcontentsline{toc}{section}{
{\bf $N$-$\phi$ interaction from lattice QCD and implication from combined analysis with femtoscopic data} \\
T.~Doi (HAL QCD Collaboration)}

\titl{$N$-$\phi$ interaction from lattice QCD and implication from combined analysis with femtoscopic data}

\name{
Takumi~Doi$^{1}$ (HAL QCD Collaboration)
}

\adr{
$^1$ Interdisciplinary Theoretical and Mathematical Sciences Program (iTHEMS), RIKEN, Wako 351-0198, Japan
}


The $N$-$\phi$ interaction has been subject of theoretical and experimental
investigations for decades,
since it is a good probe to extract the medium effects on hadron properties
and the interaction between a color-dipole and a nucleon.
Possible existence of a $N$-$\phi$ and/or nucleus-$\phi$ bound state
has been discussed as well.

In this talk,
we present the studies of the $N$-$\phi$ interaction
(1) from the lattice QCD simulation and
(2) from the combined analysis of lattice QCD and femtoscopic data.

In the first part of the talk,
we show the lattice QCD results of the $N$-$\phi$ interaction
in the $J=3/2$ channel. In this channel, the effects of open channels
such as $\Lambda K, \Sigma K$ are suppressed by the $D$-wave,
and the $N$-$\phi$ single channel analysis is performed from lattice QCD.
The simulations were performed in ($2+1$)-flavor lattice QCD
with nearly physical quark masses, $m_\pi = 146$ MeV, using the HAL QCD method [1,2].
The $N$-$\phi$ potential appears to be a combination of a  short-range attractive core 
and a long-range attractive tail, the latter of which is found to be
consistent with the two-pion exchange (TPE).
The resultant scattering length and effective range for $m_{\pi}=$ 146.4 MeV are $ a^{(3/2)}_0=-1.43(23)_{\rm stat.}\left(^{+36}_{-06}\right)_{\rm syst.} {\rm fm}$ and  $ r^{(3/2)}_{\rm eff}=2.36(10)_{\rm stat.}\left(^{+02}_{-48}\right)_{\rm syst.} {\rm fm}$, respectively [3].

In the second part of the talk,
we present the results of $N$-$\phi$ interaction in the $J=1/2$ channel,
obtained from the combined analysis of lattice QCD data and femtoscopic data.
The direct lattice QCD simulation in the $J=1/2$ channel is challenging
due to the ($S$-wave) coupled channel effects of $\Lambda K, \Sigma K$,
while the femtoscopic study cannot perform the spin projection at present.
However, by a novel combination of lattice QCD data ($J=3/2$) and
the femtoscopic data ($J=1/2, 3/2$ average)
of $N$-$\phi$ correlation function measured by ALICE,
we can extract the $N$-$\phi$ interaction in the $J=1/2$ channel.
The corresponding scattering length and effective range are
$({\rm Re}\, a_0^{(1/2)}, {\rm Im}\, a_0^{(1/2)})
=\left(
1.54\left(^{+0.53}_{-0.53}\right)_{\rm stat.}
\left(_{-0.16}^{+0.09}\right)_{\rm syst.},
0.00\left(_{-0.35}^{+0.00}\right)_{\rm stat.}
\left(_{-0.16}^{+0.00}\right)_{\rm syst.}
\right)$~fm
and
$({\rm Re}\, r_{\rm eff}^{(1/2)}, {\rm Im}\, r_{\rm eff}^{(1/2)}) =$ \\
$
\left(
0.39\left(^{+0.09}_{-0.09}\right)_{\rm stat.}
\left(^{+0.02}_{-0.03}\right)_{\rm syst.},
0.00(^{+0.00}_{-0.04})_{\rm stat.}(^{+0.00}_{-0.02})_{\rm syst.}
\right)$~fm,
respectively.
The results imply the appearance of a $N$-$\phi$ bound state
with an estimated binding energy in the range of $12.8-56.1$ MeV [4].
%
%
%
%
%
%
Finally, future prospects from lattice QCD simulations at the physical point
are presented [5].

\vfill  

\noindent{\bf References }
\begin{description}
\setlength\itemsep{-3pt}
\item{[1]} N.~Ishii, S.~Aoki and T.~Hatsuda, Phys. Rev. Lett. {\bf 99} (2007) 022001.
\item{[2]} S.~Aoki and T.~Doi, Front. Phys. {\bf 8} (2020) 307.
\item{[3]} Y.~Lyu, T.~Doi, T.~Hatsuda, Y.~Ikeda, J.~Meng, K.~Sasaki and T.~Sugiura, Phys. Rev. D {\bf 106} (2022) 074507.
\item{[4]} E.~Chizzali, Y.~Kamiya, R.~D.~Grande, T.~Doi, L.~Fabbietti, T.~Hatsuda and Y.~Lyu, Phys. Lett. B {\bf 848} (2024) 138358.
\item{[5]}
T.~Aoyama, T.~M.~Doi, T.~Doi, E.~Itou, Y.~Lyu, K.~Murakami and T.~Sugiura, arXiv:2406.16665 [hep-lat].

\end{description}

\stepcounter{count}
\clearpage

\phantomsection
\addcontentsline{toc}{section}{
{\bf Flavor probe of cosmology} \\
M.~Endo}

\titl{Flavor probe of cosmology}

\name{
Motoi Endo$^{1,2}$
}

\adr{
$^1$ KEK Theory Center, Tsukuba, Ibaraki 305--0801, Japan \\
$^2$ Graduate Institute for Advanced Studies, SOKENDAI, Tsukuba, Ibaraki 305--0801, Japan
}

The Standard Model of particle physics was established by the success of the Higgs boson discovery and has been the best theory to describe the Universe.
Nevertheless, it is widely known that the Standard Model is not the theory of everything, in particular, cannot explain phenomena that happened in the early stages of the Universe such as generations of the baryon asymmetry of the Universe.
The asymmetry, quantified as a baryon-to-photon ratio, has been determined by the measurement of the Cosmic Microwave Background as well as the Big Bang Nucleosynthesis to be $\eta \equiv n_B/n_\gamma = (6.12 \pm 0.04) \times 10^{-10}$~[1].
Inflation indicates that the asymmetry should be generated dynamically in the early stage of the Universe, necessitating a mechanism of baryogenesis.

In this proceeding, we will review recent studies on the feasibility of testing baryogenesis. 
In general, physics beyond the Standard Model can contribute to flavor measurements via quantum effects and be observed as inconsistencies between experimental data and Standard Model predictions.
The measurements enable us to probe physics at energy scales higher than beam collision energy and may have a possibility to test baryogenesis mechanisms.  
In the following, we will overview two baryogenesis scenarios. 

The first scenario we consider is leptogenesis, which is related to the unresolved origin of neutrino masses. 
In the type-I see-saw model, Majorana neutrinos are introduced.
Their out-of-equilibrium decay can generate the baryon asymmetry of the Universe.
Recently, a model with 3 quasi-degenerate Majorana spectra was studied~[2].
Heavy neutrinos are effectively coupled with electroweak gauge bosons, contributing to charged lepton-flavor violations (cLFV).
It was shown that future cLFV experiments, such as COMET, Mu2e, and PRISM/PRIME, can probe significant parameter regions of the model.

Let us next consider the electroweak baryogenesis. 
In the Standard Model, the Higgs boson is too heavy to realize the first-order phase transition strong enough for the scenario.
In the recent study~[3], a light scalar particle is introduced to modify the Higgs potential in high temperatures. 
The particle is coupled with the top quark, contributing to the $K$ meson decay. 
KOTO experiment is powerful enough to probe such a light particle, which is produced at the decay and behaves as a missing momentum in the detector.
It was argued that significant parameter regions can be covered in future.

New light particles may also be produced at the beam target. 
Ref.~[4] studied the possibility of using KOTO experiment to search for axionlike particles.
The study has shown that the experiment can probe new regions in the parameter space even without changing the main analysis steps.

In conclusion, current and future flavor experiments provide an opportunity to test a class of baryogenesis mechanisms.

-----------

\vfill  

\noindent{\bf References }
\begin{description}
\setlength\itemsep{-3pt}
\item{[1]} S.~Navas \textit{et al.} [Particle Data Group], Phys. Rev. D \textbf{110} (2024) 030001.
\item{[2]} A.~Granelli, J.~Klari\'c and S.~T.~Petcov, Phys. Lett. B \textbf{837} (2023), 137643.
\item{[3]} H.~Davoudiasl, Phys. Rev. D \textbf{103} (2021) no.8, 083534.
\item{[4]} Y.~Afik, B.~D\"obrich, J.~Jerhot, Y.~Soreq and K.~Tobioka, Phys. Rev. D \textbf{108} (2023) no.5, 5.
\end{description}

\stepcounter{count}
\clearpage

\phantomsection
\addcontentsline{toc}{section}{
{\bf $\Lambda$'s beta decay in a nucleus} \\
M.~Fujita$^*$, K.~Kamada, C.~Seong, H.~Tamura}

\titl{$\Lambda$'s beta decay in a nucleus}

\name{
M. Fujita$^{1}$, 
K. Kamada$^{2}$, 
C. Seong$^{2}$, 
H. Tamura$^{2,1}$
}

\adr{
$^1$ ASRC, JAEA, Tokai-mura, 319-1195, Japan \\
$^2$ Tohoku Univ., Sendai, 980-8578, Japan
}


The properties of nucleons in the free space are investigated well as fundamental information in nuclear physics. 
Nucleons have been treated as if their properties are not changed even in nucleus.
However, since nucleons are composite systems made of quarks and gluons, so their properties can change depending on their surrounding environment.
Therefore, understanding the properties of baryons in nuclei is crucial for a realistic comprehension of nuclear physics.

\begin{wrapfigure}{r}{8.5cm}
		\vspace{-0.7\baselineskip}
		\begin{center}
		\includegraphics[width=8cm]{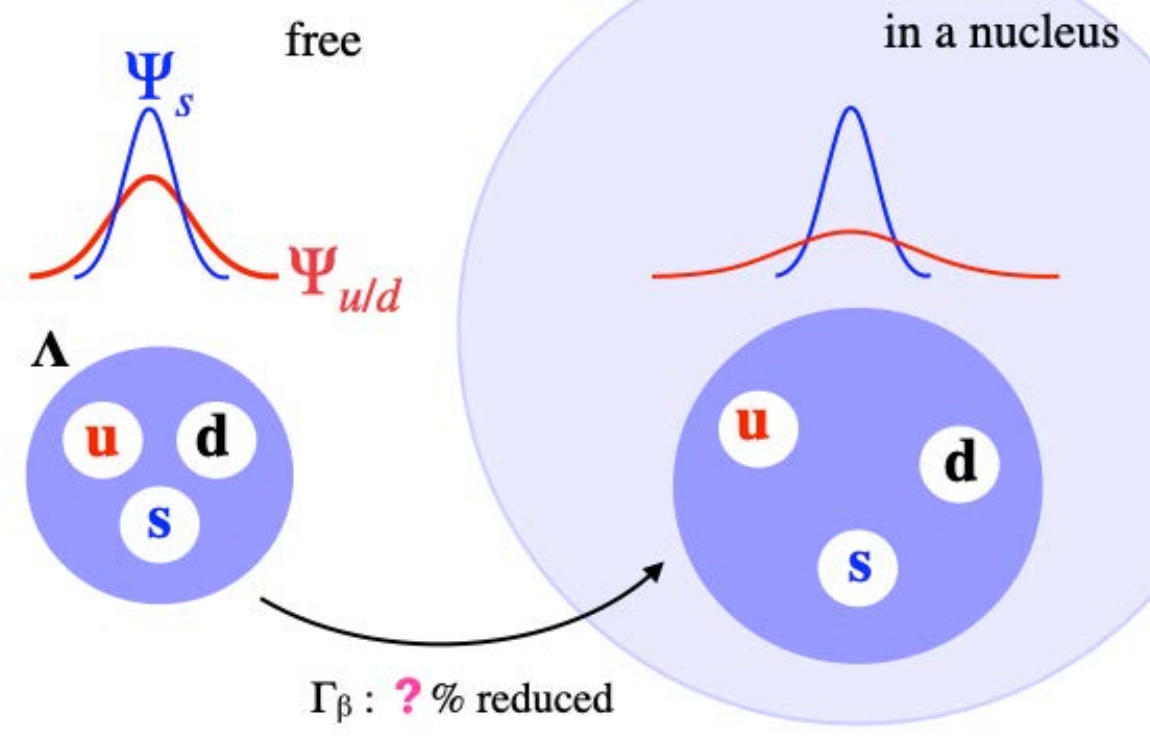}
		\caption{$\Lambda$ in free/in a nucleus}
		\label{pic1}
		\end{center}
		\vspace{-1.3\baselineskip}
 \end{wrapfigure}

Assuming the quark meson coupling (QMC) model, the wave functions of up and down quarks in baryons (nucleons and hyperons) may spread in nuclei due to the influence of the meson field.
In contrast, the wave function of the strange quark changes little within the nucleus (see Fig.\ref{pic1}). 
We are exploring methods to investigate the structure of baryons in nuclei using a strange quark. 
One proposed method is to measure the beta decay rate of $\Lambda$ particles in nuclei.
In the beta decay of $\Lambda$ particles, a strange quark decays into an up quark. This decay rate depends on the overlap of the wave functions of the up and strange quarks. 
Theoretical calculations using the QMC model suggest that the $\Lambda$ beta decay rate could decrease by up to 20$\%$ in nuclei because of their smaller overlap in nuclei as shown in Fig.\ref{pic1} right~[1]. 
Therefore, measuring the decay rate change could offer insights into the spatial modification of up quarks in hyperons inside nuclei.

To achieve this measurement, there are two important challenges;
firstly, it is necessary to accurately evaluate changes in the beta decay rate that occur independently of the spreading of the $u$ quark's wave function.
The quenching effect of the Gamow-Teller beta-decay rates, or the axial vector coupling constant, $g_A$, is known to arise from nuclear many-body effects and meson exchange current effects, and is observed in ordinary nuclei~[2].
This quenching effect is more pronounced in heavier nuclei, leading to a significant suppression of beta decay. 
Additionally, the effects of Pauli suppression in the $\Lambda$'s beta decay should be theoretically evaluated, similarly to the case of the $\Lambda$'s mesonic decays in which the Pauli effects largely suppress these decays.

Secondly, it is necessary to consider the feasibility such experiments.
To determine the beta decay rate, we measure the beta decay branching ratio ($BR_{\beta}$) and the lifetime of the hypernucleus. 
The most challenging part is measuring the $BR_{\beta}$. 
$\Lambda$ particles primarily decay via mesonic/non-mesonic decays, with the branching ratio of the beta decay being very small, on the order of $10^{-4}$ (see Table~\ref{tab1}). 
Therefore, suppressing huge backgrounds from the other decay modes is an experimental challenge.

\begin{table}[h]
\centering
\caption{The $\Lambda$'s decay branching ratio in the free space~[3].}
\vspace{0.7\baselineskip}
\begin{tabular}{ccl}
\hline
mode       & \multicolumn{1}{l}{}    & \multicolumn{1}{c}{fraction ($\Gamma_i/\Gamma$)} \\ \hline
$\Gamma_1$ & $p\pi^-$                & 64.1\%                                           \\
$\Gamma_2$ & $n\pi^0$                & 35.9\%                                           \\
$\Gamma_3$ & $n\gamma$               & $8.3\times10^{-4}$                               \\
$\Gamma_4$ & $p\pi^-\gamma$          & $8.5\times10^{-4}$                               \\
$\Gamma_5$ & $p e^-\bar{\nu_e}$      & $8.34\times10^{-4}$                              \\
$\Gamma_6$ & $p\mu^-\bar{\nu_{\mu}}$ & $1.51\times10^{-4}$                              \\ \hline
\end{tabular}
\label{tab1}
\end{table}

\begin{figure}[h]
	\begin{center}
	\includegraphics[height=75mm]{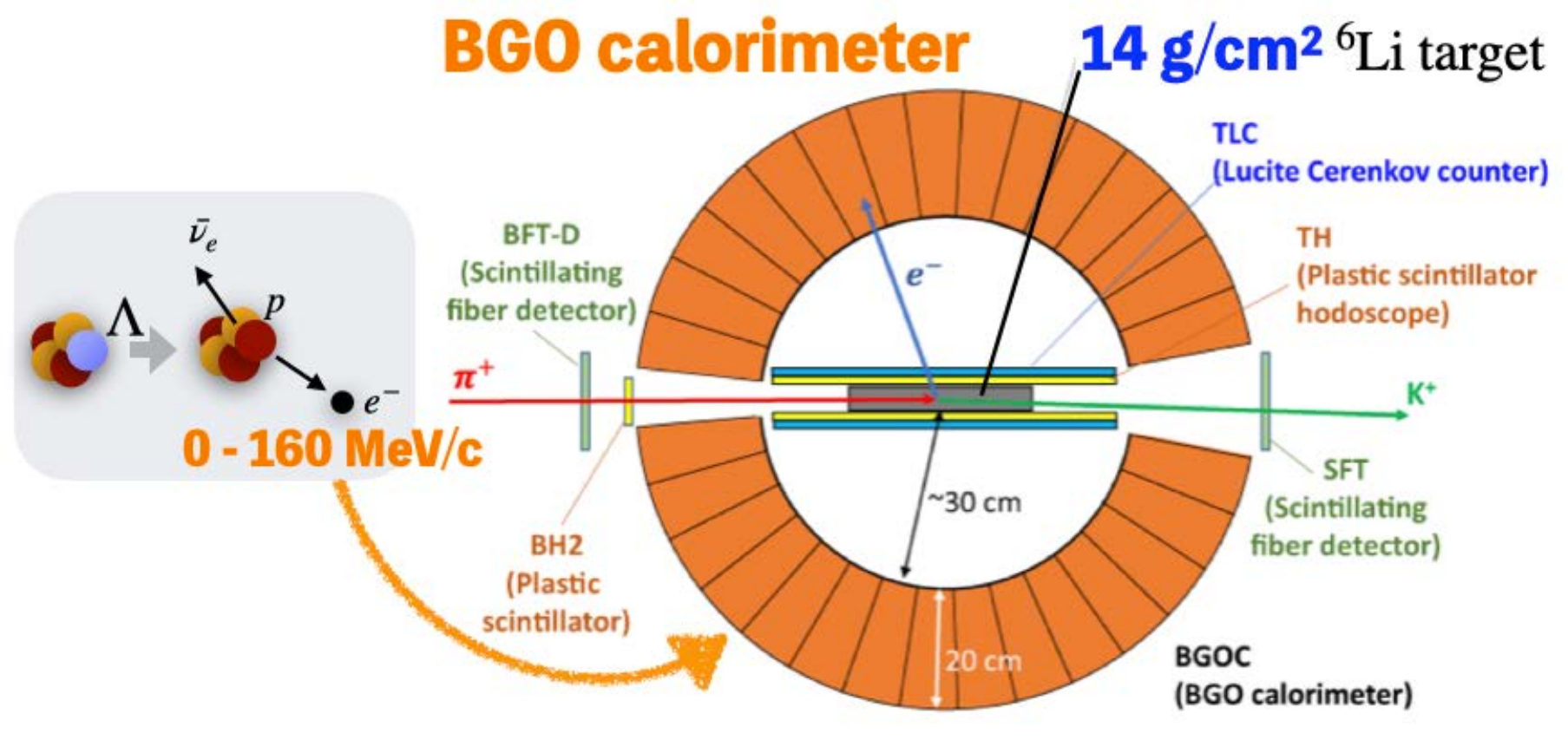}
	\caption{The detector setup around the target~[4].}
	\label{pic2}
	\end{center}
	\vspace{-1.3\baselineskip}
\end{figure}

We have devised an experimental setup a the segmented BGO calorimeter surrounding the target region, as shown in Fig.\ref{pic2}.
We are currently designing the BGO calorimeter, developing cluster analysis methods, and designing additional particle identification detectors around the target.
 It has been shown that the background can be sufficiently suppressed with this experimental setup using GEANT4 simulation~[4].
To improve the accuracy of the reaction processes in simulation, we plan to measure the response of the BGO detector to pions and neutrons~[5].

Further consideration is necessary regarding which $\Lambda$ hypernucleus to measure, but one candidate is $^5_\Lambda$He.
$^5_\Lambda$He may allow for precise few-body calculations. 
In this case, $^5_\Lambda$He will be produced in the $^6$Li$(\pi^+, K^+)^6_\Lambda$Li, $^6_\Lambda$Li$\rightarrow$$^5_\Lambda$He$+p$ using a 1.05 GeV/$c$ $\pi^+$ beam at J-PARC K1.1 beamline.
Measurements of $^5_\Lambda$He's lifetime, $\tau$, and beta decay branching ratio are estimated to require 5 days (for $\tau$) + 60 days (for $BR_{\beta}$), using a $\pi^+$ beam of $3\times10^7$/spill~[6]. 

This experiment is challenging, but we believe it will be a milestone in understanding nucleon structure. We welcome further theoretical and experimental discussions to realize this experiment.

\vfill  

\noindent{\bf References }
\begin{description}
\setlength\itemsep{-3pt}
\item{[1]}~P. A. M. Guichon, A. W. Thomas, Phys. Lett. B{\bf 773} (2017) 332.
\item{[2]}~W.T. Chou $et~al.$, Phys. Rev. C{\bf 47} (1993) 163.
\item{[3]}~S. Navas et al. (Particle Data Group), to be published in Phys. Rev. D 110, 030001 (2024). 
\item{[4]}~K. Kamada, Master thesis, Tohoku University (2022).
\item{[5]}~C. Seong, Master thesis, Tohoku University (2024). 
\item{[6]}~K. Kamada, M. Fujita, H. Tamura,  Letter of Intent for J-PARC 50 GeV Synchrotron, \\
``Modification of baryon structure in nuclear matter studied from beta-decay rate of a $\Lambda$ hypernucleus", (2021)\\ 

\end{description}



\stepcounter{count}
\clearpage

\phantomsection
\addcontentsline{toc}{section}{
{\bf Baryon-baryon interactions in\\ covariant chiral effective field theory} \\
Z.W.~Liu, Y.~Xiao, J.X.~Lu and L.S.~Geng$^*$}

\titl{Baryon-baryon interactions in\\ covariant chiral effective field theory}

\name{
Zhi-Wei Liu$^1$, Yang Xiao$^{2,1}$, Jun-Xu Lu$^1$
and Li-Sheng Geng$^{1,3,4,5}$
}

\adr{
$^1$ School of Physics, Beihang University, Beijing 102206, China\\
$^2$ School of Space and Environment, Beihang University, Beijing 102206, China\\
$^3$ Peng Huanwu Collaborative Center for Research and Education, Beihang University, Beijing 100191, China\\
$^4$ Beijing Key Laboratory of Advanced Nuclear Materials and Physics, Beihang University, Beijing 102206, China \\
$^5$ Southern Center for Nuclear-Science Theory (SCNT), Institute of Modern Physics, Chinese Academy of Sciences, Huizhou 516000, China
}

Understanding the non-perturbative strong interaction is a longstanding goal in particle and nuclear physics. In the standard model of particle physics, quantum chromodynamics (QCD) serves as the fundamental theory for describing strong interactions. However, due to the non-perturbative nature of QCD in the low-energy regime, rigorous perturbative calculations are unfeasible for the low-energy baryon-baryon interactions.\\ 

In a groundbreaking work in 1990, Weinberg suggested utilizing chiral effective field theory (ChEFT) to describe the nuclear force~[1]. It has been extremely successful and has become the \textit{de facto} standard input in modern \textit{ab initio} nuclear studies~[2]. However, Weinberg's chiral nuclear force met several challenges, including relatively slow convergence in the chiral expansions and the incompatibility with renormalization group invariance. Recently, we developed a new relativistic scheme to construct the nucleon-nucleon interaction in the framework of covariant ChEFT. The relativistic chiral interaction is formulated up to the next-to-next-to-leading order (N$^2$LO) with covariant power counting and a manifestly Lorentz invariant chiral Lagrangian~[3]. We found that the N$^2$LO relativistic chiral nuclear force can describe the $np$ phase shifts with an accuracy comparable to the non-relativistic next-to-next-to-next-to-leading order (N$^3$LO) results. Additionally, it exhibits a better renormalization group invariance (particularly in the puzzling $^3P_0$ channel)~[5] and enhanced theoretical self-consistency in the $^1S_0$ channel~[6].\\

As natural extensions of the nuclear force to the $u$, $d$, and $s$ flavor sector, hyperon-nucleon (YN) and hyperon-hyperon (YY) interactions are crucial for understanding the role of strange quarks in particle and nuclear physics. Based on the covariant ChEFT, we have constructed relativistic chiral YN and YY interactions with strangeness $S=-1$ to $-4$ at the leading order. Comparing the results with the latest experimental data and lattice QCD simulations, we verified the reliability of the constructed interactions. For the $S=-1$ system, the 12 low-energy constants were determined by fitting the 36 low-energy YN scattering data, and a quite satisfactory description of these scattering data is obtained, comparable to the next-to-leading order (NLO) heavy baryon approach~[7]. In addition, we found that although the LO relativistic YN interaction is only constrained by the low-energy data, it can describe the latest J-PARC E40 data in the high-energy region reasonably well~[8]. For the $S=-2$ system, we determined the YN and YY $S$-wave interactions by fitting the state-of-the-art lattice QCD simulations, considering all the coupled channels~[9]. Based on the strict SU(3) flavor symmetry, we predicted an attractive interaction between $\Sigma\Sigma(I=2)$, but its strength is insufficient to generate a bound state. With the so-obtained strong interactions and considering quantum statistical effects, the Coulomb interaction, and all coupled channels, we computed the $\Lambda\Lambda$ and $\Xi^-p$ femtoscopic correlation functions with a spherical Gaussian source. The agreement between the theoretical descriptions and the ALICE measurements was very good, demonstrating the reliability of the relativistic chiral YN and YY interactions with $S=-2$. For the $S=-3,-4$ systems, the covariant ChEFT describes the $S$-wave phase shifts well and predicts the phase shifts for $D$-wave and mixing angle~[10]. Furthermore, significant SU(3) flavor symmetry breaking was found among the baryon-baryon interactions of different strangeness, such as the gradual weakening of the attractiveness of the interactions from $NN(I=1)$ to $\Sigma N(I=3/2)$, $\Sigma\Sigma(I=2)$, $\Xi\Sigma(I=3/2)$, and $\Xi\Xi(I=1)$, which all belong to the same SU(3) irreducible representation ``27''. Finally, we proposed that a quantitative examination of SU(3) flavor symmetry and its breaking can be achieved by studying $pp$, $\Sigma^+p$, $\Sigma^+\Sigma^+$, $\Xi^-\Sigma^-$, and $\Xi^-\Xi^-$ femtoscopic correlation functions.\\

Based on the success of covariant ChEFT in describing baryon-baryon interactions, we plan to construct relativistic chiral YN and YY interactions up to NLO and even N$^2$LO, incorporating the latest experimental achievements, in particular, the differential cross-section data from J-PARC~[11]. In addition, we will construct relativistic chiral baryon-antibaryon interactions~[12]. On the other hand, based on the relativistic chiral potentials and the relativistic Brueckner-Hartree-Fock theory, we plan to investigate dense nuclear matter~[13], offering new insights into baryon-baryon interactions in the nuclear medium.

-----------

\vfill  

\noindent{\bf References }
\begin{description}
\setlength\itemsep{-3pt}
\item{[1]} S. Weinberg, Phys. Lett. B {\bf 251} (1990) 288-292. \href{https://doi.org/10.1016/0370-2693(90)90938-3}{Doi:10.1016/0370-2693(90)90938-3.}
\item{[2]} R. Machleidt, Few Body Syst. {\bf 64} (2023) 77. \href{https://doi.org/10.1007/s00601-023-01857-2}{Doi:10.1007/s00601-023-01857-2.}
\item{[3]} J.X. Lu, C.X. Wang, Y. Xiao, and et al., Phys. Rev. Lett. {\bf 128} (2022) 142002. \href{https://doi.org/10.1103/PhysRevLett.128.142002}{Doi:10.1103/PhysRevLett.128.142002}
\item{[5]} C.X. Wang, L.S. Geng, and B.W. Long, Chin. Phys. C {\bf 45} (2021) 054101. \href{https://doi.org/10.1088/1674-1137/abe368}{Doi:10.1088/1674-1137/abe368}
\item{[6]} X.L. Ren, C.X. Wang, K.W. Li, L.S. Geng, and J. Meng,
Chin.Phys.Lett. {\bf 38} (2021) 062101 \href{https://iopscience.iop.org/article/10.1088/0256-307X/38/6/062101}
{Doi:10.1088/0256-307X/38/6/062101}

\item{[7]} K.W. Li, X.L. Ren, L.S. Geng, and et al., Chin. Phys. C {\bf 42} (2018) 014105. \href{https://doi.org/10.1088/1674-1137/42/1/014105}{Doi:10.1088/1674-1137/42/1/014105}
\item{[8]} J. Song, Z.W. Liu, K.W. Li, and et al., Phys. Rev. C {\bf 105} (2022) 035203. \href{https://doi.org/10.1103/PhysRevC.105.035203}{Doi:10.1103/PhysRevC.105.035203}
\item{[9]} Z.W. Liu, K.W. Li, and L.S. Geng, Chin. Phys. C {\bf 47} (2023) 024108. \href{https://doi.org/10.1088/1674-1137/ac988a}{Doi:10.1088/1674-1137/ac988a}
\item{[10]} Z.W. Liu, J. Song, K.W. Li, and et al., Phys. Rev. C {\bf 103} (2021) 025201. \href{https://doi.org/10.1103/PhysRevC.103.025201}{Doi:10.1103/PhysRevC.103.025201}
\item{[11]} K. Miwa, and et al. (J-PARC E40 Collaboration), Phys. Rev. Lett. {\bf 128} (2022) 072501. \href{https://doi.org/10.1103/PhysRevLett.128.072501}{Doi:10.1103/PhysRevLett.128.072501}
\item{[12]} Y. Xiao, J.X. Lu, and L.S. Geng, arXiv:2406.01292. \href{https://arxiv.org/abs/2406.01292v1}{https://arxiv.org/abs/2406.01292v1}
\item{[13]} W.J. Zou, J.X. Lu, P.W. Zhao, and et al., Phys. Lett. B {\bf 854} (2024) 138732. \href{https://doi.org/10.1016/j.physletb.2024.138732}{Doi:10.1016/j.physletb.2024.138732}
\end{description}

\stepcounter{count}
\clearpage

\phantomsection
\addcontentsline{toc}{section}{
{\bf Hadron Physics at EIC} \\
T.~Gunji}

\titl{Hadron Physics at EIC}

\name{
T. Gunji$^{1}$
}

\adr{
$^1$ Center for Nuclear Study, Graduate School of Science, the University of Tokyo, Hongo Campus. 7-3-1, Hongo, Bunkyo-ku, Tokyo, 113-0033, Japan
}


The Electron-Ion Collider (EIC) is a highly advanced facility that will have 
the capability to collide high-energy electron beams with high-energy 
polarized protons, polarized beams of light-ions (deuterons and $^3$He) and heavy-ion beams. 
The EIC is the major project in the field of nuclear physics and is the only new collider 
to be designed and built in the world in the next decade.
EIC will be hosted at the Brookhaven National Laboratory (BNL) and 
will start its operation in 2030s. 
The main design requirements are~[1]:
\begin{itemize}
\item{Highly polarized electron ($\sim$80\%) beams from 5-18 GeV and proton ($\sim$70\%) beams from 40-275 GeV} 
\item{Ion beams from deuterons to heavy nuclei such as gold, lead, or uranium from 40-110 GeV per nucleon}
\item{Variable center-of-mass energies from 20-140 GeV}
\item{High collision electron-nucleon luminosity 10$^{33}$ - 10$^{34}$ cm$^{-2}$s$^{-1}$, corresponding to $\times$100-1000 larger than HERA at DESY}
\item{Possibility to have more than one interaction region}
\end{itemize}

The EIC will unlock the secrets of the strongest force in nature. 
The EIC covers very board and rich science from low-energy 
nuclear physics to high-energy particle physics. 
Main questions to be addressed by the EIC are~[1]:
\begin{itemize}
\item{How are the quarks and gluons distributed in space and momentum inside the nucleon?}
\item{How do the nucleon properties like mass,  spin, radius, etc emerge from quarks and their interactions?}
\item{How does the nuclear environment affect the distributions of quarks and gluons and their interactions inside nuclei?}
\item{What is the nature of dense gluon matter (Color Glass Condensate)?}
\item{How the short range nucleon-nucleon force ($NN$, $NNN$, etc) emerges from quark and gluon interactions?}
\item{In what manner do color-charged quarks and gluons interact with the nuclear medium and how do the confined hadronic states emerge? }
\end{itemize}

\begin{figure}[ht]
   \vspace{-2mm}
    \centering
    \includegraphics[width=0.975\linewidth]{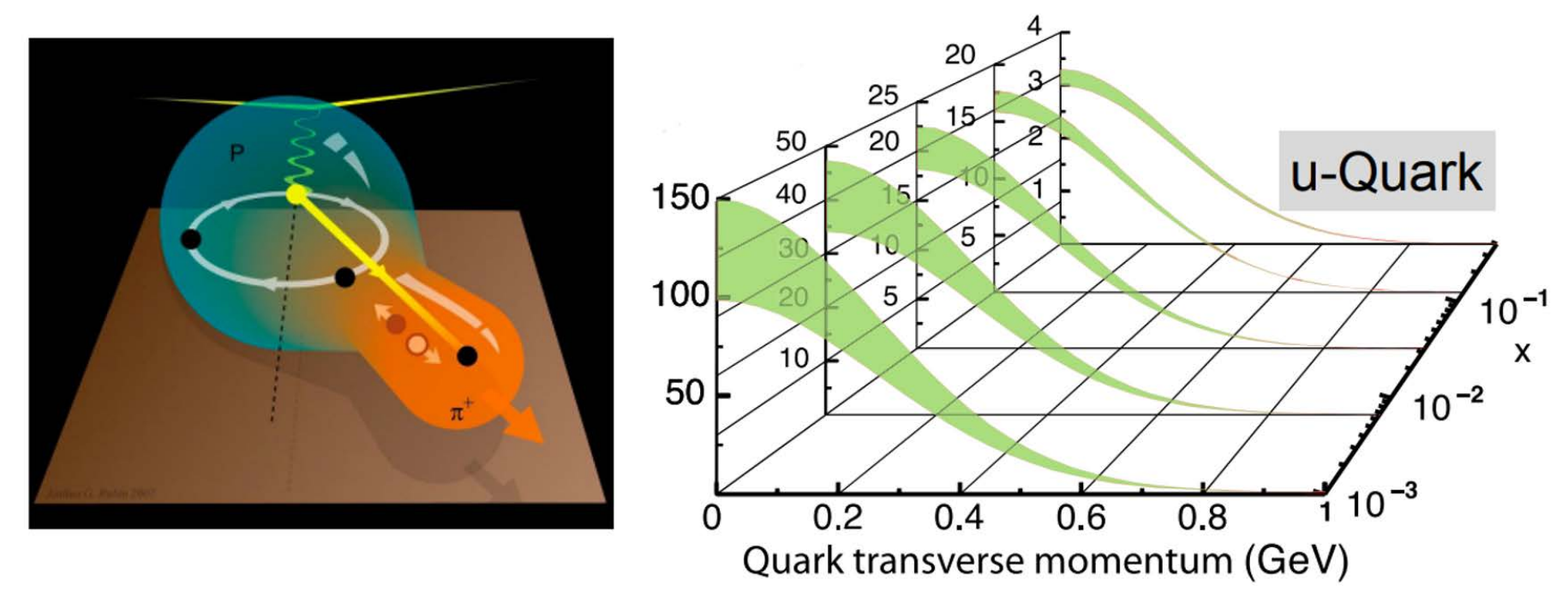}
    \vspace{-4mm}
    \caption{Left: Image of SIDIS process. Right: Example of quark transverse momentum distribution extracted from Sivers asymmetry measurements in SIDIS process~[1].}
    \label{fig:gunji_fig1}
\end{figure}

By providing access to regions within 
nucleons and nuclei, where their structure is primarily dominated by gluons, 
the EIC will elucidate the 3D internal structure of nucleon and nucleus. 
The 3D parton structure of hadrons in momentum space is encoded in transverse momentum
dependent parton distributions (TMDs). 
For a spin 1/2 hadron, a total of 8 leading-twist TMDs exist 
for both quark and gluon sectors~[2]. Those TMD functions 
represent different correlations between spins and transverse momenta and 
reveal different insights into the dynamics of nucleons. 
TMDs can be measured via semi-inclusive deep-inelastic scattering (SIDIS), 
where one detects an identified hadron in addition to the scattered lepton.
Figure~\ref{fig:gunji_fig1} shows 
example of Silvers effects (correlation between nucleon transverse spin 
and transverse momentum of partons) in SIDIS process and extracted quark transverse 
momentum distributions at different $x$ (momentum fraction of quarks carrying over protons). 
 
\begin{figure}[ht]
   \vspace{-2mm}
    \centering
    \includegraphics[width=0.95\linewidth]{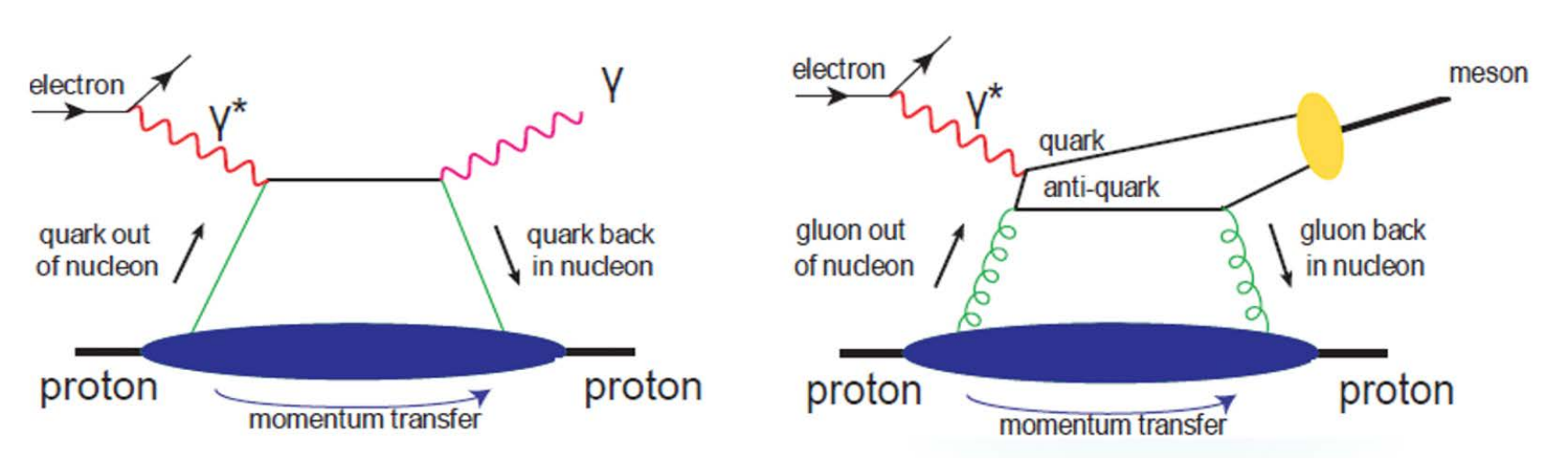}
    \vspace{-4mm}
    \caption{Left: deeply virtual Compton scattering (DVCS). Right: deeply virtual meson production (DVMP)}
    \label{fig:gunji_fig2}
\end{figure}

\begin{figure}[ht]
   \vspace{-2mm}
    \centering
    \includegraphics[width=0.93\linewidth]{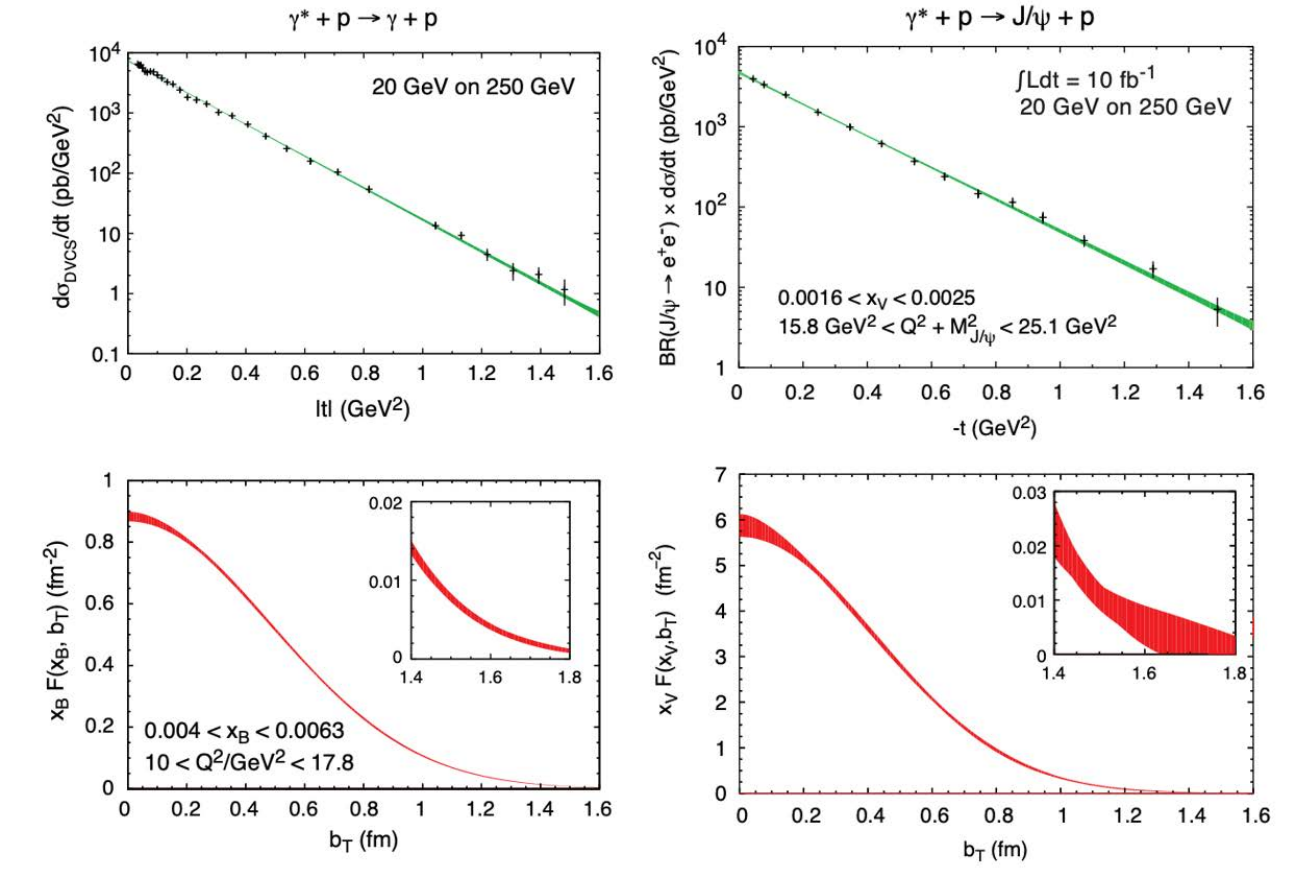}
    \vspace{-4mm}
    \caption{The differential cross-section 
for the momentum transfer $t$ for DVCS (upper left) and DVMP (upper right) and 
The spatial distributions of quarks (bottom left) and gluons (bottom right) extracted from
the Fourier transform of the differential cross-section~[1]}
    \label{fig:gunji_fig3}
\end{figure}

As in the case of the transverse momentum distribution of quarks and gluons 
inside a hadron, EIC can also provide quark and gluon distributions in the 
transverse spatial dimensions, combined with the information about the 
longitudinal momentum fraction $x$.
The non-perturbative quantities that encode the spatial distributions are called generalized parton distributions (GPDs)~[3]. 
In addition to the fundamental role of GPDs concerning the spatial distribution of partons
inside hadrons, the second moment of GPDs provides insight into the total angular 
momentum of quarks and gluons in the proton, which is crucial to understand 
the origin of nucleon spin. 
GPDs and the energy-momentum tensor are related to each other 
and thus potentially deepen the understanding of the nucleon mass, the pressure 
and shear forces inside hadrons.
Determination of GPDs experimentally requires a particular category of measurements, 
that of exclusive reactions. 
Examples are deeply virtual Compton scattering (DVCS) and deeply virtual meson production (DVMP), as shown in Fig.~\ref{fig:gunji_fig2}.  
For those reactions, the proton remains intact, and a photon or a meson is produced. Those exclusive measurements need all final-state products to be detected, such as 
the scattered electron, the produced photon or meson, and the scattered proton. 
The spatial distributions of quarks and gluons are extracted 
from the Fourier transform of the differential cross-section 
for the momentum transfer $t$ between the incoming and the scattered proton. The example is shown in Fig.~\ref{fig:gunji_fig3}.\\

\begin{figure}[ht]
   \vspace{-2mm}
    \centering
    \includegraphics[width=0.95\linewidth]{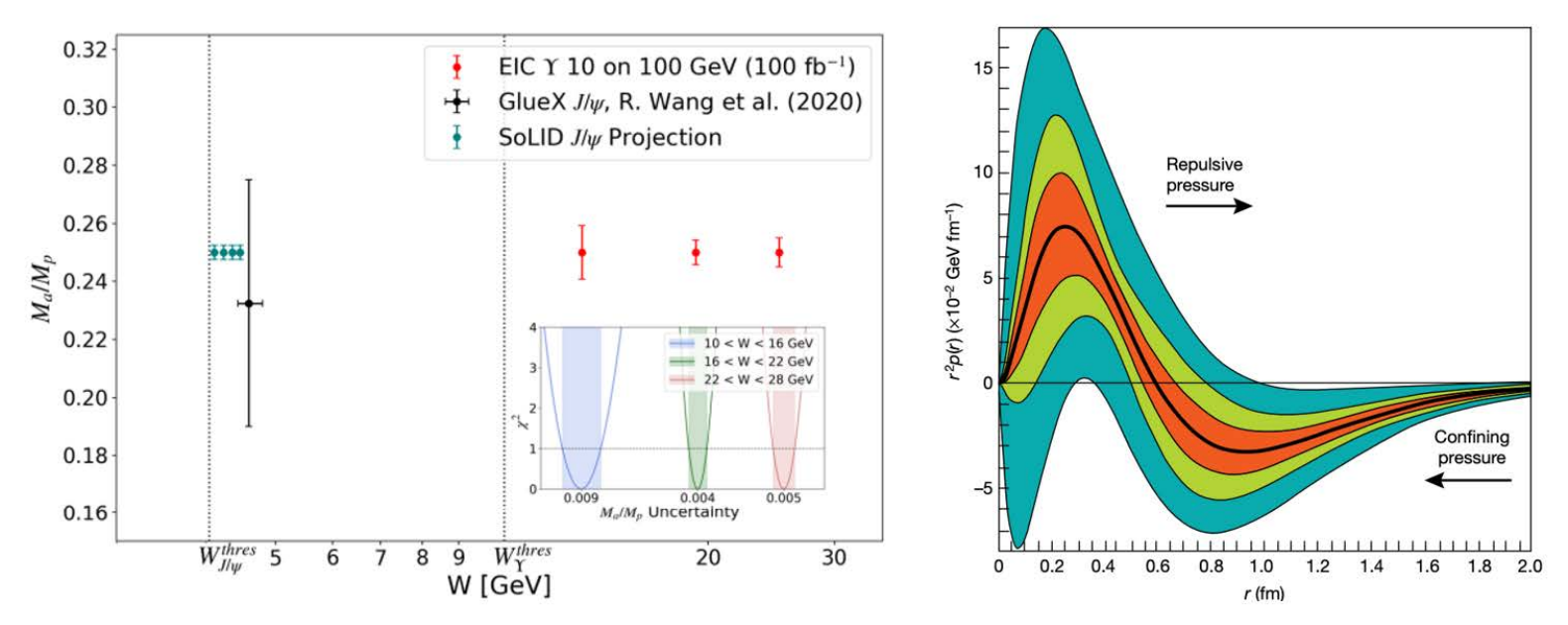}
    \vspace{-4mm}
    \caption{Left: Projection of the trace anomaly contribution to the proton mass with
$\Upsilon$ photoproduction on the proton at the EIC~[1]. Right: The pressure distribution $r^{2}p(r)$ as a function of the radial distance $r$ from the centre of the proton~[5].}
    \label{fig:gunji_fig6}
\end{figure}

The Higgs mechanism provides mass to the fundamental constituents of matter but 
can only explain a small fraction of the nucleon mass. 
The rest of the mass is the result of the equilibrium reached 
through dynamical processes between quarks and gluons. Understanding how the 
hadron mass emerges in QCD is therefore of utmost importance.
The gravitational form factors (GFF) which are obtained from the matrix elements of the energy-momentum 
tensor provide information about internal distributions of mass, energy, pressure, and shear forces. 
The gravitational form factors are accessible as the second moments of GPDs. 
Figure~\ref{fig:gunji_fig6} shows the projection of the trace anomaly contribution, which is originated from gluon condensation, to the proton mass and 
shows pressure distribution inside proton constrained by the measurements at JLab, 
where one can see the negative pressure inside proton as radius becomes large and that is closely related to the confinement of quarks inside hadrons.\\


The nucleus is a complex structure made of bound nucleons. 
Understanding the formation of nuclei in terms of QCD degrees-of-freedom 
is an ultimate goal of nuclear physics. 
With EIC's broad kinematic reach, capability to accelerate a variety of
nuclei, and ability to measure inclusive, semi-inclusive, and exclusive DIS 
measurements, the EIC will produce new opportunities 
for exploring the internal 3-dimensional sea quark and gluon structure of a nucleus at low $x$.  Furthermore, the nucleus itself can be a QCD laboratory 
for studying the possible collective behavior of high-dense gluonic matter, 
for studying the propagation of fast-moving color charges in a nuclear medium,  
and for studying the quark-gluon origin 
of short range nucleon-nucleon correlations in the nuclei.\\

\begin{figure}[ht]
   \vspace{-5mm}
    \centering
    \includegraphics[width=0.88\linewidth]{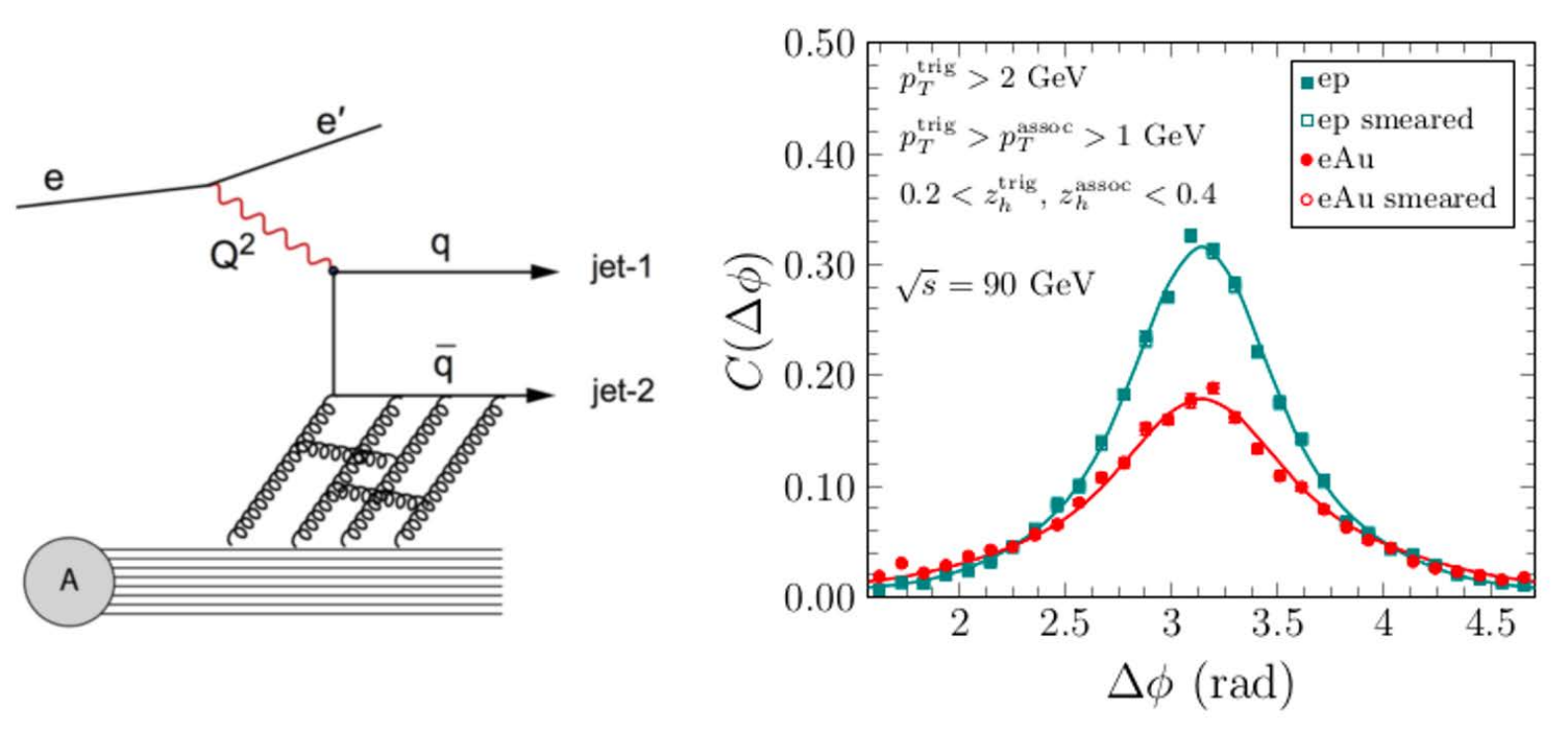}
    \vspace{-4mm}
    \caption{Left: back-to-back hadron(jet) - hadron or hadron(jet) production to probe the gluon saturation. Right: Expected depletion of away-side peak due to occurance of gluon saturation in Au nuclei~[1]. }
    \label{fig:gunji_fig4}
\end{figure}

The study of emergent properties of the high-dense gluonic matter is an important topic
in high-energy nuclear physics. Such properties are important to further understand the 
initial conditions and early-time dynamics of high-energy nucleus-nucleus collisions 
and the evolution of the Quark-Gluon Plasma. 
It is known that a large number of low-momentum gluons exist 
inside high energy nucleons and nuclei. When the gluon density is sufficiently high, 
the gluon density is expected to saturate (gluon saturation) as a balance between 
gluon splitting ($g \rightarrow gg$) and gluon recombination ($gg \rightarrow g$). 
The gluon saturation is characterized by the saturation scale ($Q_s$), where 
$Q_s$ is interpreted as the average transverse momentum of saturated gluons. 
If saturation scale, $Q_s$, significantly exceeds the QCD scale $\Lambda_s$,
the dynamics of gluon saturation can be described by by weak-coupling many-body approach such as the Color Glass Condensate (CGC). The CGC predicts that $Q_s$ is proportional 
to $A^{1/3}$ and that $Q_s$ becomes larger than 1 GeV with Au ions at the EIC, meaning that the saturation can be studied especially well in large nuclei. 
Figure~\ref{fig:gunji_fig4} shows the projection of the measurements of 
back-to-back hadron(jet) - hadron or hadron(jet), which is sensitive to the gluon saturation. Due to the interaction within high-density gluon fields, 
away-side peak is quenched in case that the gluon saturation is realized. \\

\begin{figure}[ht]
   \vspace{-4mm}
    \centering
    \includegraphics[width=0.85\linewidth]{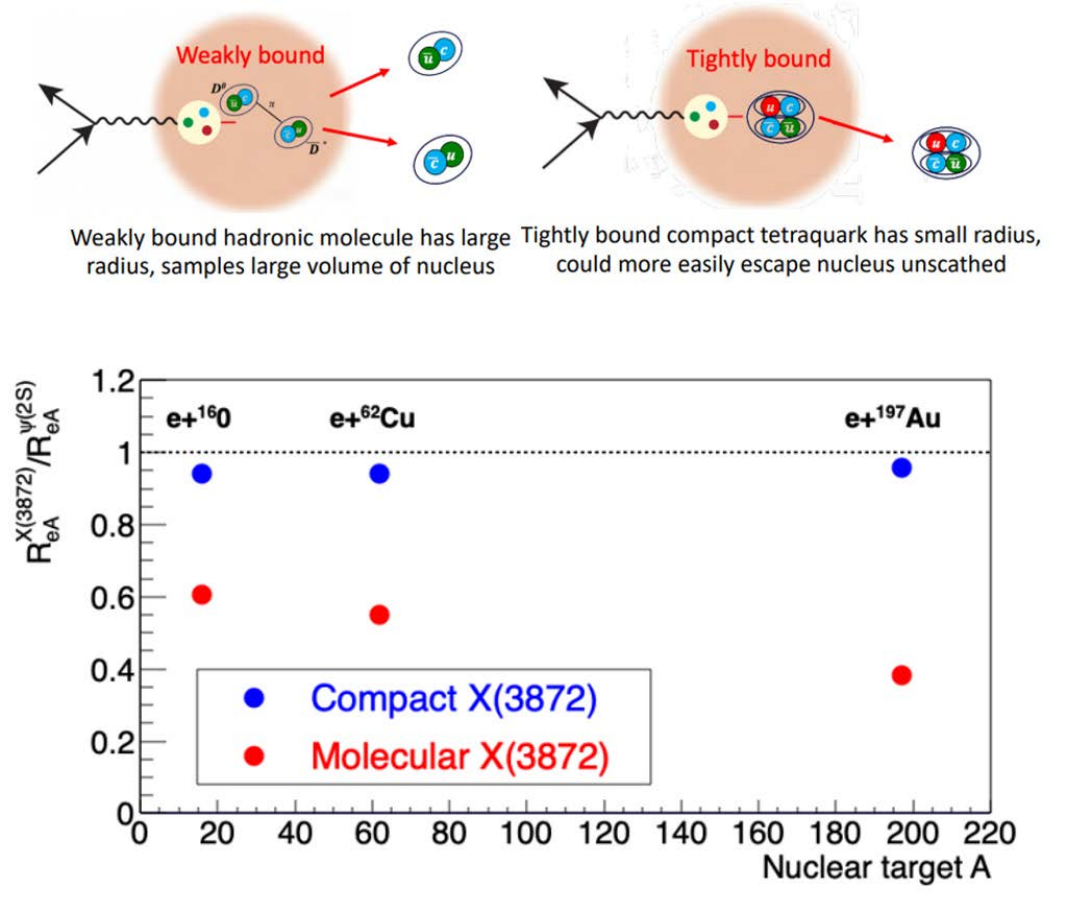}
    \vspace{-4mm}
    \caption{Top: Examples of weakly and tightly bound tetraquark state propagation in a large nucleus. Bottom: the ratio of nuclear modification factors $R_{eA}$ for
$X(3872)$ to $\psi(2S)$, for two different assumptions of the $X(3872)$ structure~[1].}
    \label{fig:gunji_fig5}
\end{figure}

The EIC will also offer the opportunity to study the exotic hadron structure. 
Multiple candidates for tetra- and pentaquark states have been proposed for such as the $X(3872)$ and the $P^{+}_{c}$ states. However, there is no firm conclusion on 
whether these states are hadronic molecules or compact multiquark states. 
$A$ dependence of those productions can provide more information on their properties 
since one would expect that large and weakly bound hadronic molecules would be destructed more significantly while traversing the nucleus than a compact state as shown in Fig.~\ref{fig:gunji_fig5}~[1].\\

\begin{figure}[ht]
   \vspace{-2mm}
    \centering
    \includegraphics[width=0.90\linewidth]{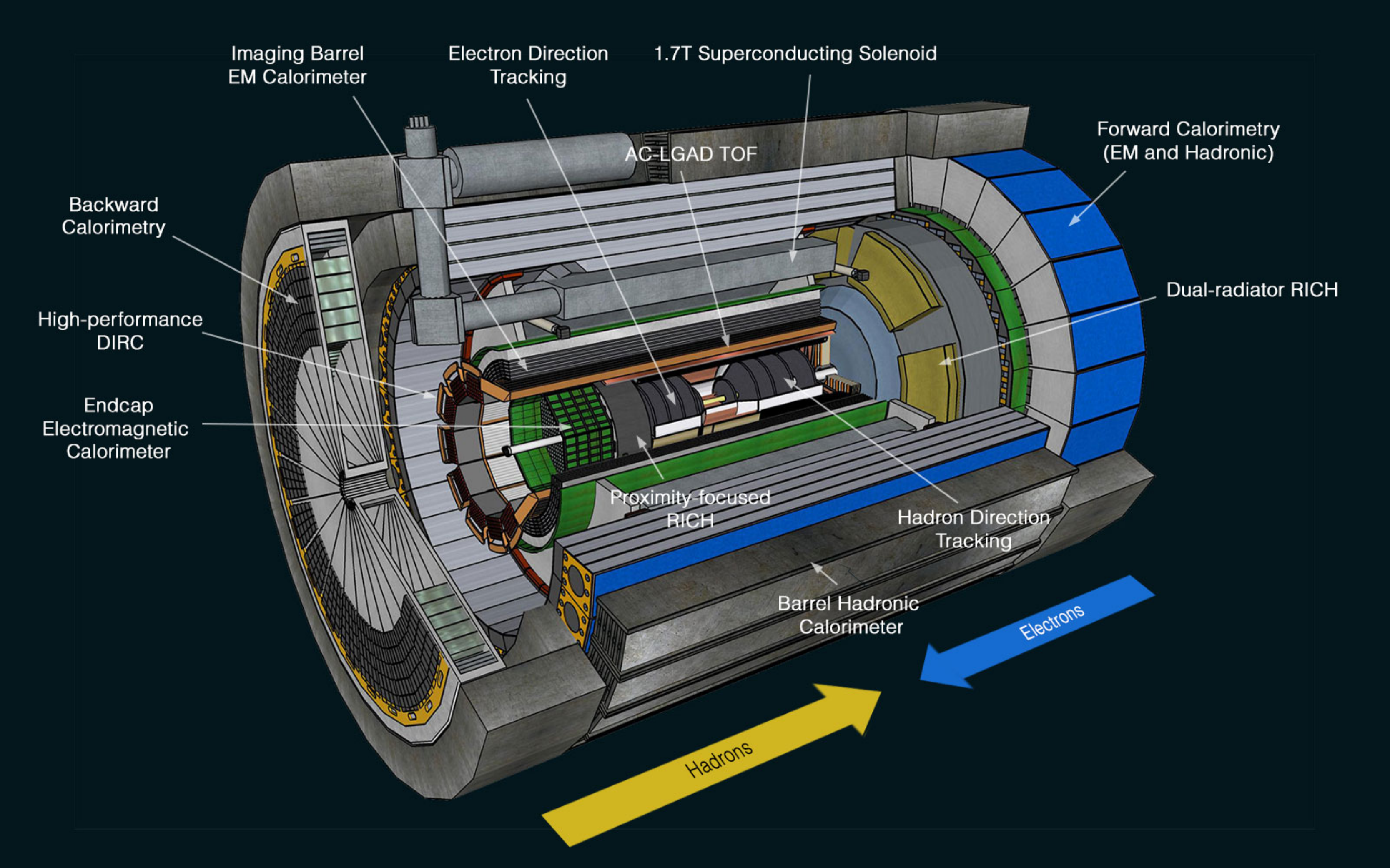}
    \vspace{-4mm}
    \caption{Cutaway view of the ePIC central barrel detector}
    \label{fig:gunji_fig5_2}
\end{figure}

Based on the details of those physics case, the required detector setup and the detector concepts at the EIC are considered and summarized in the Yellow Report~[1].
The international ePIC collaboration is formed to further develop
the detector technologies and design towards its construction~[5].
The ePIC experiment will be an approximately 10-meter long cylindrical barrel detector and far-forward and backward detectors that will be located around 45~m away from interaction point. The central barrel detector of the ePIC experiment is shown in Fig.~\ref{fig:gunji_fig5_2}.
The ePIC experiment will use cutting-edge technologies 
and the leading candidates for the detector technologies are under investigation.
The ePIC experiment will deploy 
high-precision silicon detectors based on MAPS technologies for tracking particles, 
high-granularity calorimeters for measuring the energy of electromagnetic particles with exquisite precision, a set of particle identification detectors 
using different technologies such as AC-LGAD silicon pixel and strip TOF detectors, 
Aerogel RICH detectors, and high-performance DIRC detector for 
determining the quark content of particles over a wide range of energies. 
EIC Physics demands $\sim$100\% acceptance for all final state particles 
including particles associated with initial ions under 500 kHz interaction rate.  
Therefore, the ePIC experiment 
will deploy a revolutionary streaming data acquisition system without the need for a traditional hardware trigger and will employ machine learning (ML) and 
artificial intelligence (AI) technologies for the real-time data processing, 
calibrations, analysis, and feedback to the detectors and systems 
for quick diagnostics and optimization of the detector. \\

The ePIC-Japan team has been formed and 10 institutes participate in the team.
Currently, the ePIC-Japan team has been working on various detector projects such as 
AC-LGAD based Time-Of-Flight system in the central barrel (bTOF), 
Zero-Degree Calorimeter (ZDC) in far-forward detector, 
and streaming data acquisition system. 
AC-LGAD development, especially AC-LGAD sensor characterization, is on-going with eRD112, 
ZDC R\&D is underway with ALICE-FoCAL project~[6], and 
development of the streaming readout is on-going and collaborative efforts are under development with the Signal Processing and 
Data acquisition Infrastructure Alliance (SPADI-Alliance)~[7].\\

\vfill  

\noindent{\bf References }
\begin{description}
\setlength\itemsep{-3pt}
\item{[1]} Science Requirements and Detector Concepts for the Electron-Ion Collider : EIC Yellow Report, Nucl.Phys.A 1026 (2022) 122447
\item{[2]} R. Boussarie, {\it et al.}, TMD Handbook, arXiv:2304.03302
\item{[3]} M. Diehl, Generalized parton distributions, Phys. Rep. 388, (2003), 41-277 
\item{[4]} A. V. Belitsky, A. V. Radyushkin, Unraveling hadron structure with generalized 
parton distributions, Phys. Rep. 418, (2005), 1-387
\item{[5]}V. D. Burkert, L. Elouadrhiri, F. X. Girod, 
The pressure distribution inside the proton, Nature volume 557, pages396–399 (2018)
\item{[6]} \url{https://www.bnl.gov/eic/epic.php}
\item{[7]} \url{https://alice-collaboration.web.cern.ch/menu_proj_items/FOCAL}
\item{[8]} \url{https://www.rcnp.osaka-u.ac.jp/~spadi/}

\end{description}

\stepcounter{count}
\clearpage

\phantomsection
\addcontentsline{toc}{section}{
{\bf The nature of the lightest positive parity open charm states and its implications} \\
C.~Hanhart}

\titl{The nature of the lightest positive parity open charm states and its implications }

\name{C. Hanhart
}

\adr{
IAS-4, Forschungszentrum J\"ulich, J\"ulich, Germany
}

The lightest positive parity open charm mesons show properties that seem incompatible with the most
simple realisation of the quark model, where open charm mesons are described as $c\bar q$ states,
with $\bar q$ for one of the light antiquarks $\bar u, \bar d$ and $\bar s$. The three major puzzles 
they pose are that $(i)$ the spin partner states $D_{s0}(2317)$ and $D_{s1}^*(2460)$ are of the order
of 100 MeV lighter than the respective quark model predictions and $(ii)$ they are both located 
about 45 MeV below the $KD$ and $KD^*$ thresholds, respectively (in quark models the spin splittings
in different multiplets are typically different). Moreover, $(iii)$ the non-strange partner states are near degenerate
to the mentioned strange ones: 
$$M_{D_{s0}}=2317\ \mbox{MeV} \ \approx \   2300\ \mbox{MeV} = M_{D_0}\ ,$$
$$M_{D_{s1}^*}=2460\ \mbox{MeV} \ \approx \   2430\ \mbox{MeV} = M_{D_1}\ .$$
In a series of papers it was demonstrated that all those puzzles can be resolved if the mentioned states
are interpreted as two-hadron states. All masses are taken from Ref.~[1].

The first two issues are resolved by the observation that for two-hadron states spin symmetry predicts
for bound systems of a light meson with a heavy one
that the binding energies are equal to leading order. The solution to the third puzzle is more involved.
Here one needs to employ a dynamical treatment of the system. In our works we used for that 
unitarised chiral perturbation theory to next-to-leading order (UChPT). The parameters of the amplitude
were fixed in Ref.~[2] via fits to light quark mass dependent scattering lengths determined in lattice QCD.
The same amplitude turned out to also describe the energy dependences determined in lattice QCD, 
demonstrated in Ref.~[3]. In this work it was also shown that in the non-strange sector there are not
one but two poles present, one around 2.1 GeV and one around 2.5 GeV ---
 the former in the flavor [$\bar 3$] and the latter in the flavor [6].
 This kind of
two-pole structure is a more general phenomenon. For a recent review on this subject see Ref.~[4]. 

The pole in the [6] was not reported in the lattice study of Ref.~[5], however, this
was explained in Ref.~[6] as being the result of this pole sitting on a hidden sheet.
In Ref.~[7] it was demonstrated that the structure observed in experiment that
lead to the name $D_0(2300)$ is in fact generated from the interplay of the two poles 
mentioned above. Moreover, the phase of the production amplitude was in Ref.~[8]
shown to be in line with that from the UChPT amplitude, but in at odds with that of a
Breit-Wigner at 2.3 GeV.

Group theory gives that in addition to the mentioned two the system also accesses the [15] multiplet,
however, here the leading chiral amplitude is repulsive and thus no bound state is expected
in this channel --- in line with the findings from lattice QCD~[9,10].

\vfill  

\noindent{\bf References }
\begin{description}
\setlength\itemsep{-3pt}
\item{[1]} R.~L.~Workman \textit{et al.} [Particle Data Group],
PTEP \textbf{2022} (2022), 083C01.
\item{[2]} 
L.~Liu, K.~Orginos, F.~K.~Guo, C.~Hanhart and U.-G.~Mei\ss ner,
Phys. Rev. D \textbf{87} (2013) no.1, 014508
[arXiv:1208.4535 [hep-lat]].
\item{[3]} M.~Albaladejo, P.~Fernandez-Soler, F.~K.~Guo and J.~Nieves,
Phys. Lett. B \textbf{767} (2017), 465-469
[arXiv:1610.06727 [hep-ph]].
\item{[4]} U.-G.~Mei\ss{}ner,
Symmetry \textbf{12} (2020) no.6, 981
[arXiv:2005.06909 [hep-ph]]. 
\item{[5]} G.~Moir, M.~Peardon, S.~M.~Ryan, C.~E.~Thomas and D.~J.~Wilson,
JHEP \textbf{10} (2016), 011
[arXiv:1607.07093 [hep-lat]].
\item{[6]}  A.~Asokan, M.~N.~Tang, F.~K.~Guo, C.~Hanhart, Y.~Kamiya and U.-G.~Mei\ss{}ner,
Eur. Phys. J. C \textbf{83} (2023) no.9, 850
[arXiv:2212.07856 [hep-ph]].
\item{[7]} M.~L.~Du, M.~Albaladejo, P.~Fern\'andez-Soler, F.~K.~Guo, C.~Hanhart, U.-G.~Mei\ss{}ner, J.~Nieves and D.~L.~Yao,
Phys. Rev. D \textbf{98} (2018) no.9, 094018
[arXiv:1712.07957 [hep-ph]].
\item{[8]} M.~L.~Du, F.~K.~Guo, C.~Hanhart, B.~Kubis and U.-G.~Mei\ss{}ner,
Phys. Rev. Lett. \textbf{126} (2021) no.19, 192001
[arXiv:2012.04599 [hep-ph]].
\item{[9]} E.~B.~Gregory, F.~K.~Guo, C.~Hanhart, S.~Krieg and T.~Luu,
[arXiv:2106.15391 [hep-ph]].
\item{[10]} J.~D.~E.~Yeo, C.~E.~Thomas and D.~J.~Wilson,
[arXiv:2403.10498 [hep-lat]].
\end{description}

\stepcounter{count}
\clearpage

\phantomsection
\addcontentsline{toc}{section}{
{\bf A chiral quark model with Hidden Local Symmetry} \\
B.R.~He, M.~Harada$^*$, B.S.~Zou}

\titl{A chiral quark model with Hidden Local Symmetry}

\name{Bing-Ran He$^1$, Masayasu Harada$^{2,3,4}$, Bing-Song Zou$^{5,6,7}$}

\adr{
\small
$^1$ Department of Physics, Nanjing Normal University, Nanjing 210023, PR China \\
$^2$ Kobayashi-Maskawa Institute for the Origin of Particles and the Universe, Nagoya University, Nagoya, 464-8602, Japan\\
$^3$ Department of Physics, Nagoya University, Nagoya, 464-8602, Japan \\
$^4$ Advanced Science Research Center, Japan Atomic Energy Agency, Tokai 319-1195, Japan\\
$^5$ CAS Key Laboratory of Theoretical Physics, Institute of Theoretical Physics, Chinese Academy of Sciences, Beijing 100190, China\\
$^6$ School of Physical Sciences, University of Chinese Academy of Sciences, Beijing 100049, China\\
$^7$ School of Physics, Peking University, Beijing 100871, China
}

In this mini-proceedings, we briefly summarize the works done in Refs.~[1,2], where we developed a new chiral quark model. (Talk in the workshop was given by M.~Harada.)

\quad
In the first paper~[1], 
we proposed a chiral quark model including the $\omega$ and $\rho$ meson contributions in addition to the
$\pi$ and $\sigma$ meson contributions. 
The vector mesons are included based on the hidden local symmetry~[3,4], which is consistent with the chiral symmetry.
We showed that the masses of the ground state baryons such as the
nucleon, $\Lambda_c$ and $\Lambda_b$ are dramatically improved in the model with the vector mesons compared with the one without them. 
This is the consequence of the $\omega$ exchange contribution  which provides the attractive force between a quark and an anti-quark and the repulsive force between two quarks.
In addition, we performed the analysis of the tetraquark $T_{cc}$, and found that 
the resultant mass is much closer to its experimental value than the result without
vector meson contribution. 
Our result indicates that $T_{cc}$ takes molecule-like structure, while $T_{bb}$ takes diquark-like structure.

\quad
In the second paper~[2],
we extended the chiral quark model to include the nonet pseudo-scalar and
vector mesons together with the singlet scalar meson based on the SU$(3)_L\times$SU$(3)_R$ chiral symmetry
combined with the hidden local symmetry, which mediate force among $u$, $d$ and $s$ quarks. 
We
fitted 22 model parameters to the masses of 46 known ground state mesons and baryons. 
We showed that the mass
spectra of those hadrons are beautifully reproduced.
We also predicted the masses of missing ground states,
one meson and twenty baryons, as shown in Table~1.
Our predictions are consistent with some results from lattice QCD analyses, and will be tested in the future experiment.
\begin{table}[htp]
\footnotesize
\begin{tabular}{ c c c c c c c c c c c c }
\hline\hline
$B_c^*$&$\Xi_{cc}^*$ &$\Xi_{bc}$ & $\Xi_{bc}'$ & $\Xi_{bc}^*$ & $\Xi_{bb}$ & $\Xi_{bb}^*$ &  
$\Omega_{cc}$&$\Omega_{cc}^*$ &$\Omega_{bc}$ & $\Omega_{bc}'$ \\
6306.9 &3698.9 & 6943.9&6958.4 &6976.4 & 10176& 10198.5& 
3763.9 & 3806.8&7044.7 &7057.7 \\
\hline
$\Omega_{bc}^*$ & $\Omega_{bb}$ & $\Omega_{bb}^*$ & 
$\Omega_{b}^*$ & $\Omega_{ccc}$ & $\Omega_{ccb}$ & $\Omega_{ccb}^*$& $\Omega_{bbc}$& $\Omega_{bbc}^*$& $\Omega_{bbb}$& \\
7076&10266.9 & 10289& 6068.1 & 4795& 8014.9& 8027.5&11191.9 & 11207.7& 14351.2& \\
\hline\hline
\end{tabular}
\caption{\label{tab:mass_spec} Predicted mass spectrum (in MeV) of ground states for mesons and baryons
which have not been experimentlally confirmed.}
\end{table}


\vfill  

\noindent{\bf References }
\begin{description}
\setlength\itemsep{-3pt}

\item{[1]}\label{ref1}
B.~R.~He, M.~Harada and B.~S.~Zou,
Phys. Rev. D \textbf{108}, no.5, 054025 (2023).

\item{[2]}\label{ref2}
B.~R.~He, M.~Harada and B.~S.~Zou,
Eur. Phys. J. C \textbf{83}, no.12, 1159 (2023).

\item{[3]}
M.~Bando, T.~Kugo and K.~Yamawaki,
Phys. Rept. \textbf{164}, 217-314 (1988).

\item{[4]}
M.~Harada and K.~Yamawaki,
Phys. Rept. \textbf{381}, 1-233 (2003).

\end{description}

\stepcounter{count}
\clearpage

\phantomsection
\addcontentsline{toc}{section}{
{\bf Experimental programs using HypTPC} \\
S.~Hayakawa on behalf of the HypTPC collaboration
}

\titl{Experimental programs using HypTPC}

\name{
  Shuhei Hayakawa$^{1}$ \\
  on behalf of the HypTPC collaboration
}

\adr{
$^1$ Department of Physics, Tohoku University, Sendai, 980-8578, Japan
}

Hyperon Spectrometer, consisting of a Helmholtz-type superconducting dipole magnet [1],
a time projection chamber known as HypTPC [2,3], and a timing hodoscope,
is utilized in a series of experiments at J-PARC focusing on exotic hadrons.
The first experiment, J-PARC E42, conducted in 2021 at K1.8 beamline,
searched for the $H$-dibaryon, an exotic hadron composed of six quarks ($uuddss$) theoretically predicted [4].
Recent calculations and experiments suggest an attractive ${\Xi}N$ interaction, 
potentially indicating a resonance state near the ${\Xi}N$ threshold.
The $^{12}\mathrm{C}(K^-, K^+)$ reaction at 1.8 GeV/$c$ measure $\Lambda p\pi^-$, $\Lambda\Lambda$, and $\Xi^-p$ systems,
, with particle identification confirming the reconstruction of $\Lambda$ and $\Xi^-$.
About half of the data has yielded 3,000 $\Lambda\Lambda$ events, and the mass spectrum will be published soon.
Other research topics include measurements of the differential cross-sections and polarization of $\Xi(1530){^-}$,
differential cross-section measurements of $K^\ast(892)^-$ using the $p(K^-, p)$ reaction, and kaonic nuclear search by the $^{12}$C$(K^-, p)$ reaction.

The upcoming J-PARC E72 experiment aims to search for a narrow resonance $\Lambda^\ast$ suggested by a past experiment [5],
measuring the differential cross-section of the $p(K^-, {\Lambda}){\eta}$ reaction using a 735 MeV/$c$ beam.
Angular distributions and $\Lambda$ polarization measurements will determine the existence, spin and parity of the new resonance.
A new gas vessel for HypTPC and a liquid hydrogen target system have been developed,
with successful testing Cherenkov counters for triggering.

J-PARC E45 will conduct spectroscopy of $N^\ast$ and $\Delta^\ast$ states using ${\pi}p\to{\pi\pi}N$ and ${\pi}p\to{KY}$ reactions
over a wide energy range from 1.5 to 2.15~GeV, gathering statistics a hundred times greater than past experiments. 
It seeks to clarify many unestablished states in the PDG, which are theoretically suggested to dominantly couple to ${\pi\pi}N$ [6].

The J-PARC E90 experiment aims to measure the ${\Sigma}N$ cusp to determine the ${\Sigma}N$ scattering length
using K1.8 and S-2S spectrometers, targeting a statistical error of less than 0.3~fm.
The ${\Sigma}N$ cusp, generated by the $d(K^-, \pi^-)$ reaction at 1.4 GeV/$c$,
indicates the scattering length near the ${\Sigma}N$ threshold, essential for understanding ${\Sigma}N$ hypernuclei and ${\Sigma}N$--${\Lambda}N$ coupling.

In summary, the J-PARC experimental program with Hyperon Spectrometer and HypTPC aims
to advance our understanding of exotic hadrons and hypernuclei.
The results promise to significantly enhance knowledge of hadronic interactions and exotic states,
highlighting the importance of J-PARC's role in nuclear physics research.
The author extends gratitude to all collaborators involved in the J-PARC E42, E72, E45, and E90 experiments utilizing HypTPC.

\vfill  

\noindent{\bf References}
\begin{description}
\setlength\itemsep{-3pt}
\item{[1]} \href{https://doi.org/10.1016/j.nima.2022.167775}
  {J.K. Ahn, et al., Nucl. Instrum. Meth. A {\bf 1047} (2023) 167775}
\item{[2]} \href{https://doi.org/10.1016/j.nima.2014.06.007}
  {H. Sako, et al., Nucl. Instrum. Meth. A {\bf 763} (2014) 65-81.}
\item{[3]} \href{https://doi.org/10.1016/j.nima.2019.06.050}
  {S.H. Kim, et al., Nucl. Instrum. Meth. A {\bf 940} (2019) 359-370.}
\item{[4]} \href{https://doi.org/10.1103/PhysRevLett.38.195}
  {R.L. Jaffe, Phys. Rev. Lett. 38 (1977) 195};
  Erratum \href{https://doi.org/10.1103/PhysRevLett.38.617}
  {Phys. Rev. Lett. 38 (1977) 617}.


\item{[5]} \href{https://doi.org/10.1103/PhysRevC.64.055205}
  {A. Starostin, et al., Crystal Ball Collaboration, Phys. Rev. C {\bf 64} (2001) 055205.}

\item{[6]} \href{https://doi.org/10.1103/PhysRevC.79.025206}
  {H. Kamano, et al., Phys. Rev. C {\bf 79} (2009) 025206.}
\end{description}

\stepcounter{count}
\clearpage

\phantomsection
\addcontentsline{toc}{section}{
{\bf Search for ${}^{9}_{\Lambda}$He via the ($\pi^{-}, K^{+}$) reaction at K1.8BR and HIHR} \\
R.~Honda$^*$, S.H.~Hayakawa, T.~Fukuda, T.~Harada, E.~Hiyama, A.~Sakaguchi, A.~Umeya}

\newcommand{\HeL}[1]{{}^{#1}_{\Lambda}\textrm{He}}
\newcommand{\LiL}[1]{{}^{#1}_{\Lambda}\textrm{Li}}
\newcommand{\HL}[1]{{}^{#1}_{\Lambda}\textrm{H}}
\newcommand{\Nucl}[2]{{}^{#1}\textrm{#2}}

\titl{Search for ${}^{9}_{\Lambda}$He via the ($\pi^{-}, K^{+}$) reaction at K1.8BR and HIHR}

\name{
R.~Honda$^{1}$, S.H.~Hayakawa$^{2}$, T.~Fukuda$^{3}$, T.~Harada$^{3}$, E.~Hiyama$^{2}$, A.~Sakaguchi$^{4}$, A.~Umeya$^{5}$
}

\adr{
$^1$ Institute of Particle and Nuclear Studies, High Energy Accelerator Research Organization, Tsukuba 305-0801, Japan\\
$^2$ Department of Physics, Tohoku University, Sendai, Miyagi 980-8578, Japan\\
$^3$ Research Center for Physics and Mathematics, Osaka Electro-Communication University, Neyagawa, Osaka, 572-8530, Japan\\
$^4$ Student Life Cycle Support Center, Osaka University, Toyonaka, Osaka 560-0043, Japan\\
$^5$ Liberal Arts and Sciences, Nippon Institute of Technology, Saitama 345-8501, Japan
}

Hypernuclei being accessible via the double-charge-exchange (DCX) reaction have a neutron excess environment, which is a noble tool to investigate a particle mixing effect in the $\Lambda N$ interaction, the $\Lambda N$-$\Sigma N$ mixing.
Observation of a single peak structure and precise measurement of $-B_{\Lambda}$ are missing peaces for the search for the neutron-rich $\Lambda$ hypernuclei via the  DCX reaction.
We are planing to perform the experiment for search for ${}^{9}_{\Lambda}$He via ${}^{9}$Be($\pi^{-}$, $K^{+}$)${}^{9}_{\Lambda}$He as the revised experiment of the J-PARC E10.

$\HeL{9}$ is a suitable hypernucleus for observing the single-peak structure since the spin parity of the $\Nucl{8}{He}$ ground state is $0^{+}$.
Some theoretical calculations have predicted $-B_{\Lambda}$ of around $-$7-9 MeV [1-3] reflecting the fact that the ground state of $\Nucl{8}{He}$ is deeply bound.
Myo and Hiyama [3] suggested that the energy difference between g.s. and excited states (${\frac{3}{2}}^{+}, {\frac{5}{2}}^{+}$) is around 4.5 MeV, being experimentally favorable for the single-peak observation.

For the production of $\Lambda$-hypernuclei via the ($\pi^{-}, K^{+}$) reaction, it is known that the one step reaction, in which virtual $\Sigma^{-}$ is a doorway, is favored through the theoretical analysis of the KEK-PS E521 experiment [4].
Both coherent and incoherent $\Lambda N$-$\Sigma N$ mixing can contribute to the production, however, the production of $\HeL{9}$ ($1/2^{+}$) through the coherent mixing will be quite small due to the small spectroscopic factor for $\Nucl{9}{Be}$ (${3/2}^{-}$) $\rightarrow p + \Nucl{8}{Li}$ ($0^{+}$).
On the other hand, the coherent mixing will contribute to the production of the exited states.
Thus, the production cross-section ratio between the ground state and excited states provides the fraction of coherent (incoherent) contribution for the naive.

We propose a staging approach for this experiment.
In stage 1, we perform the experiment at K1.8 with 10-20 days beam time and reveal the $\Sigma^{-}$ admixture probability of coherent/incoherent mixing from the cross-section ratio.
In this stage, a clear peak structure may not be observed due to a lack of yield.
In stage 2, we carry out the experiment at K1.8 or HIHR beamlines according to the stage-1 result with longer beam time.
If we carry out the experiment at HIHR, it is expected to yield 200 events for g.s. with 32-days beam time even with the quite small cross section of 2.5 nb/sr.

The experimental method for the investigation of neutron-rich $\Lambda$ hypernuclei via the DCX reaction will be established at the end of success of this experiment.
It is an important milestone for the future series study of neutron-rich $\Lambda$ hypernuclei.

\vfill  

\noindent{\bf References }
\begin{description}
\setlength\itemsep{-3pt}
\item{[1]} L. Majling, Nucl. Phys. A 585, 211c (1995). 
\item{[2]} A. Gal and D.J. Millener, PLB725, 445 (2013).
\item{[3]} T. Myo and E. Hiyama, PRC107 054302 (2023).
\item{[4]} T. Harada, A. Umeya, and Y. Hirabayashi, PRC79, 014603 (2009).
\end{description}

\stepcounter{count}
\clearpage

\phantomsection
\addcontentsline{toc}{section}{
{\bf J-PARC-HI Project (Hadron physics in the J-PARC-HI project} \\
Y.~Ichikawa for the J-PARC-HI collaboration}

\titl{J-PARC-HI Project \\(Hadron physics in the J-PARC-HI project)}

\name{
Yudai Ichikawa for the J-PARC-HI collaboration
}

\adr{
Tohoku University, Sendai 980-8578, Japan \\
Japan Atomic Energy Agency, Tokai, Ibaraki 319-1195, Japan
}


The main goal of the J-PARC-HI project is to search for QCD critical point and 
study high-density matter using heavy ion collisions.
The world's highest heavy-ion beam intensity with J-PARC-HI opens a new era in this field.
The expected interaction rate is as high as 100 MHz, providing us with high statistics.
This event rate will provide the equivalent of one year's worth of statistics at AGS at BNL within five minutes of running at J-PARC-HI. 
The detail of J-PARC-HI project is described in Ref. [1]. 
The high statistics will enable us to usher in a new era, not only in the study of QCD phase structure but also in hadron 
and strangeness nucleus physics, which are main subjects at the J-PARC Hadron Hall facility.

Here, we propose three main topics related to hadron and strange nuclear physics in the J-PARC-HI experiment:
\begin{quote}
 \begin{itemize}
  \item Femtoscopy
  \item Hypernucleus measurement at mid-rapidity region
  \item Exotic hadron
 \end{itemize}
\end{quote}
In this document, we focus on one of the new topics, Femtoscopy. 

Great effort has been made to investigate hadron-hadron interactions through recent Femtoscopic studies. 
Nuclear collisions at relativistic energies are abundant sources of various particle species. 
The presence of approximately one hundred particles at mid-rapidity makes it feasible to study their interactions. 
The produced particles may endure the final state interactions, and the resulting correlations in momentum 
space can be studied to test the underlying dynamics using correlation functions. 
Recently, Femtoscopy has been successfully applied by the RHIC STAR and LHC ALICE Collaborations 
[2, 3, 4]. 
They measured correlation functions of various hadronic pairs such as $\Lambda \Lambda$, $p \Omega^-$, 
$p \Xi^{-}$, $p K^{-}$, and $p \phi$ and provide the constraints for their interactions. 

Experimentally, the two-particle  correlation fuction is fined as:  
\begin{equation}
C(k^{*}) = N\frac{A(k^*)}{B(k^*)}, 
\end{equation}
where $k^{*}$ is relative momentum of two particles evaluated in the pair rest frame~[5]. 
The term $A$, reffered to as the ``signal'' is constructed from the same event samples, while the term 
$B$ is constructed from the different (mixed) event samples to represent the uncorrelated 
reference distribution. 
The paramter $N$ is a normalization parameter determined such that for large $k^{*}$, $C(k^*) \to 1$ 
as $ (k^* \to \infty)$ [5]. 

Theoretically, the correlation function is given by so-called Koonin-Pratt formula [6,~7]: 
\begin{equation}~\label{eq:Correlation function}
C(k^{*}) = \int S(r^*) |\Psi(k^*, r^*)|^2 d^3 r^*, 
\end{equation}
where $S(r^*)$ represents the source emission function, and $\Psi (k^*, r^*)$ is the pair wave function, 
which depends on the two-particle interaction.
When only the strong interaction is present without Coulomb interaction, 
the correlation function can be simplified as [5, 8]: 
\begin{equation}\label{eq:LL}
C(k^*) = 1 + \sum_{I, S} \rho_{(I, S)} \bigg[\frac{1}{2} \bigg|\frac{f(k^*)}{R}\bigg| ^2 \bigg(1- \frac{d_0}{2\sqrt{\pi}R} \bigg) 
+ \frac{2\Re f(k^*)}{\sqrt{\pi} R} - \frac{\Im f(k^*)}{R}F_2(2k^*R) \bigg].
\end{equation} 
Here, the sum is over all pair-spin ($S$) and pair-isospin ($I$) configurations with weights $\rho_{(I, S)}$ 
originating from Clebsch-Gordan coefficients, 
$F_1(z) = \int_{0}^{z} (e^{x^2 - z^2}) dx$,  
$F_2(z) = (1 - e^{-z^2})/z$, and R is the one-dimensional source size. 
This explanation refers to the Lednick\'{y}-Lyuboshits analytical model.  
In this model, the scattering amplitude, $f(k^*)$, is included, which can be expressed as a function of scattering parameters 
such as the scattering length $f_0$ and the effective range $d_0$ through the effective range expansion. 
Consequently, fitting the measured correlation function with Eq.~\ref{eq:LL}, it averages over pair-spin ($S$) and isospin ($I$) states. 
Thus, the scattering parameters obtained from the fitting with the Lednick\'{y}-Lyuboshits model represent 
averages over pair-spin and isospin of two-particle system. 
This implies that it is difficult to determine the scattering parameters for each pair-spin and isospin state using this method. 
Therefore, the measured $C(k^*)$ spectrum is often compared with the calculations inputting the scattering parameters 
provided by theoretical models.

To address this issue, J-PARC-HI femtoscopy aims to measure the correlation function with 
$\alpha$ ($^4$He) and various hadrons, such as $\alpha-\Xi$ and $\alpha-\bar{K}$. 
Since both spin and isospin of $\alpha$ are 0, the pair-spin and isospin should be uniquely determined. 
Therefore, there is no need to consider averaged interactions. 
Here, the production rate of $\alpha$ particles in the mid-rapidity region reaches its maximum at the
 J-PARC-HI energy [9]. 
This represents a significant advantage over the RHIC and LHC facilities.

Recently, Kamiya and Hyodo investigated the compositeness ($X$) of near-threshold states to explore 
the internal structure of exotic states~[10]. They extended the weak-binding relation, 
originally derived by Weinberg~[11], to more general cases within the framework of effective field theory. 
According to Ref.~[10], the scattering length $f_0$ can be expressed as:
\begin{equation}\label{eq:weak_binding}
f_0 = R \left( \frac{2X}{1+X} + \mathcal{O}\left(\frac{R_{\rm typ}}{R}\right) \right)
\end{equation}
The radius $R$ of the wave function is given by $R \equiv \sqrt{2\mu B}$, where $\mu$ is the reduced mass 
and $B$ is the binding energy. $R_{\rm typ}$ represents the interaction range. 
Therefore, once we successfully determine the scattering length through femtoscopy 
and the binding energy through ordinal spectroscopy independently, 
we may be able to discuss the compositeness ($X$) of this system using the weak-binding relation 
(Eq.~\ref{eq:weak_binding}).
The $\Xi$-hypernucleus presents an intriguing system, as it may contain components 
resembling an H-dibaryon like structure. 
By measuring the correlation function between $\Xi^-$ and $\alpha$ particles in J-PARC-HI Femtoscopy, 
we may gain insights into the internal structure of the $\Xi$-hypernucleus.

As described above, the current Femtoscopic techniques derive pair-spin (isospin) averaged scattering parameters. 
Isospin decomposition, easier than spin decomposition, is achievable by measuring different isospin combinations; 
for example, $\Xi^- p$ can have $I = 0$ and 1, while $\Xi^0 p$ has only $I = 1$. 
Spin decomposition, however, is challenging, similar to situations in scattering experiments where only 
total cross-section ($\sigma$) is measured. 
Obtaining each partial wave requires additional observables like angular distribution ($d\sigma / d\Omega$) and spin observables such 
as analyzing power and depolarization. 
Similarly, decomposing the correlation function in Eq.~\ref{eq:Correlation function} 
may be possible by measuring spin observables. 
For $\Lambda p$, with $I = 1/2$, the total spin can be $S = 0$ or 1 in S-wave, 
suggesting potential for spin decomposition through $\Lambda p$ spin-spin correlation measurements.
Such a new technique to decompose the correlation function will be studied for the J-PARC-HI Femtoscopy. 

In the J-PARC-HI experiment, we plan to install a large acceptance spectrometer to measure the leptons and hadrons with
high statistics. 
Therefore, we are able to study not only the main topics, searching the QCD critical point and studying the high-density 
matter, but also the other topics such as hadron and strange nuclear physics. 
The world's highest heavy-ion beam intensity at J-PARC-HI opens a new era of the nuclear and hadron physics. 

-----------

\vfill  

\noindent{\bf References }
\begin{description}
\setlength\itemsep{-3pt}
\item{[1]} \url{https://j-parc.jp/researcher/Hadron/en/pac$\_$1607/pdf/LoI$\_$2016-16.pdf}
\item{[2]} {STAR Collaboration, Phys. Rev. Lett., {\bf 114}, 022301 (2015).}
\item{[3]} {ALICE Collaboration, Nature, {\bf 588}, 232 (2020).}
\item{[4]} {ALICE Collaboration, Phys. Rev. Lett., {\bf 123}, 112002 (2019).}
\item{[5]} {ALICE Collaboration, Phys. Lett. B, {\bf 802}, 135223 (2020).}
\item{[6]} {S.E. Koonin, Phys. Lett. B, {\bf 70}, 43 (1977).}
\item{[7]} {S. Pratt, T. Csorgo, and J. Zimanyi, Phys. Rev. C, {\bf 42}, 2646 (1990).}
\item{[8]} {R. Lednick\'{y} and V. Lyuboshits, Sov. J. Nucl. Phys.(Engl. Transl.);(United States) (1982).}
\item{[9]} {A. Andronic, Phys. Lett. B, {\bf 697}, 203 (2011).}
\item{[10]} {Y. Kamiya and T. Hyodo Prog. Theor. Exp. Phys. {\bf 2017}, 023D02 (2017). }
\item{[11]} {S. Weinberg, Phys. Rev. {\bf 137}, B672 (1965)}
\end{description}

\stepcounter{count}
\clearpage

\phantomsection
\addcontentsline{toc}{section}{
{\bf Clustering and deformation of $\Lambda$ hypernuclei with antisymmetrized molecular dynamics} \\
M.~Isaka}

\titl{Clustering and deformation of $\Lambda$ hypernuclei with antisymmetrized molecular dynamics}

\name{
Masahiro Isaka$^{1}$
}

\adr{
$^1$ Hosei University, 2-17-1 Fujimi, Chiyoda, Tokyo 102-8160, Japan
}


Since a $\Lambda$ particle is unaffected by the nuclear Pauli principle and the $\Lambda N$ interaction 
is different from the nuclear force, it can be regarded as an impurity in hypernuclei. 
Using the $\Lambda$ particle, it is possible to probe the structure of the core nuclei. 
In particular, the $\Lambda$ particle in \textit{p} orbit is of interest, because it would be 
sensitive to the core structure due to the spatial anisotropy of the \textit{p} orbit.
The aim of this work is to reveal the difference of the $\Lambda$ binding energy ($B_\Lambda$) on 
the core structure in both the ground and \textit{p} states. 
Since the excitation energy of the \textit{p} states is determined by the difference of $B_\Lambda$ 
between these states, it reflects the structure dependence of $B_\Lambda$. 
In this work, I focus on the Be hypernuclei whose core nuclei have the $2\alpha + Nn$ cluster structure, 
where the degree of the 2$\alpha$ clustering is dependent on the number of the extra neutron $N$. 
To investigate it, I have applied the HyperAMD [1,2] to $^{9}_\Lambda$Be, $^{10}_\Lambda$Be, 
$^{11}_\Lambda$Be and $^{13}_\Lambda$Be.  
In Fig. \ref{fig}, it is clearly seen that the core nuclei have the different quadrupole deformations 
which essentially come from the difference of the 2$\alpha$ clustering. 
It is found that the $B_\Lambda$ of the ground state becomes smaller 
as the nuclear quadrupole deformation decreases, 
whereas that in the \textit{p} states increases as the deformation enhances. 
As results, the excitation energies of the \textit{p} states becomes lower as the core deformation is enhanced. 
Therefore, the small excitation energy of the \textit{p} state can be a signature of the core deformation.

\begin{figure}[h]
\includegraphics[width=0.9\hsize]{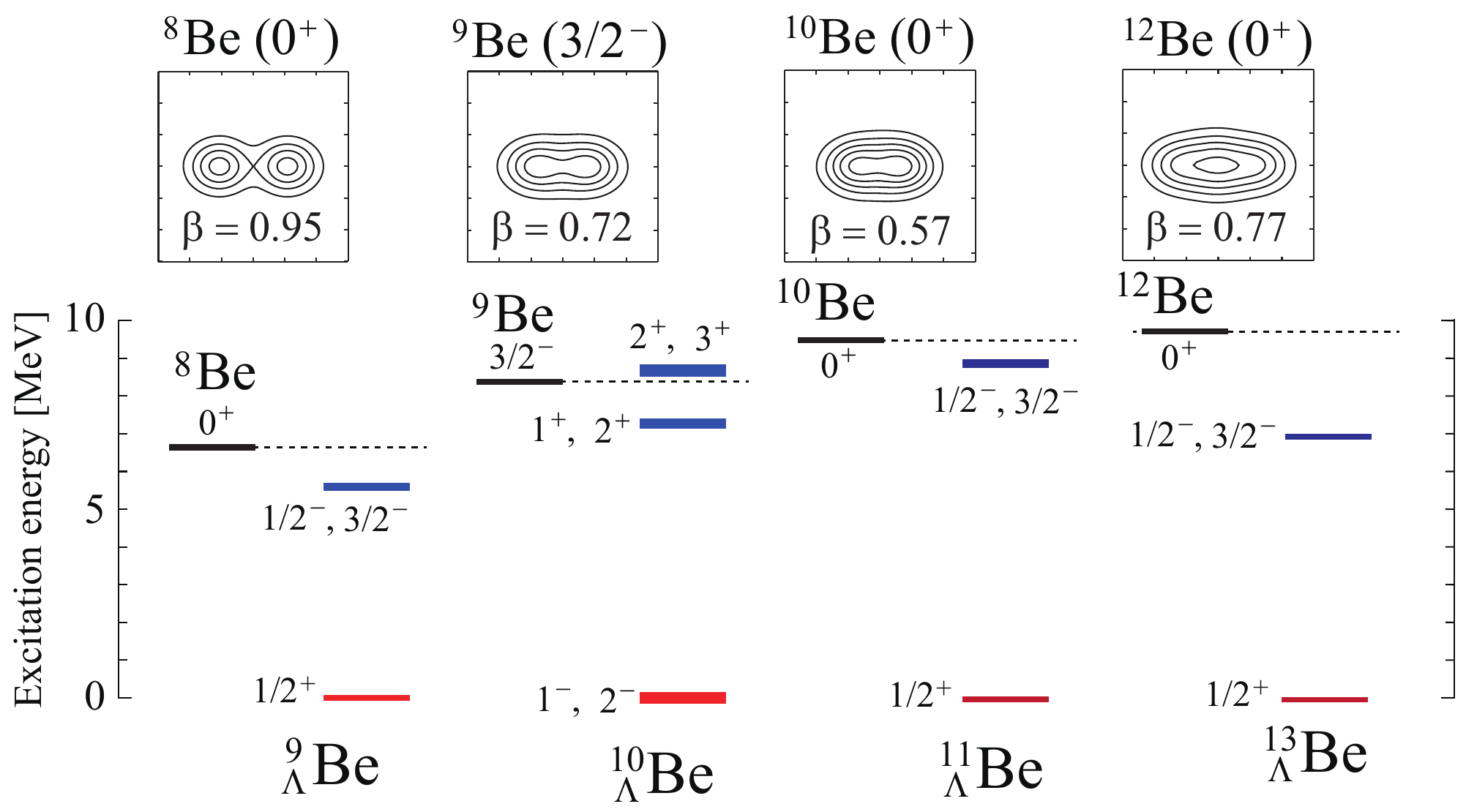}
 \caption{
 Excitation spectra of the ground (red line) and $p$ states (blue line) 
 in $^{9}_\Lambda$Be, $^{10}_\Lambda$Be, $^{11}_\Lambda$Be and $^{13}_\Lambda$Be,
 where the $^{A-1}$Be $+ \Lambda$ threshold energies are also shown by black dotted line. 
 Density distribution and nuclear quadrupole deformation $\beta$ of the intrinsic state 
 that has the largest contribution to the core ground state is also presented in top panels. 
 } 
 \label{fig}
\end{figure}

\vfill  

\noindent{\bf References }
\begin{description}
\setlength\itemsep{-3pt}
\item{[1]} M. Isaka, M. Kimura, A. Dote and A. Ohnishi, Phys. Rev. C{\bf 83} (2011) 044323. 
\item{[2]} M. Isaka, M. Kimura, A. Dote and A. Ohnishi, Phys. Rev. C{\bf 83} (2011) 054303.
\end{description}

\stepcounter{count}
\clearpage

\phantomsection
\addcontentsline{toc}{section}{
{\bf Low-energy $K^{+}N$ scattering revisited and in-medium strange quark condensate} \\
D.~Jido}

\titl{Low-energy $K^{+}N$ scattering revisited and in-medium strange quark condensate}

\name{
Daisuke Jido$^{1}$
}

\adr{
$^1$ Department of Physics, Tokyo Institute of Technology, Megro, Tokyo 152-8551, Japan
}


It is important to investigate the in-medium quark condensates to understand the mechanism of the spontaneous breaking of chiral symmetry. The up and down quark condensates in nuclear medium are studied by pionic atoms and low energy pion nucleus scattering, and it is found that the magnitude of the ud quark condensates may be reduced by 30\% at the nuclear saturation density [1]. This is known as partial restoration of chiral symmetry in nuclear medium. For a systematic study of partial restoration of chiral symmetry, it is interesting to see how the strange quark condensate behaves in nuclear matter. 

The chiral ward identity connects the in-medium quark condensate to the soft limit value of a correlation function of the pseudoscalar fields evaluated in nuclear medium [2]. For the strange quark condensate, one considers the correlation function of the pseudoscalar fields with strangeness. The correlation function describes in-medium propagation of kaon and it is obtained phenomenologically by kaon-nucleon scattering in the low density approximation. Thus, one can learn the behavior of the quark condensate in nuclear medium from phenomenologically determined kaon-nucleon scatterings at the soft limit. 

In this work [3] we describe the kaon-nucleon scattering amplitude in chiral perturbation theory and its twelve low energy constants are determined by existent $K^{+}N$ scattering data. The theoretical reproduction of the $K^{+}p$ scattering amplitude with $I=1$ is rather good up to $p_{\rm lab}=800$ MeV/c, where inelasticity becomes significant, while the $I=0$ amplitudes are not well reproduced. Especially the differential cross sections of the $K^{+}n$ elastic scattering, which are not used for fitting, are poorly described. This is because due to the lack of reliable data of $K^{+}n$ scattering in energies lower than $p_{\rm lab}=400$ MeV/c the theoretical scattering amplitude cannot be constrained well. Performing analytic continuation of the scattering amplitude obtained by chiral perturbation theory, we take the soft limit of the scattering amplitude and we evaluate the in-medium quark condensate $\langle \bar uu + \bar ss \rangle^{*}$ with a low density expansion. We find that the size of the quark condensate is reduced in nuclear medium. For further quantitative discussion, $I=0$ $KN$ scattering data in low energies are definitely important to constrain the theoretical parameters  and perform extrapolation of the amplitude to the soft limit. For one of the theoretical efforts, direct theoretical calculation of $K^{+}d \to KNN$ scattering is extremely desirable. 

\vfill  

\noindent{\bf References }
\begin{description}
\setlength\itemsep{-3pt}
\item{[1]} For example, T.~Nishi {\it et al}., N.\ Phys.\ 19, 788 (2023).

\item{[2]} D.~Jido, T.~Hatsuda, T.~Kunihiro, Phys.\ Lett.\ B 670, 109 (2008). 

\item{[3]} Y.~Iizawa, D.~Jido, S.~H\"ubsch, PTEP 2024, 053D01 (2024).

\end{description}

\stepcounter{count}
\clearpage

\phantomsection
\addcontentsline{toc}{section}{
{\bf $\Lambda$ potential at high densities studied by heavy-ion collision and hypernuclei} \\
A.~Jinno$^*$, K.~Murase, Y.~Nara, and A.~Ohnishi}

\titl{$\Lambda$ potential at high densities studied by heavy-ion collision and hypernuclei}

\name{
Asanosuke Jinno$^{1}$,
Koichi Murase$^{2}$,
Yasushi Nara$^{3}$, and
Akira Ohnishi$^{4,\dagger}$
}

\adr{
$^1$ Department of Physics, Faculty of Science, Kyoto University, Kyoto 606-8502, Japan

$^2$ Department of Physics, Tokyo Metropolitan University, Hachioji 192-0397, Japan

$^3$ Akita International University, Yuwa, Akita-city 010-1292, Japan

$^4$ Yukawa Institute for Theoretical Physics, Kyoto University, Kyoto 606-8502, Japan
}


The hyperon puzzle of neutron stars is the problem that most equations of state with hyperons are too soft to support the observed massive neutron stars.  One promising solution is three-baryon forces that are sufficiently repulsive to prevent the appearance of lambda hyperons ($\Lambda$) in neutron stars.  From chiral effective field theory ($\chi$EFT) with three-baryon forces estimated by the decuplet dominance approximation [1], a $\Lambda$ potential aligning with the scenario without $\Lambda$'s has been obtained.

In this proceedings, we shall examine whether this scenario is consistent with the $\Lambda$ binding energy of hypernuclei and the directed flow of $\Lambda$ in heavy-ion collisions.
We also discuss whether the difference in the experimental observables can be found when employing the repulsive and attractive $\Lambda$ potentials.

\begin{figure}[hbp]
\centering
\begin{minipage}[b]{.4\textwidth}
\includegraphics[width=\hsize]{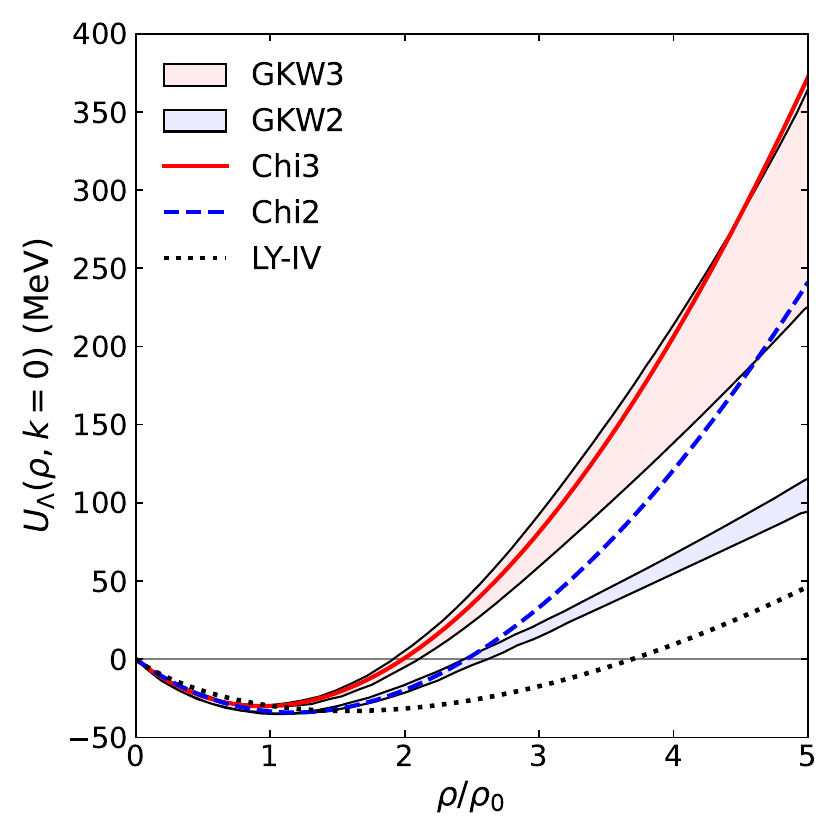}
\end{minipage}
\begin{minipage}[b]{.5\textwidth}
\includegraphics[width=\hsize]{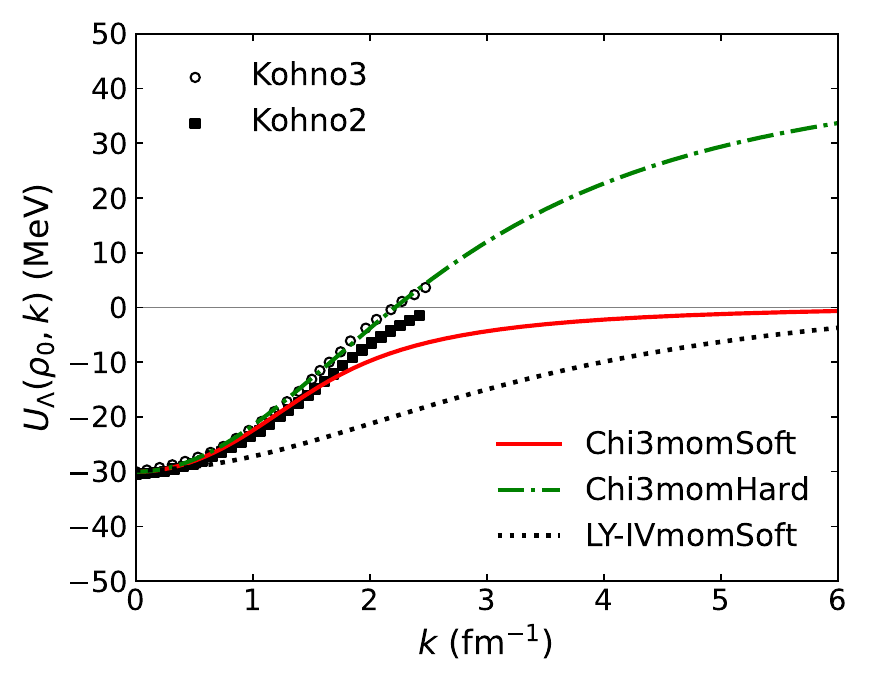}
\end{minipage}
\caption{(left panel) Density dependence of the $\Lambda$ potential.  GKW3 represents the results from $\chi$EFT with the two- and three-body forces~[1].  GKW2 is the results from $\chi$EFT without the three-body force~[1].  Chi3 (solid line) and Chi2 (dashed line) are fitted to GKW2 and GKW3 up to $\rho/\rho_0<1.5$, respectively.  LY-IV (dotted line) is a conventional $\Lambda$ potential [3].
(right panel) Momentum dependence of the $\Lambda$ potential.  Kohno3 represents the results from $\chi$EFT with the two- and three-body forces~[2].  Kohno2 is the results from $\chi$EFT without the three-body force~[2].  Chi3momSoft (solid line) and Chi3momHard (dash-dotted line) are constructed to reproduce Kohno3 up to $2.5~\text{fm}^{-1}$ and $1.0~\text{fm}^{-1}$, respectively.  LY-IVmomSoft (dotted line) is fitted to the momentum dependence of LY-IV up to $1.0~\text{fm}^{-1}$.
}
\label{fig:potentials}
\end{figure}

We employ three $\Lambda$-potential models and their variations.
We constructed the Chi3 potential by fitting the result of $\chi$EFT with the two- and three-body force [1,2] to the Skyrme-type $\Lambda$ potential
\begin{align}
    \label{eq:ULam}
    U_\Lambda(\rho,k) &= a \rho + b  \rho^{4/3} + c \rho^{5/3} + U_m(\rho, k), \\
    U_m(\rho,k) &= a^\Lambda_2 k^2 \rho,
\end{align}
where $\rho$ is the density of nuclear matter, $k$ is the momentum of $\Lambda$, and $a,~b,~c$, and $a^\Lambda_2$ are fitting parameters.
For reference, the Chi2 potential was similarly constructed without the three-body force.
The LY-IV potential is a conventional attractive $\Lambda$ potential [3], with which $\Lambda$'s appear in dense neutron star matter.  
The density dependence of the $\Lambda$ potentials is plotted in the left panel of Fig.~\ref{fig:potentials}.

For simulation of the heavy-ion collisions, we extrapolate the momentum-dependent potential $U_m$ in Eq.~\eqref{eq:ULam} to a high momentum region by assuming the Lorentzian form:
\begin{align}
    \label{eq:Um_HIC}
    U_m(\rho(x),k) = \dfrac{C}{\rho_0} \int d^3 k' \dfrac{f(x,k')}{1+\left[(\boldsymbol{k}-\boldsymbol{k'})/\mu\right]^2},
\end{align}
where $C$ and $\mu$ are fitting parameters, $\rho_0 = 0.168~\text{fm}^{-3}$ is the nuclear saturation density, and $f(x,k)$ is the single-particle distribution function.
In the actual heavy-ion simulations, we implement the momentum-dependent potential as the Lorentz vector $U^\mu_m$ [4].
Since $\chi$EFT cannot be applied above the momentum cutoff of $550~\text{MeV} \simeq 2.8~\text{fm}^{-1}$~[2], we prepared two variations: Chi3momHard and Chi3momSoft are constructed to reproduce the $\chi$EFT result [2] up to $2.5~\text{fm}^{-1}$ and $1.0~\text{fm}^{-1}$, respectively.
LY-IVmomSoft is constructed to reproduce the momentum dependence of LY-IV with Eq.~\eqref{eq:Um_HIC} up to $1.0~\text{fm}^{-1}$.
The momentum dependence of the $\Lambda$ potentials is plotted in the right panel of Fig.~\ref{fig:potentials}.
We note that the density dependence of Chi3momSoft and Chi3momHard is almost identical to that of Chi3, as is LY-IVmomSoft to LY-IV.

\begin{figure}[hbp]
\begin{minipage}[h]{.5\textwidth}
\includegraphics[width=\hsize]{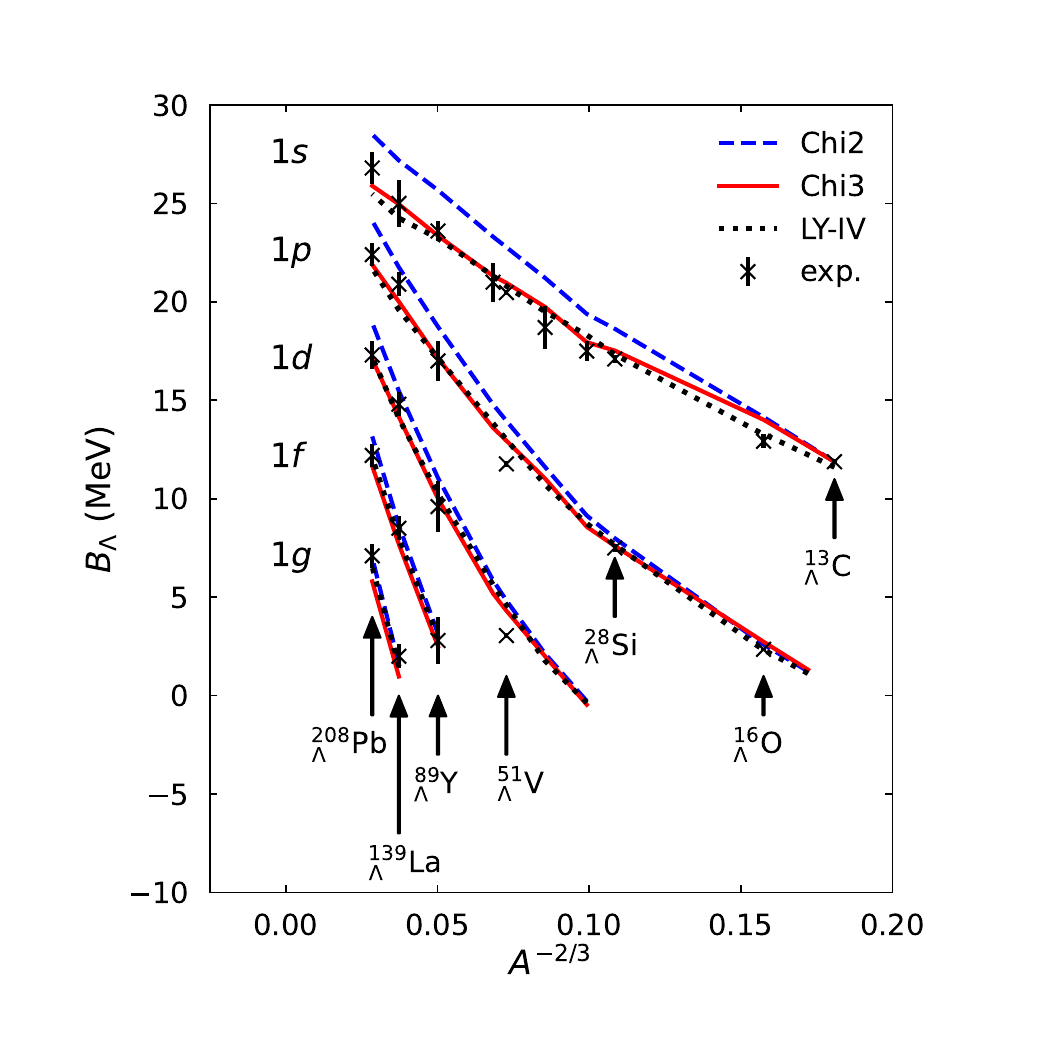}
\end{minipage}
\begin{minipage}[h]{.5\textwidth}
\includegraphics[width=\hsize]{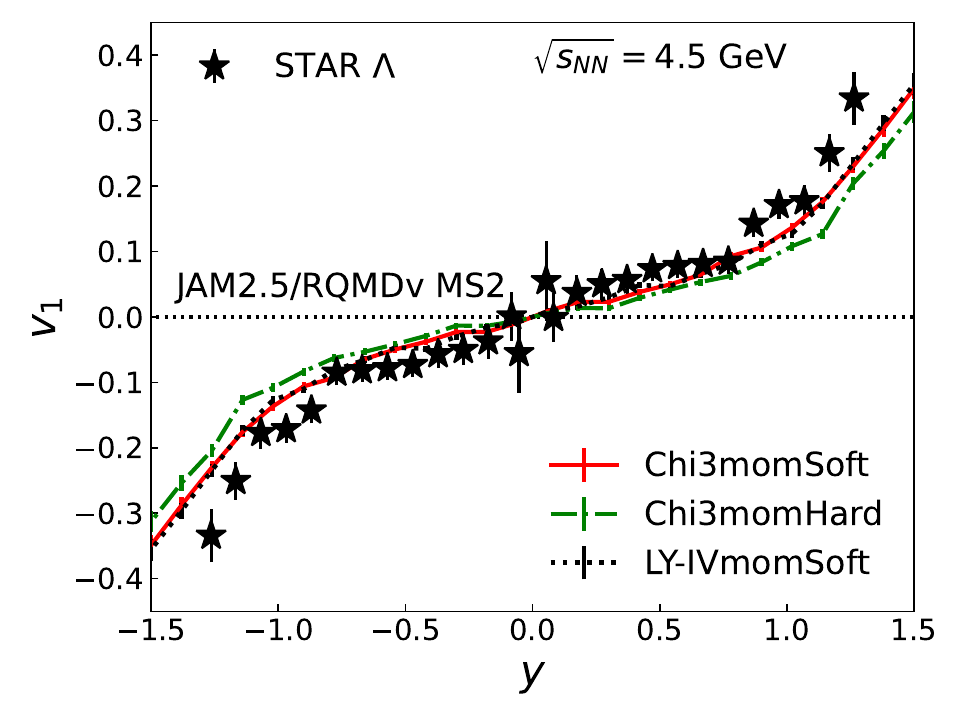}
\end{minipage}
\caption{(left panel) $\Lambda$ binding energy of the $\Lambda$ hypernuclei~[5]. Experimental data (cross) can be found in Ref.~[5].  (right panel) Directed flow of $\Lambda$ in mid-central Au+Au collisions at $\sqrt{s_{NN}}=4.5~\text{GeV}$ (right panel)~[4]. STAR data are taken from Ref.~[8].
}
\label{fig:results}
\end{figure}

We compare the binding energies of the hypernuclei [5] by employing the Skyrme-Hartee-Fock method using the above-mentioned three different $\Lambda$ potentials in Fig.~\ref{fig:results}. 
One parameter that cannot be determined from the uniform-matter results is tuned to reproduce the $\Lambda$ binding energy data of $^{13}_{~\Lambda} \mathrm{C}$\@.  We found that Chi2 overbounds by several $\text{MeV}$ due to the excessive potential depth at $\rho_0$.  In contrast, Chi3 reproduces the data as accurately as LY-IV\@.  Thus, Chi2 can be excluded, yet we need other data to distinguish the repulsive and attractive $\Lambda$ potentials.

Next, we consider the rapidity dependence of the $\Lambda$ directed flow,
\begin{equation}
v_1=\langle \cos \phi\rangle
=\biggl\langle \frac{p_x}{\sqrt{p_x^2+p_y^2}}\biggr\rangle,
\end{equation}
in heavy-ion collisions [4], where $\phi$ is the azimuthal angle measured from the reaction plane and $p_x$ and $p_y$ are the transverse momenta of a particle. We use the Lorentz vector version of the relativistic quantum molecular dynamics (RQMDv) model [6] implemented in the \texttt{JAM2} transport code [7].

The results of $v_1$ of $\Lambda$ in mid-central Au + Au collisions at $\sqrt{s_{NN}}=4.5~\text{GeV}$ are shown in the right panel of Fig.~\ref{fig:results} and compared with the STAR data~[8].
We found that both Chi3momSoft and LY-IV reproduce $v_1$ of $\Lambda$  with equal accuracy, implying that $v_1$ of $\Lambda$ is not so sensitive to the density dependence of the $\Lambda$ potential.  On the other hand, Chi3momHard underestimates $v_1$ of $\Lambda$, indicating that $v_1$ of $\Lambda$ is sensitive to the momentum dependence of the $\Lambda$ potential. 
Experimental information on the optical potential of $\Lambda$ may be useful for reducing the model uncertainty.

In these calculations, we used the $\Lambda$ potential for all other hyperons, including their resonance states. However,
large numbers of $\Sigma$ hyperons and hyperon resonances are populated during the evolution of  heavy-ion collisions.  For future studies, we plan to include different potentials for different hyperons to explore their effects on $v_1$ of $\Lambda$.

\vfill  

\noindent{\bf References }
\begin{description}
\setlength\itemsep{-3pt}
\item{[1]} D. Gerstung, N. Kaiser, and W. Weise, Eur.\ Phys.\ J.\ A {\bf 56} (2020) 175.
\item{[2]} M. Kohno, Phys. Rev. C {\bf 97} (2018) 035206.
\item{[3]} D. E. Lanskoy and Y. Yamamoto, Phys. Rev. C {\bf 55} (1997) 2330.
\item{[4]} Y. Nara, A. Jinno, K. Murase, and A. Ohnishi, Phys. Rev. C {\bf 106} (2022) 044902.
\item{[5]} A. Jinno, K. Murase, Y. Nara, and A. Ohnishi, Phys. Rev. C {\bf 108} (2023) 065803.
\item{[6]} Y. Nara and A. Ohnishi, Phys. Rev. C {\bf 105} (2022) 014911.
\item{[7]} Y. Nara, \url{https://gitlab.com/transportmodel/jam2.}
\item{[8]} J.~Adam \textit{et al.} [STAR],
Phys. Rev. C \textbf{103}, no.3, 034908 (2021).

\end{description}

\stepcounter{count}
\clearpage

\phantomsection
\addcontentsline{toc}{section}{
{\bf Hadron femtoscopy in heavy-ion collisions with memories of Prof. A.~Ohnishi} \\
Y.~Kamiya}

\titl{Hadron femtoscopy in heavy-ion collisions with memories of Prof. A.~Ohnishi }

\name{Yuki Kamiya
}

\adr{Helmholtz Institut f\"ur Strahlen- und Kernphysik and Bethe Center for Theoretical Physics, Universit\"at Bonn, D-53115 Bonn, Germany
}






The two-particle momentum correlation functions from high-energy nuclear collisions is beginning to be used to study hadron-hadron interaction. 
This technique, so called femtoscopy, enables us to directly access the interactions of the short-lived hadrons, which are difficult to measure in  usual scattering experiments~[1]. 
Furthermore, the correlation function is so sensitive to the low-energy interaction that 
it is useful to study the nature of near-threshold resonances and the underlying mechanism of the interaction. 
Professor Akira Ohnishi, who passed away prematurely on  May 16, 2023, has made a great effort to  develop  the femtoscopic technique for the study of hadron interactions from the very beginning. 
This talk has been devoted to review recent femtoscopic studies and his significant contributions. 

The correlation function is theoretically calculated with the Koonin-Pratt formula given as~[2, 3] 
\begin{align}
		C(\bm{q}) = \int d^3 r S(\bm{r})\left|\Psi^{(-)}(\bm{q};\bm{r})\right|^2, \label{eq:KP}
\end{align}
with the relative momentum in the pair rest frame $\bm{q}$, the normalized source function $S(\bm{r})$, and  the relative wave function with out-going boundary condition $\Psi^{(-)}(\bm{q};\bm{r})$.
Given the source function $S(\bm{r})$, the property of the interaction can be derived through the  wave function $\Psi^{(-)}(\bm{q};\bm{r})$.

Analyzing data from one particular source is obviously important, but investigating the source size dependence of the correlation using different sources can be very helpful to investigate the nature of  a bound state or a resonance. 
One of the most interesting pairs which shows the significant source size dependence is the  $p\Omega$ correlation function.
The $p\Omega$ correlation function from the large source with the heavy-ion collisions by the STAR collaboration shows the suppression~[4] while the data from small source obtained measured by the ALICE collaboration with the $pp$ collisions show the strongly enhanced correlation~[5]. 
This behavior of correlation is in good agreement with the theoretical calculations with the strongly attractive potential which supports a $N\Omega$ bound state~[6]. 
Such observations of  significant signal of a bound state in exotic channels encourage further experimental and theoretical studies to elucidate the nature of the bound state and the  hyperon interactions.



Recently, a new technique to study the $\Lambda N$ interaction at finite densities using the $\Lambda$-${}^4\mathrm{He}(\alpha)$  correlation function is proposed~[7]. 
Considering that the central density in $\alpha$ reaches about twice the normal nuclear density, 
the short range part of the $\Lambda\alpha$ interaction may reflect the behavior of the $\Lambda N$ interaction at high densities.
In fact, the repulsive nature of the $\Lambda$ potential at high densities induces the repulsive core in the $\Lambda\alpha$ interaction at short range.
The first theoretical predictions of the $\Lambda\alpha$ correlation function are calculated with the 
 $\Lambda\alpha$ potentials constructed from the Skyrme-type $\Lambda $ potentials and the phenomenological $\Lambda N$ potentials.
It is found that the $\Lambda\alpha$ correlation from the small sources is suppressed in the order of the repulsive strength of the $\Lambda\alpha$ potential at short range.
This indicates that the $\Lambda\alpha$ potential at short range can be constrained by measuring the $\Lambda\alpha$ correlation function.
In experiments, the high production rate of $\Lambda$ and $\alpha$ particles, which is required to the femtoscopic study,  can be achieved in the heavy ion collisions with the medium-collision energies.
Thus, future correlation data  from the experimental facilities such as J-PARC~HI is expected to give the important constraint on the $\Lambda $ potential in nuclear medium.




\vfill  


\begin{description}
\setlength\itemsep{-3pt}

\item{[1]} 	ExHIC, S.~Cho {\em et~al.},
	\newblock Prog. Part. Nucl. Phys. {\bf 95},  (2017) 279.
	
\item{[2]} 	S.~E. Koonin,
	\newblock Phys. Lett. {\bf 70B},  (1977) 43.
	
\item{[3]} 	S.~Pratt, T.~Csorgo, and J.~Zimanyi,
	\newblock Phys. Rev. {\bf C42},  (1990) 2646.
	
\item{[4]} 	STAR, J.~Adam {\em et~al.},
	\newblock Phys. Lett. {\bf B790},  (2019) 490. 
	
\item{[5]} 	ALICE, A.~Collaboration {\em et~al.},
	\newblock Nature {\bf 588}, (2020) 232,
	\newblock [Erratum: Nature 590, (2021)  E13].
	
\item{[6]} 	K.~Morita {\em et~al.},
	\newblock Phys. Rev. C {\bf 101}, (2020)  015201.
	
\item{[7]} 	A.~Jinno, Y.~Kamiya, T.~Hyodo, and A.~Ohnishi,
	\newblock Phys. Rev. C {\bf 110}, (2024)  014001.

\end{description}

\stepcounter{count}
\clearpage

\phantomsection
\addcontentsline{toc}{section}{
{\bf A study of neutron star matter based on a parity doublet model with $a_0(980)$ meson effect} \\
Y.K.~Kong$^*$, M.~Harada}

\titl{A study of neutron star matter based on a parity doublet model with $a_0(980)$ meson effect}

\name{
Yuk Kei Kong$^{1}$ and Masayasu Harada$^{2,1,3}$
}

\adr{
$^1$ Department of Physics, Nagoya University, Nagoya 464-8602, Japan
}
\adr{
$^2$ Kobayashi-Maskawa Institute for the Origin of Particles and the Universe, Nagoya University, Nagoya, 464-8602, Japan
}
\adr{
$^3$ Advanced Science Research Center, Japan Atomic Energy Agency, Tokai 319-1195, Japan
}


Spontaneous chiral symmetry breaking (SCSB) is a crucial property in low-energy hadron physics as it provides some clues to  understand the complicated properties of hadrons and its underlying dynamics. For example, SCSB contributes to hadron masses and causes the mass difference between chiral partners. Specifically, exploring the origin of nucleon mass based on the chiral symmetry structure is particularly intriguing. In recent decades, increasing interest is focused on the hadronic and qaurk matter at finite temperature and/or density which is accessible through experimental and astrophysical studies. We expect many interesting phenomena
in these matter in extreme conditions such as partial chiral restoration or color-superconductivity in super dense cold QCD matter. By studying these yet well-understood phenomenon, it will further our understanding to the QCD and its underlying dynamics.
\newline

Parity doublet model (PDM) proposed in Ref.~[1] is one of the promoting models to study the low-energy hadron physics. 
In the PDM, an excited nucleon such as $N(1535)$ is regarded as the chiral partner to ordinary nucleon. The SCSB 
generates the mass difference between them. If the chiral symmetry was not broken, their masses would be degenerated into, so called, the chiral invariant mass $m_0$. The existence of the chiral invariant mass and its possible relation to the parity doubling of nucleons is supported by the lattice QCD simulation such as in Refs.~[2,3].
In addition, recent analysis based on the QCD sum rules~[4] also supports the existence which suggests that the origin of the chiral invariant mass is the gluon condensate. Therefore, quantitative and qualitative study of the chiral invariant mass will help us to understand the origin of hadron masses.
\newline

In this talk, we have presented our work~[5] on the effect of the isovector scalar $a_0(980)$ meson, which is sometimes called the $\delta$ meson, on the properties of nuclear matter and the neutron star (NS) matter by constructing a parity doublet model with including the $a_0(980)$ meson based on the chiral U(2)$_L \times$ U(2)$_R$ symmetry with explicit symmetry breaking due to current quark masses of up and down quarks and U(1)$_A$ anomaly. 
The isovector scalar meson account for the attractive force in the isospin asymmetric matter and withstand the repulsive force of $\rho$ meson, and therefore should be considered when studying isospin asymmetric matter such as neutron star matter. 
\newline

\begin{figure}[h]
\centering
\begin{minipage}{.48\linewidth}
    \includegraphics[scale=0.48]{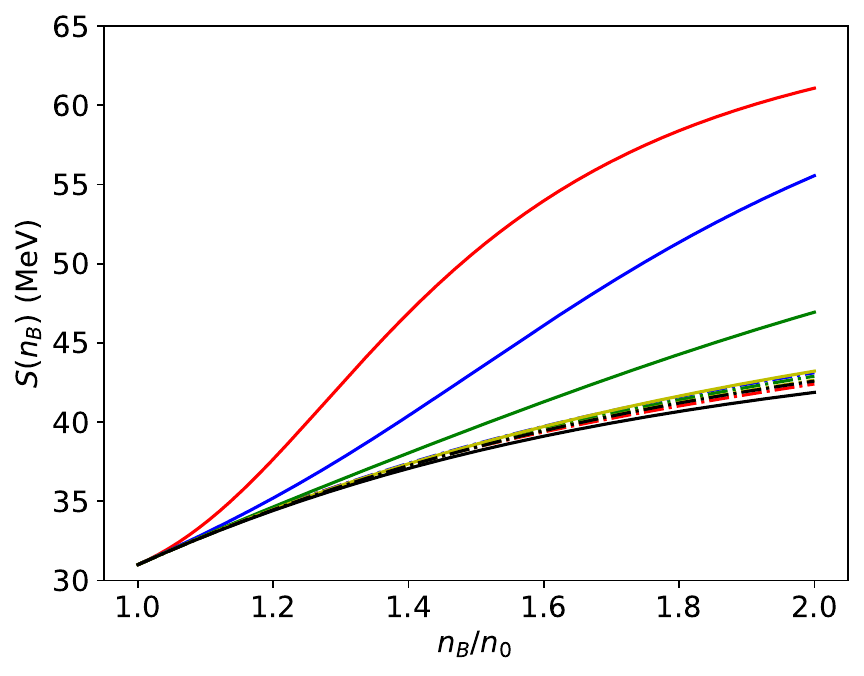}
    \caption{Symmetry energy $S(n_B)$ 
for 
$m_0 = 500$-$900$\,MeV, with the slope parameter $L_0 = 57.7$\,MeV. Solid curves represent the $S(n_B)$ of the model including $a_0(980)$, while the dash-dot curves show the results of the model without $a_0(980)$. Details of the model can be found in Ref.~[5].}
    \label{Scompare with VM L0060}
\end{minipage}
\hfill
\begin{minipage}{.48\linewidth}
    \includegraphics[scale=0.46]{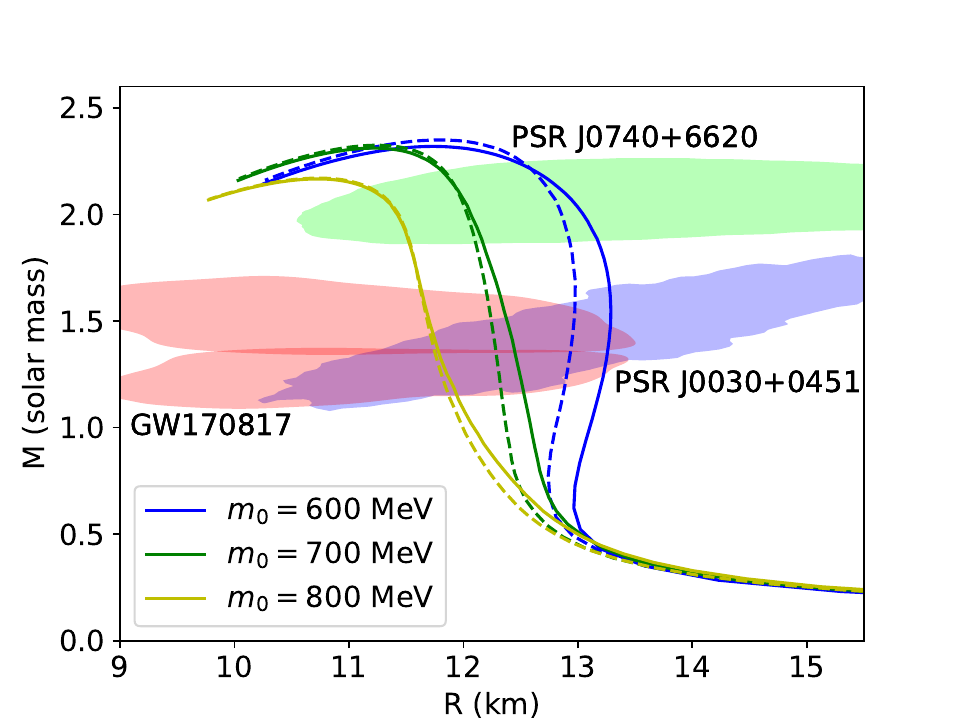}
    \caption{$M$-$R$ relations 
for $m_0 = 600$-$800$\,MeV with $L_0=57.7$\,MeV. Solid curves represent the $M$-$R$ relations from the model with $a_0(980)$ meson and dashed curves the ones of the model without $a_0(980)$.
More details of the parameters can be found in Ref.~[5].}
    \label{30}
\end{minipage}
\end{figure}

In our work, we first study the nuclear matter properties and find that the presence of $a_0(980)$ meson enlarges 
the symmetry energy depending on the value of $m_0$ (See \autoref{Scompare with VM L0060}). The result can be understood as a result of the competition between the repulsive $\rho$ meson interaction and the attractive $a_0(980)$ interaction, in addition to the kinetic contribution from the nucleons. On the other hand, in the model without $a_0$ meson, only repulsive contributions exist. Since the symmetry energy at the saturation density is fixed as $S_0=31$\,MeV in both models with and without $a_0(980)$ meson, the $\rho$ meson coupling $g_{\rho NN}$ is strengthened by the existence of the attractive $a_0(980)$ contribution in the model with $a_0$ comparing to the model without $a_0$.
\newline

In addition, we also studied the effect of the $a_0(980)$ meson to the neutron star properties by extending the PDM-NJL crossover model developted in Refs.~[6,7], in which 
the equation of state (EoS) of PDM is smoothly connected to an NJL-type quark model following the construction introduced in Ref.~[8], to include the $a_0(980)$ meson.
\autoref{30} shows the comparison of the neutron star mass-radius ($M$-$R$) relation 
in the model with $a_0(980)$ to the relation in the model withtout $a_0$. 
We observed that the inclusion of isovector scalar meson $a_0(980)$ increases the radius of NS by around 1\,km depending on the model parameters. By comparing the model predictions to the recent NS observational data, we constrained the chiral invariant mass of the nucleon as
 $ 580 \text{ MeV} \lesssim m_0 \lesssim 860 \text{ MeV}$ for $L_0=57.7$\,MeV.
\newline



\vfill  

\noindent{\bf References }
\begin{description}
\setlength\itemsep{-3pt}

\item{[1]} C. DeTar and T. Kunihiro, Phys. Rev. D {\bf 39}, 2805 (1989).

\item{[2]} G. Aarts et al., Phys. Rev. D {\bf 92}, 014503 (2015).

\item{[3]} G. Aarts, et al., Journal of High Energy Physics 2017, 34 (2017).

\item{[4]} J. Kim and S. H. Lee, Phys. Rev. D {\bf 105}, 014014 (2022).

\item{[5]} Y. Kong, T. Minamikawa, and M. Harada, Phys. Rev. C {\bf 108}, 055206 (2023).

\item{[6]} 
T.~Minamikawa, T.~Kojo and M.~Harada,
Phys. Rev. C \textbf{103}, 045205 (2021).

\item{[7]}
T.~Minamikawa, B.~Gao, T.~Kojo and M.~Harada,
Symmetry \textbf{15}, 745 (2023).

\item{[8]}
G. Baym et al., Reports on Progress in Physics 81, 056902 (2018).


\end{description}

\stepcounter{count}
\clearpage

\phantomsection
\addcontentsline{toc}{section}{
{\bf ($\pi^{-}, K^{0}$) spectroscopy with S-2S and HIHR} \\
K.~Miwa}

\titl{($\pi^{-}, K^{0}$) spectroscopy with S-2S and HIHR}

\name{
Koji Miwa$^{1}$
}

\adr{
$^1$ Tohoku University, Sendai, 980-8578, Japan
}


{\bf Introduction : }
From the viewpoint of the study of two-body $\Lambda N$ interactions, it is important to study $^6_{\Lambda}$He, the only bound $\Lambda$ hypernucleus in the $A=6$ system, 
because $^6_{\Lambda}$He has an $\alpha+(\Lambda+n)$ cluster structure. 
Especially, $^6_{\Lambda}$He is an ideal nucleus for determining the spin-spin force of the two-body $\Lambda N$ interaction. 
Studies of $\Lambda$ hypernuclei have been done for hypernuclei with three-body clusters of $\Lambda NN$ such as $^7_{\Lambda}$Li ($\alpha+p+n+\Lambda$)  so far [1].
Therefore, the $\Lambda NN$ interaction might contribute a lot to the energy splitting of $\Lambda$ hypernuclear spin-doublets.
If the energy splitting of the ground spin doublets of $^6_{\Lambda}$He is much smaller than that expected from the hypernuclear studies so far, 
it would indicate that the $\Lambda N$ interaction is strongly influenced by three-body forces. 
It is known that $^6_{\Lambda}$He has a bound ground state from the emulsion measurement [2].
Although its spin and parity are not known, the ground state is predicted to be the 1$^{-}$ state in which the spin of the $\Lambda N$ system is 0 [3]. 
Since $^6$Li has an $\alpha+d$ cluster structure, when $^6_{\Lambda}$He is generated using the ($\pi^-, K^0$) reaction, which does not change the spin, 
it is possible to populate the 2$^-$ state in which $\Lambda N$ is paired with spin 1. 
By comparing the $B_{\Lambda}$ value of this 2$^-$ state with the value measured in the emulsion, we can verify the spin dependence of the $\Lambda N$ interaction.

In addition, establishing the experimental technique of the ($\pi^-, K^0$) spectroscopy that converts a proton into $\Lambda$ is extremely important for studying neutron-rich $\Lambda$ hypernuclei, and also for studying charge-symmetric $\Lambda$ hypernuclei by combining it with the ($\pi^+, K^+$) reaction. 
Therefore, we have investigated a method to realize ($\pi^-, K^0$) spectroscopy at J-PARC.

{\bf Consideration of realistic ($\pi^-, K^0$) spectroscopy : }
We consider ($\pi^-, K^0$) spectroscopy with a $\pi^-$ beam momentum of 1.05 GeV/$c$. 
$K^{0}_{S}$ with momentum of about 0.7 GeV/$c$ emitted in a forward angle is identified from the $K^{0}_{S} \to \pi^+\pi^-$ decay. 
When $\pi^-$ is emitted forward with high momentum of about 0.7 GeV/$c$, $\pi^+$ with low momentum of about 0.1 GeV/$c$ is emitted at a large angle over 70 degrees. 
We believe that high-resolution ($\pi^-, K^0$) spectroscopy will be possible by performing momentum analysis of the particles emitted forward with a high-resolution magnetic spectrometer such as S-2S or HIHR, and detecting the low-momentum $\pi^+$ emitted in the sideward direction with a calorimeter.
We consider S-2S as the forward spectrometer, and assume the momentum resolution to be 0.05\% in FWHM. 
Assuming that the energy resolution of the side calorimeter is 1 MeV ($\sigma$) for 100 MeV, a missing mass resolution of 0.66 MeV ($\sigma$) can be expected when the thickness of the $^6$Li target is 1 cm. 
The detection efficiency of $K^0$ is approximately 0.8\%, taking into account the S-2S acceptance, the side detector acceptance, the $\pi$ survival rates, and the proportion of $K^0_S$ of 50\%. 
With a $\pi^-$ beam of $10^7$ /spill, 125 $^6_{\Lambda}$He events can be detected in 20 days of beamtime, and the statistical precision of determining the center of the $\Lambda$ binding energy is approximately 60 keV in $\sigma$. 
The yield can be increased by thickening the target from 1cm, but the energy loss of the $\pi^-$ beam in target directly affects the missing mass resolution.
Therefore the yield and resolution must be balanced.
The masses of $\Lambda$ and $^{12}_{\Lambda}$B, which can be produced using a CH$_2$ target, can be used for calibration of $B_{\Lambda}$. 

{\bf Possibility of ($\pi^-, K^0$) spectroscopy with HIHR : }
From here, we will discuss the possibility of ($\pi^-, K^0$) spectroscopy using HIHR. 
The advantage of using HIHR is that by using a high-intensity $\pi^-$ beam of $10^8$ /spill, 
sufficient yield can be expected even with a thin target (about 1 mm if $^6$Li is used). 
By making the target thinner, it is expected that the resolution would improve to about 300 keV ($\sigma$). 
Since the same level of resolution can be expected for both the ($\pi^+, K^+$) and ($\pi^-, K^0$) reactions, 
it would be possible to systematically investigate the same states of the charge-symmetric $\Lambda$ hypernuclei.
Then we can discuss the effect of charge symmetry breaking in the $\Lambda N$ interaction. 
In addition, by combining the ($\pi^-, K^+$) reaction, it will be possible to systematically measure the isotope dependence of the $\Lambda$ hypernuclei.
However, the current detector configuration of HIHR cannot measure the momenta of the beam and the scattered particles, 
so ($\pi^-, K^0$) spectroscopy cannot be performed with the present setup. 
In order to realize the ($\pi^-, K^0$) reaction, it is necessary to get the momentum, angle, and position information of $\pi^-$ from $K^0$ decay at the target position. 
To do this, a position detector must be installed between the two dipole magnets D1S and D2S of the forward spectrometer. 
By combining this with the information from the focal plane detector, the momentum of the scattered particle, its position  and its angle on the target can be determined. 
It is also possible to measure the momentum of the beam from the horizontal position on the target.
Then we can make conventional missing mass spectroscopy possible. 
By measuring the $\pi^+$ emitted in the large angle in the $K^0$ decay with a calorimeter or another magnetic spectrometer, 
($\pi^-, K^0$) spectroscopy becomes possible. 
Although matrix tuning  would be required, the excellent resolution of 300 keV ($\sigma$) is great advantage of HIHR.

-----------

\vfill  

\noindent{\bf References }
\begin{description}
\setlength\itemsep{-3pt}
\item{[1]} H. Tamura {\it et al.}, Phys. Rev. Lett. 84, 5963 (2000)
\item{[2]} M. Juric {\it et al.}, Nucl. Phys. B52 (1973) 1
\item{[3]} T. Myo and E. Hiyama, Phys. Rev. C 107, 054302 (2023)
\end{description}

\stepcounter{count}
\clearpage

\phantomsection
\addcontentsline{toc}{section}{
{\bf Hadron Physics with Photon Beam at SPring-8/LEPS2 and JLab/GlueX} \\
K.~Mizutani}

\titl{Hadron Physics with Photon Beam at SPring-8/LEPS2 and JLab/GlueX}

\name{
Keigo Mizutani$^{1}$ \\
}

\adr{
$^1$ Research Center for Nuclear Physics, Osaka University, Japan \\
}

Hadron physics experiments using photon beams are conducted at SPring-8/LEPS2 in Japan and JLab/GlueX in the USA. These facilities provide unique opportunities to study the structure and interactions of hadrons using high-energy photon beams.

SPring-8/LEPS2, located in Hyogo, Japan, utilizes backward Compton scattering to produce photon beams with energies around 2 GeV. This facility has been instrumental in investigating threshold production processes, which provide insights into hadron structure and dynamics. In contrast, JLab/GlueX, located in Virginia, USA, employs coherent bremsstrahlung to produce photon beams with energies around 9 GeV. The GlueX experiment aims to map out the spectrum of light quark mesons and search for exotic states, providing crucial information about the nature of confinement in Quantum Chromodynamics (QCD). The photon beam at JLab/GlueX, characterized by well-defined energy, flux, and polarization, serves as an ideal probe for studying hadron structure and its production mechanisms.

One of the key aspects of these experiments is the study of the photoproduction of $\phi$ and $J/\psi$ mesons. In $\phi$ photoproduction, significant non-monotonic structures have been observed in the energy dependence of the cross-section near the threshold [1, 2], which cannot be explained by current theoretical models. Similarly, experimental results suggest the presence of non-monotonic structures in $J/\psi$ photoproduction [3], which are indicative of complex underlying processes. Studying $J/\psi$ photoproduction near threshold is particularly motivated by its potential to probe the gluon distribution inside the proton, providing deeper insights into its internal structure.

To explain these experimental results, it is necessary to identify the production mechanisms using decay angular distribution measurements. This approach will help in understanding the dynamics of these interactions and contribute to refining theoretical models.

In conclusion, photon beam experiments at SPring-8/LEPS2 and JLab/GlueX continue to advance our understanding of hadron physics. These facilities provide essential data for exploring the properties of hadrons and their interactions, contributing to the ongoing efforts to unravel the complexities of the strong force that binds quarks together.

\vfill  

\noindent{\bf References }
\begin{description}
\setlength\itemsep{-3pt}
\item{[1]} T. Mibe \textit{et al.} (LEPS Collaboration), Phys. Rev. Lett. {\bf 95}, 182001 (2005).
\item{[2]} K. Mizutani \textit{et al.} (LEPS Collaboration), Phys. Rev. C {\bf 96}, 062201(R) (2017).
\item{[3]} S. Adhikari \textit{et al.} (GlueX Collaboration), Phys. Rev. C {\bf 108}, 025201 (2023).
\end{description}

\stepcounter{count}
\clearpage

\phantomsection
\addcontentsline{toc}{section}{
{\bf Lambda(1405) in the flavor SU(3) limit from lattice QCD} \\
K.~Murakami$^*$, S.~Aoki for HAL QCD Collaboration}

\titl{Lambda(1405) in the flavor SU(3) limit from lattice QCD}

\name{
Kotaro~Murakami$^{1, 2}$, Sinya~Aoki$^{3}$ \\ for HAL QCD Collaboration
}

\adr{
$^1$ Department of Physics, Tokyo Institute of Technology, 2-12-1 Ookayama, Megro, Tokyo 152-8551, Japan \\
$^2$ Interdisciplinary Theoretical and Mathematical Sciences Program (iTHEMS), RIKEN, Wako 351-0198, Japan \\
$^3$ Yukawa Institute for Theoretical Physics, Kyoto University, Kitashirakawa Oiwakecho, Sakyo-ku, Kyoto 606-8502, Japan
}


We study $\Lambda(1405)$ by analyzing S-wave meson-baryon interactions in lattice QCD. 
One of the promising properties of $\Lambda(1405)$ is the two-pole structure~[1, 2]; the peak in the spectrum corresponding to $\Lambda(1405)$ is composed of two poles rather than a single pole. 
Our goal is to examine the existence and mechanism of such a two-pole structure using the HAL QCD method~[3-5], in which we extract the scattering phase shifts from lattice QCD via the interaction potentials.
As a first step towards the goal, we investigate $\Lambda(1405)$ in the flavor SU(3) limit. 
In our study, we focus on the S-wave meson-baryon systems in the singlet channel and the two octet channels (denoted by $8_{s}$ and $8_{a}$), where the poles corresponding to $\Lambda(1405)$ have been predicted in the chiral unitary model~[2].
In our numerical calculation, we use gauge configurations in the flavor-SU(3) limit on $32^4$ lattice volume~[6],
whose lattice spacing $a=0.121(2)~\textrm{fm}$.
The meson and baryon masses in our setup are $m_M=671.2(1.5)~\textrm{MeV}$, $m_{B}=1488.8(3.9)~\textrm{MeV}$, respectively. 
In this analysis, we ignore the coupling of $8_{s}$ and $8_{a}$ channels and perform the single-channel analysis in each channel.

\begin{figure}[h]
    \begin{center}
        \includegraphics[width=0.40\textwidth]{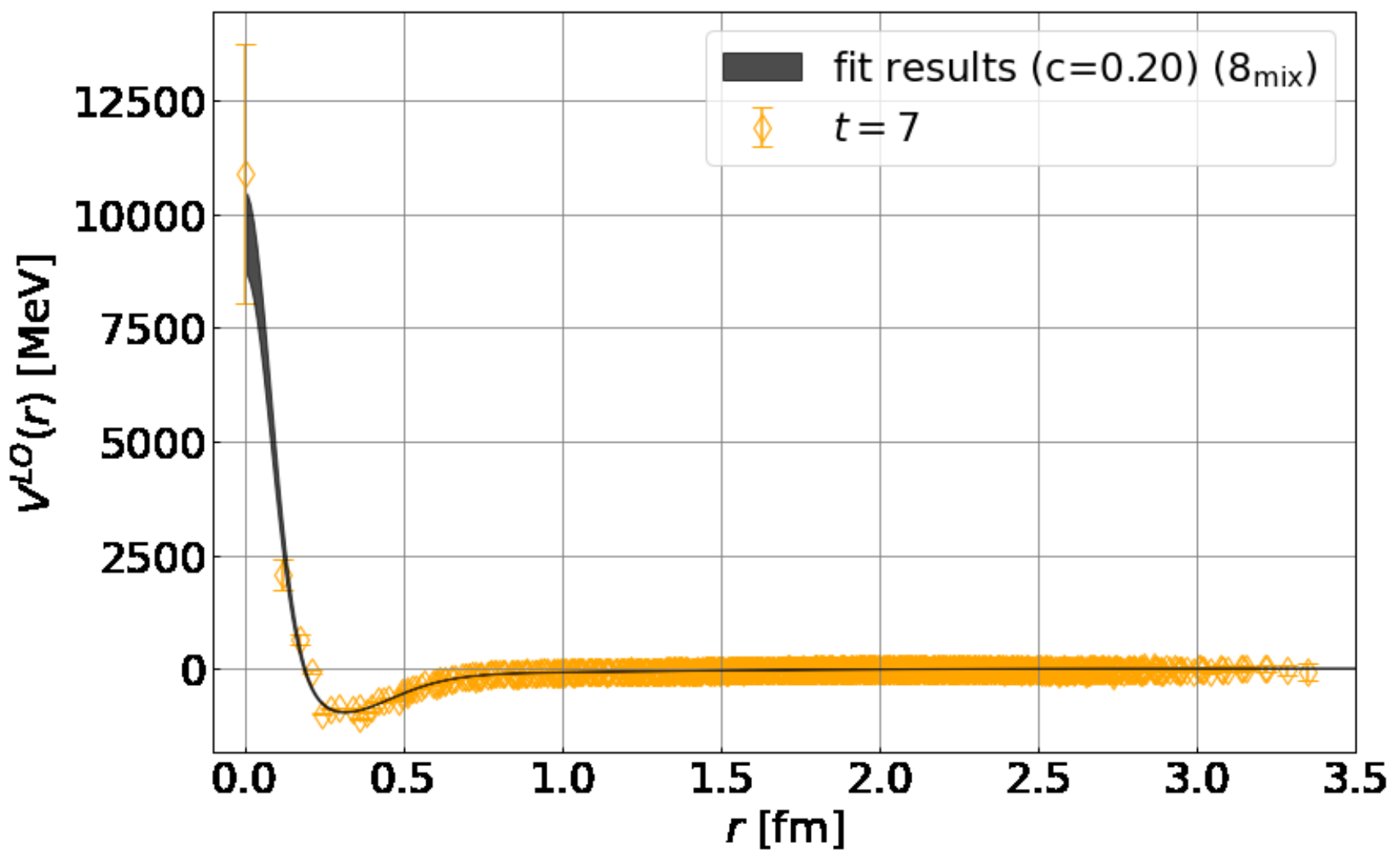}
        \includegraphics[width=0.40\textwidth]{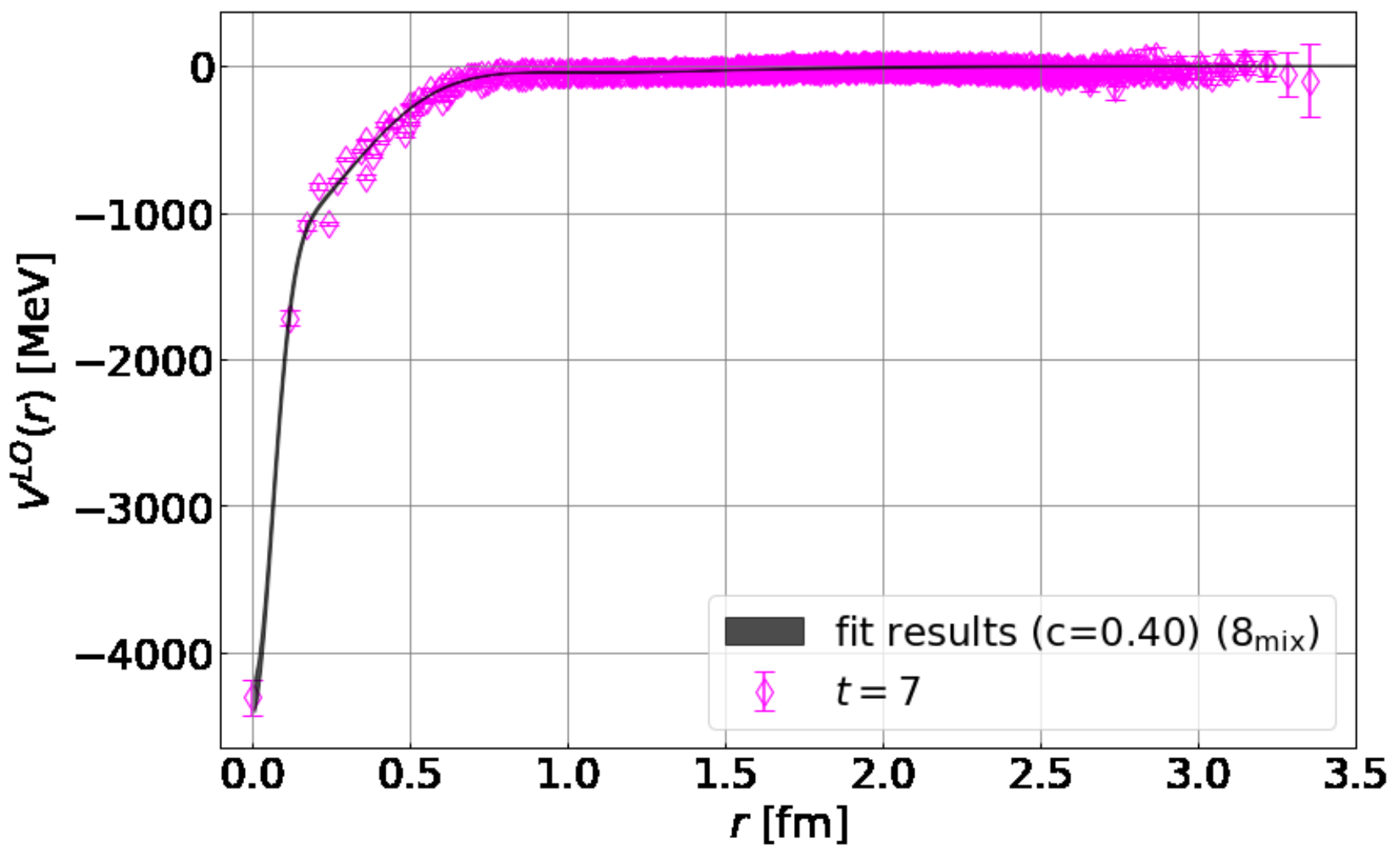}
        \includegraphics[width=0.40\textwidth]{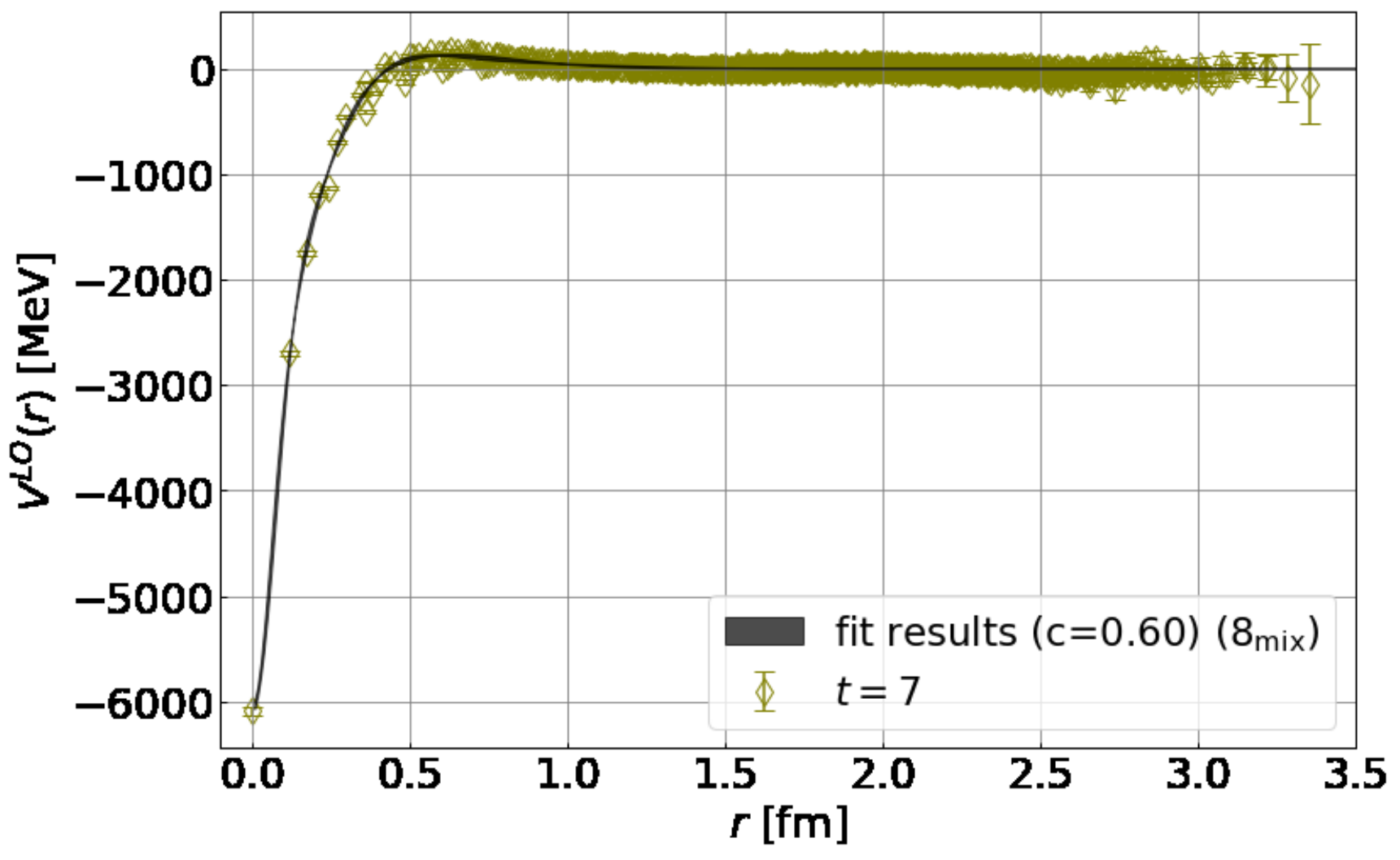}
        \includegraphics[width=0.40\textwidth]{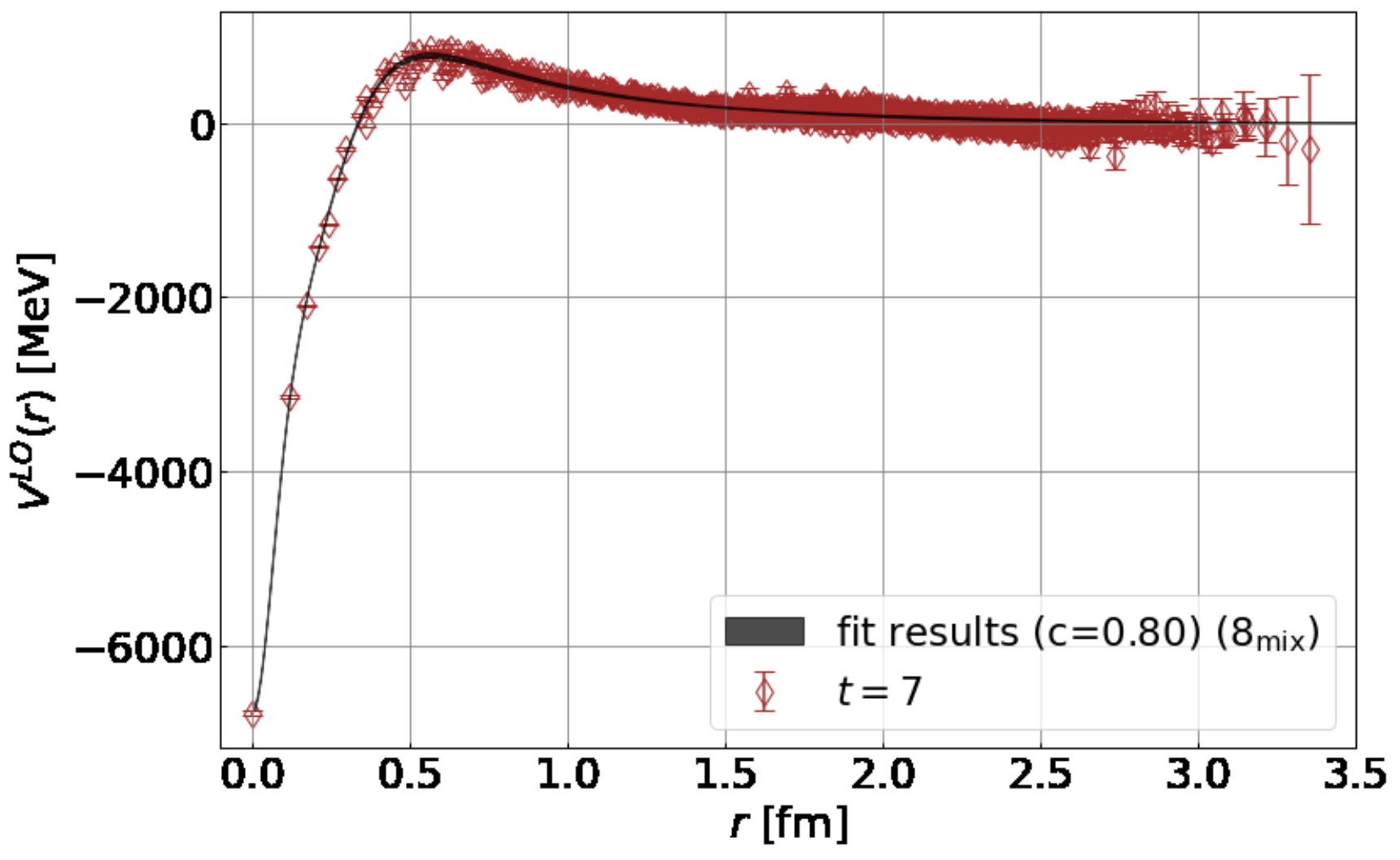}
	    \caption{Leading order potentials for different relative weights $c$. Black bands correspond to the fit results. 
        }
	\label{fig:results_pot_mix}    
    \end{center}
\end{figure}
Our results of the $R$-correlator (a kind of wave function) have a vanishing point in all channels.
Thus, the leading order potential obtained by the single channel analysis has a singular point, which makes it difficult to obtain reliable binding energies. 
To overcome this problem, assuming that states in the $8_{s}$ and $8_{a}$ channels approximately degenerate at low energy, we take a linear combination of two octet $R$-correlators as $R^{8_s}({\bf r},t) - c R^{8_a}({\bf r},t)$ with a constant $c$, such that it does not cross zero.
As seen in Figure.~\ref{fig:results_pot_mix}, the potential calculated from the linear combination shows strong attraction without singularities, though its shape depends on the relative weight $c$.

To extract the binding energy, we first fit the obtained potential by the multi-Gaussian function and then solve the Schr\"{o}dinger equation. 
The fit results are represented by black bands in Figure.~\ref{fig:results_pot_mix}.
We find that binding energies for different $c$ have similar values despite different shapes of potentials, which indicates that they generate almost identical dynamics at least in the low-energy regime. 
Our estimation of the binding energy in the octet channel is $E^{8}_{\textrm{bind}}=163(7)(^{+16}_{-64})~\textrm{MeV}$, where the first and second parentheses correspond to the statistical error and systematic error estimated from the $c$ dependence.
The final result is consistent with the estimation from the two-point correlation function with the single $\Lambda$ operator, $156(8)~\textrm{MeV}$, within errors. 
This indicates that approximation and assumption used in our analysis are reasonable.

Nevertheless, it is difficult to draw a physical interpretation of the dynamics, since shapes of the obtained potentials depend on the relative weight $c$, which is introduced to avoid zero in $R$-correlators. 
The appearance of zero in $R$-correlators may indicate strong non-local effects in interactions, which we ignore in this study. 
Such non-locality, together with effects from the mixing and the breaking of the degeneracy for the bound states between the two octet channels, should be examined in future.
Furthermore, we cannot apply the linear combination used for the octet channel to the singlet channel since we have only one $R$-correlator.
Therefore, another technique to avoid the singular point of the potential is necessary for the singlet channel, which is left for future works.

We would like to thank the members of the HAL QCD Collaboration for fruitful discussions.
We also appreciate Profs. D.~ Jido and M.~Oka for their useful comments. 
We thank Prof. T.~Inoue and ILDG/JLDG~[7] for providing us with gauge configurations used in this paper. 
We use the lattice QCD code of Bridge++ (http://bridge.kek.jp/Lattice-code/)~[8] and its optimized version by Dr. I. Kanamori~[9].
Our research uses computational resources of Wisteria/BDEC-01 Odyssey (the University of Tokyo), provided by the Multidisciplinary Cooperative Research Program in the Center for Computational Sciences, University of Tsukuba.
K.~M. is supported in part by JST SPRING, Grant Number JPMJSP2110, by Grants-in-Aid for JSPS Fellows (Nos.\ JP22J14889, JP22KJ1870), and by JSPS KAKENHI Grant No.\ 22H04917.
This work is also supported in part by JPMXP1020230411.

-----------

\vfill  

\noindent{\bf References }
\begin{description}
\setlength\itemsep{-3pt}
\item{[1]} J.~A.~Oller and U.~G.~Meissner, Phys. Lett. B \textbf{500}, 263-272 (2001).
\item{[2]} D.~Jido, J.~A.~Oller, E.~Oset, A.~Ramos and U.~G.~Meissner, Nucl. Phys. A \textbf{725}, 181-200 (2003).
\item{[3]} N.~Ishii, S.~Aoki and T.~Hatsuda, Phys. Rev. Lett. \textbf{99}, 022001 (2007). 
\item{[4]} S.~Aoki, T.~Hatsuda and N.~Ishii, Prog. Theor. Phys. \textbf{123}, 89-128 (2010).
\item{[5]} N.~Ishii \textit{et al.} [HAL QCD], Phys. Lett. B \textbf{712}, 437-441 (2012).
\item{[6]} T.~Inoue [HAL QCD], PoS \textbf{CD15}, 020 (2016).
\item{[7]} T.~Amagasa \textit{et al.} J. Phys. Conf. Ser. \textbf{664}, no.4, 042058 (2015).
\item{[8]} S.~Ueda, \textit{et al.} J. Phys. Conf. Ser. \textbf{523}, 012046 (2014).
\item{[9]} I.~Kanamori and H.~Matsufuru, Lect. Notes Comput. Sci. {\bf 10962}, 456–471 (2018).
\end{description}

\stepcounter{count}
\clearpage

\phantomsection
\addcontentsline{toc}{section}{
{\bf Current Status and Future Prospects of Slow Extraction} \\
R.~Muto}

\titl{Current Status and Future Prospects of Slow Extraction}

\name{
Ryotaro Muto$^{1}$
}

\adr{
$^1$ High Energy Accelerator Research Organization (KEK) and
J-PARC Center, Japan
}

\vspace{7mm}
{\Large\bf Current Status}
\vspace{4mm}

There are two major challenges for the slow extraction of the J-PARC Main Ring (MR):
increasing the beam power and improving the spill structure.
After achieving 64~kW beam power for slow extraction in 2021
the MR entered the long shutdown for the device upgrades (MR Upgrade)
which aimed at shortening the acceleration time from 1.4~s to 0.65~s~[1].
Various components such as main magnet power supplies, RF cavities,
fast extraction devices, and MR collimators were upgraded in this MR Upgrade.
Despite encountering several issues
during the startup of these upgraded devices,
each problem was addressed one by one,
leading to the success of the fast extraction operation
with the shortened acceleration time
by the end of 2023.

With the acceleration time reduced,
the repetition period for slow extraction operations will decrease
from 5.2~s to 4.24~s.
Consequently,
the beam power is expected to increase by approximately 1.23 times,
assuming the number of particles per pulse remains constant.
Beam tuning with the shortened repetition period is scheduled to begin in March 2024.

Regarding the spill structure,
fluctuations in the betatron tune
caused by current ripples in the main magnet power supplies are
creating large time structures in the extracted beam.
To address this,
two systems are being utilized: the spill feedback system
and the transverse RF system.
During operations in 2021,
the spill duty factor,
an indicator of the spill structure defined as
$\langle I \rangle ^2 / \langle I^2 \rangle$
(where $I$ denotes the beam current and
brackets represent the time average during extraction),
was approximately 75\%.

\vspace{7mm}
{\Large\bf Future Plans}
\subsection*{Beam Power}

To increase the beam power,
two critical issues must be addressed:
reducing beam loss and suppressing beam instability at the debunch timing.

\subsubsection{Beam Loss Reduction}

To reduce beam loss, we will implement the following measures:

\begin{enumerate}
\item {\bf Beam Diffusers:} 
The beam diffusers installed upstream of the ESS
will be put into practical use as soon as possible through user operations.
A beam test of these beam diffusers was conducted in 2021,
and the results showed that the predicted beam loss reduction,
as estimated by simulations, was almost achieved~[2].

\item {\bf Bent Silicon Crystals:} 
We plan to utilize bent silicon crystals~[3].
When charged particles are injected into a bent silicon crystal,
they are deflected along the curvature of the crystal
by a process called channeling.
By utilizing this channeling effect,
protons colliding with the ESS ribbons are deflected,
thereby reducing beam loss.
We will continue developing this device in the future.
\end{enumerate}

\subsubsection{Beam Instability}
To address beam instability at the debunch timing,
we are exploring further possibilities of RF manipulations,
which are the current mitigation methods.
Additionally, we are considering the implementation of optics
with large slippage
and the introduction of VHF cavities as future plans.

\subsection*{Spill Structure}

We will continue our efforts to reduce the current ripples
in the main magnet power supplies.
Additionally,
we are considering future plans to improve the spill structure:

\begin{enumerate}
\item {\bf Ripple Canceller:} 
We plan to add a ripple canceller~[4] that inputs information
on the current fipples directly in to the spill feedback system.

\item {\bf Transverse RF Feedback Control:} 
We are also looking into inplementeing feedback control
on the transverse RF to further improve the spill structure.
\end{enumerate}

\vfill  

\noindent{\bf References }
\begin{description}
\setlength\itemsep{-3pt}

\item{[1]} S. Igarashi {\it et al.}, ``Accelerator design for 1.3-MW beam
power operation of the \mbox{J-PARC} Main Ring,''
{\it Progress of Theoretical and Experimental Physics},
vol.~2021, no.~3, 033G01, 2021.\\
\url{doi:10.1093/ptep/ptab011b}

\item{[2]} R. Muto {\it et al.}, ``Simulation study on double diffuser for loss
reduction in slow extraction at J-PARC Main Ring,'' in
{\it Proc. IPAC’21}, Campinas, Brazil, May 2021, pp.~3069–3072.\\
\url{doi:10.18429/JACoW-IPAC2021-WEPAB192}

\item{[3]} F. M. Velotti {\it et al.}, ``Septum shadowing by means of a bent
crystal to reduce slow extraction beam loss,''
{\it Phys. Rev. Accel. Beams}, vol.~22, p.~093~502, 2019.\\
\url{doi:10.1103/PhysRevAccelBeams.22.093502}

\item{[4]} D. Naito {\it et al.}, ``Real-time correction of betatron tune ripples
on a slowly extracted beam,'' {\it Phys. Rev. Accel. Beams}, vol.~22,
p.~072~802, 7~2019.\\
\url{doi:10.1103/PhysRevAccelBeams.22.072802}
\end{description}

\stepcounter{count}
\clearpage

\phantomsection
\addcontentsline{toc}{section}{
{\bf High-resolution Decay Pion Spectroscopy of hypernuclei at MAMI, JLab and J-PARC} \\
S.~Nagao}

\titl{High-resolution Decay Pion Spectroscopy of hypernuclei at MAMI, JLab and J-PARC}

\name{
S.~Nagao$^{1}$
}

\adr{
$^1$ The University of Tokyo, 7-3-1 Hongo Bunkyo, Tokyo, 113-0033, Japan
}

Abstract

-----------

Decay Pion Spectroscopy (DPS) was developed as a high-resolution, high-precision mass spectroscopy technique of $\Lambda$ hypernuclei.
In 2015, a successful proof-of-principle experiment was conducted, resulting in the determination of the mass of the $^4_\Lambda$H ground state with an accuracy of less than 100 keV/$c^2$.
The results provided crucial data for discussing the matter of charge symmetry breaking in A$=$4 $\Lambda$ hypernuclei.
This experimental technique can be applied to other accelerator facilities, and further hypernucleus measurements at JLab and J-PARC are currently being planned.
Further advancement of the discussion of hypernuclei is expected to result from more precise hypernuclear measurements.


-----------

The $\Lambda$ hypernucleus is a quantum many-body system composed of $u, d$, and $s$ quarks.
It is an important probe for studying nuclear structure as well as the Baryon-Baryon interaction based on SU$_f$(3). 
The $\Lambda$ binding energy (B$_\Lambda$) of a hypernucleus, defined as the binding energy at which the $\Lambda$ is bound to its core nucleus, is an important variables for characterizing hypernuclei.
In many $s-$ and $p-$shell hypernuclei, the B$_\Lambda$ has been measured by the emulsion [1].
Recently, new B$_\Lambda$ have been reported for some hypernuclei by the $(e,e^\prime K^+)$ reaction spectroscopy and invariant mass spectroscopy by heavy-ion collisions [2, 3, 4, 5].
These findings have prompted discussions on the charge symmetry breaking (CSB) effect and hypertriton puzzles.

A novel method, Decay Pion Spectroscopy (DPS), has been developed to achieve good precision in the determination of B$_\Lambda$.
This method deduces the mass of the hypernucleus by precisely measuring the momentum of a monochromatic momentum $\pi^-$ which is emitted from the hypernuclear mesonic weak decay (Figure 1).

\begin{figure}[ht]
  \centering
  \includegraphics[width=8cm]{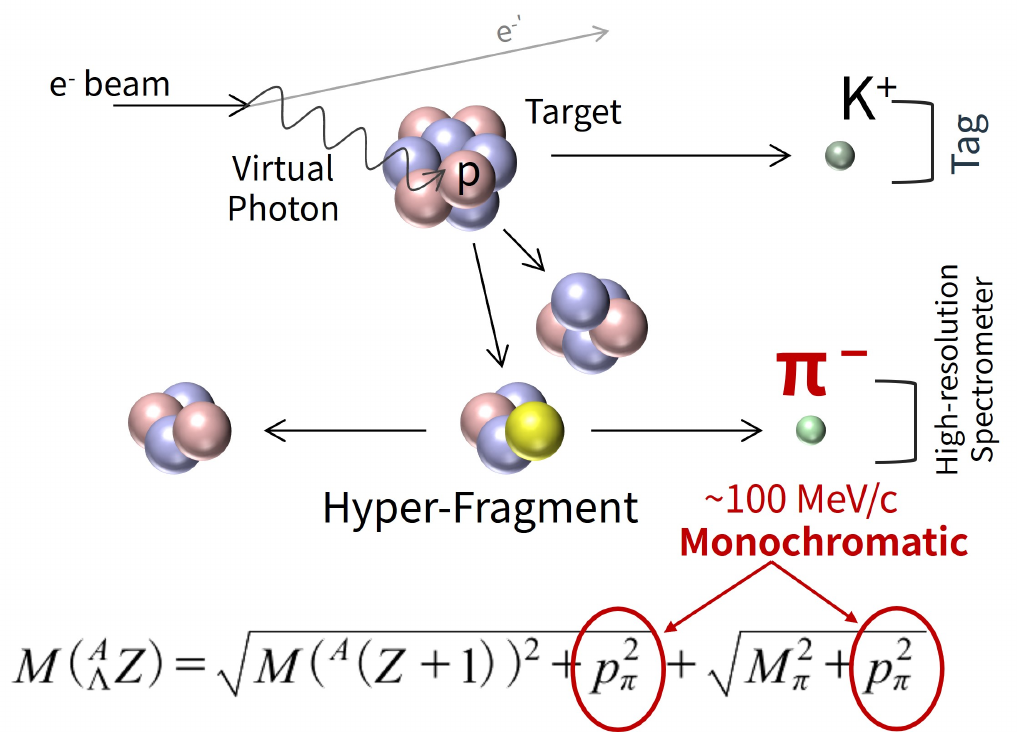}
  \caption{Principle of the hypernuclear decay pion spectroscopy.}
  \label{fig:decaypi}
\end{figure}

The $\pi^-$ peak can be observed with a resolution of approximately 100 keV/$c$ in FWHM using a high-resolution $\pi^-$ spectrometer.
The high-resolution spectroscopy enables the precise measurement of B$_\Lambda$ with an accuracy of less than 100 keV.
The $K^+$ identification (KID) as a $s$ quark production filter is also crucial for the selection of $\pi^-$s from the large amount of background, thus improving the signal-to-noise ratio.

The experiments have been performed at Mainz Microtron (MAMI) of Johannes Gutenberg University Mainz, Germany.
The Be foil target, the high-resolution pion spectrometer ``Spek-C'', and the $K^+$ tagger ``Kaos'' were utilized in the experiment.
In 2015, a $\pi^-$ peak from the two-body weak decay of $^4_\Lambda$H was successfully observed, and the $\Lambda$ binding energy was determined to be B$_\Lambda(^4_\Lambda{\rm H}(0^+)) = 2.12 \pm 0.01({\rm stat.}) \pm 0.09({\rm syst.})$ MeV [6].
Subsequent experiments have resulted in B$_\Lambda(^4_\Lambda{\rm H}(0^+)) = 2.157 \pm 0.005({\rm stat.}) \pm 0.077({\rm syst.})$ MeV [7].
Although these experiments were able to measure the B$_\Lambda$ with small statistical errors, relatively large systematic errors were introduced due to the uncertainty of momentum calibration on the $\pi^-$ spectrometer.
The momentum calibration has been conducted using the scattered electron from electron-nuclear elastic scattering, however the uncertainty of the beam electron energy attributed non-negligible error.

In 2022, data collection was conducted for the measurement of $^3_\Lambda$H using a Li target, which is anticipated to the higher yield of $^3_\Lambda$H.
In the spectrometer momentum calibration conducted in 2024, absolute beam energy measurement using the synchrotron radiation interferometry was introduced for the purpose of suppressing the systematic error of the B$_\Lambda$ to 10-20 keV [8, 9].
The data analysis is currently underway.

The Decay Pion Spectroscopy, which was initiated at MAMI, is a highly effective new method for determining the B$_\Lambda$, and its expansion is currently planned.
Preparations for the $(e,e^\prime K^+)$ reaction spectroscopy experiment are currently on-going at JLab, and the High-resolution Kaon Spectrometer (HKS) will be installed.
We plan to collect decay pions from hypernuclei with a newly installed $\pi^-$ spectrometer ``Enge''  while tagging $K^+$ in HKS.
In the experiment, the high-energy and high-intensity electron beams from JLab CEBAF and the highly efficient KID of HKS will provide statistics approximately 30 times higher than those in the MAMI experiments.
A Letter of Intent was submitted to JLab PAC51 in 2023 [10], and we aim to approve the experiment as soon as possible.

DPS experiments have been developed using electron beams.
In DPS, not only electron beams but also pions and Kaons can be used as production beams of hypernuclei.
The current pion beam intensity in J-PARC is not strong enough for DPS; however, if more intense beamlines are constructed in the J-PARC Hadron Hall, DPS with the $(\pi^+, K^+)$ reaction will be possible.
This will enable event selection for hypernuclei on the missing mass deduced by the $(\pi^+, K^+)$ reaction, which will further improve the signal-to-noise ratio compared to DPS with electron beams.

The DPS method, which was pioneered at MAMI, will be applied to JLab and J-PARC in the future.
The high-resolution and high-precision hypernucleus spectroscopy enabled by DPS will open the door to new developments in the study of the hypernuclear physics.

-----------

\vfill  

\noindent{\bf References }
\begin{description}
\setlength\itemsep{-3pt}
\item{[1]} M.~Juri\'{c} {\it et al.}, Nucl. Phys. B {\bf 55} (1973) 1.
\item{[2]} T.~Gogami {\it et al.}, Phys. Rev. C {\bf 93} (2016) 034314.
\item{[3]} T.~Gogami {\it et al.}, Phys. Rev. C {\bf 94} (2016) 021302(R).
\item{[4]} The STAR Collaboration, Nature Phys. {\bf 16} (2020) 409.
\item{[5]} S.~Acharya {\it et al.}, Phys. Rev. Lett. {\bf 131} (2023) 102302.
\item{[6]} A.~Esser {\it et al.}, Phys. Rev. Lett. {\bf 114} (2015) 232501.
\item{[7]} F.~Schulz {\it et al.}, Nucl. Phys. A {\bf 954} (2016) 149.
\item{[8]} P.~Klag {\it et al.}, Nucl. Inst. Meth. A {\bf 910} (2018) 147.
\item{[9]} P.~Klag {\it et al.}, J. Phys.: Conf. Ser. {\bf 2482} (2022) 012016.
\item{[10]} S.~Nagao {\it et al.}, Letter of Intent LOI12-23-011, PAC51, Jefferson Lab (2023).
\end{description}


\stepcounter{count}
\clearpage

\phantomsection
\addcontentsline{toc}{section}{
{\bf Future experiments on kaonic nuclei at K1.8BR} \\
T.~Nanamura for the J-PARC E80 and P89 collaboration}

\titl{Future experiments on kaonic nuclei at K1.8BR}

\name{
T. Nanamura$^{1}$ for the J-PARC E80 and P89 collaboration
}

\adr{
$^1$ Riken, Wako, Saitama, 351-0198, Japan
}


 Attractive $\bar{K}N (I=0)$ interaction and quasi-bound states formed by the attractive interaction, such as $\Lambda(1405)$ and kaonic nuclei,  have been getting a lot of attention.  At the K1.8BR beamline in the Hadron Experimental Facility, many physics programs on $\bar{K}N$ interaction have been performed, X-ray spectroscopy of kaonic atoms [E57 and E62 experiment], spectroscopy of $\Lambda(1405)$ [E31 experiment], and search for the lightest kaonic nuclei, ``$K^-pp$" [E15 experiment].

J-PARC E15 experiment [1] successfully observed event concentration interpretted as  $``K^- pp"$ bound state by reconstructing not only the $\Lambda p$ invariant-mass but also momentum transfer to the $\Lambda p$ system.  Binding energy of``$K^-pp$" was $42 \pm3(\text{stat})^{+3}_{-4}(\text{syst})$ MeV and width was $\Gamma =100 \pm 7(\text{stat})^{+19}_{-9}(\text{syst})$. This experiment had two excellent points. First, $``K^- pp"$ was produced via $^3\text{He}(K^-, n)$ reaction, where reaction process is clear and mass region below the $\bar{K}NN$ threshold can be investigated. Second, all particles from the ``$K^-pp" \to \Lambda p$ were detected, which enabled an exclusive analysis for the $\Lambda pn$ final state.

 In order to apply this experimental method to other (light) kaonic nuclei and decay channels, where the number and kinds of decay particles from kaonic nuclei increase, we are planning to upgrade the K1.8BR beamline and spectrometer system. For the spectrometer side, we are developping solenoidal spectrometer system with larger acceptance. New spectrometer system consists of a superconducting solenoidal magnet, a 2.6m-length cylindrical drift chamber, 3m-length plastic scintillators for charged particles/neutron detection, and a vertex fiber tracker surrounding the target. The geometrical acceptance of the new spectrometer is 1.6 times larger than present one and the detection efficiency for neutron (proportional to the thickness of plastic scintillators) will be 4 times larger than present one. 
Development of the new spectrometer system is going well with secured budgets. Production and performance evaluation of the magnet and each detector will be completed in JFY 2025 and installation to beamline will start in JFY 2026. For beamline side, we are considering to shorten the beamline by 3.7 m removing a bending magnet. With this modification, $K^- $ beam intensity would be 1.6 times larger than without it.

 There are two proposed experiments using the new spectrometer system and modified beamline: J-PARC E80 [2] and P89 [3] experiments.
J-PARC E80 experiment aims to systematic investigation of the light kaonic nuclei. As a first step, $\bar{K}NNN$ produced by the in-flight $^4\text{He}(K^-,n)$ reaction will be studied. From the binding energy and branching ratio, structure of the $\bar{K}NNN$ will be revealed. This experiment will start after the commissioning of the new spectrometer, in JFY 2026.
J-PARC P89 experiment aims to investigate the spin parity of the $\bar{K}NN$ state. In this experiment, spin-correlation between $\Lambda$ and $p$ from $``K^-pp" \to \Lambda p$ decay is measured and the spin-parity of $``K^-pp"$ is determined whether $0^-$ or $1^-$.  The search for $``\bar{K^0} nn"$, isospin partner of the $``K^-pp"$, can be also performed. This experiment will make our understanding of ``$K^- pp$" as a nuclear bound state more solid.

\vfill  

\noindent{\bf References }
\begin{description}
\setlength\itemsep{-3pt}
\item{[1]} T. Yamaga et al., Phys. Rev. C, 102, 044002 (2020).
\item{[2]} F. Sakuma et al., J-PARC proposal P80.
\item{[3]} T. Yamaga et al., J-PARC proposal P89.
\end{description}

\stepcounter{count}
\clearpage

\phantomsection
\addcontentsline{toc}{section}{
{\bf KOTO and KOTO II to study $K_L\to\pi^0\nu\overline{\nu}$} \\
H.~Nanjo for the KOTO collaboration}

\titl{KOTO and KOTO II to study $K_L\to\pi^0\nu\overline{\nu}$}

\name{
Hajime Nanjo$^{1}$ for the KOTO collaboration
}

\adr{
$^1$ Department of Physics, Osaka University, Toyonaka, Osaka 560-0043, Japan
}

The rare kaon decay $K_L\to \pi^0 \nu\overline{\nu}$ is being searched for
in the KOTO experiment~[1,2]
at the hadron experimental facility of J-PARC.
The branching ratio of the decay is 
$3 \times 10^{-11}$ with the theoretical uncertainty less than 2\%~[3] in the Standard Model (SM).
Owing to the suppressed and precise value in the SM,
a small deviation from the branching ratio expected from the SM could be detected. 
Therefore, it is sensitive to new physics beyond the SM. 

In the KOTO experiment,
$K_L$ is generated at the production target
with primary 30-GeV protons.
The $K_L$ beam is collimated with two collimators 
in the 20-m long KL beamline
into the solid angle of $\SI{7.8}{\micro sr}$.
The production angle is 16 degrees
from the primary proton beam,
and the peak of the $K_L$ momentum spectrum is 1.4~GeV/c.
Short-lived particles decay out, and 
charged particles are swept out with a magnet in the beamline.
Finally,
long-lived neutral particles $\gamma$, neutron, and $K_L$ remain.

The KOTO detector starts at 21.5 m from the production target,
which is the origin of the coordinate
with the z axis along the $K_L$ beam direction.
In order to identify the signal $K_L\to \pi^0\nu\overline{\nu}$,
two photons from the $\pi^0$ decay
are detected with a calorimeter 1.9 m in diameter
at $6.4~\mathrm{m}$ in z.
The decay vertex ($Z_\mathrm{vtx}$) and the transverse momentum ($P_\mathrm{T}$)of the $\pi^0$
are reconstructed with the energyies and positions of the two photons
at the calorimeter by assuming the vertex on the z axis. 
The signal decay region is set from 3 to 5 m in z, and
a hermetic detector system surrounding the decay region
ensures no other detectable particles in the decay.
The signal is searched for by applying veto conditions for the detector, 
and imposing selections
for the two photons and the reconstructed $\pi^0$.

The current upper limit of the branching ratio
$3.0\times 10^{-9}$ at 90\% confidence level (CL)
was set by KOTO
with the data taken in 2015~[4].
In the analysis of the data taken in 2016-2018,
the single event sensitivity is $7.2\times 10^{-10}$
with three events observed in the signal region~[5].
The number of observed events
is consistent to the number of the estimated background.
The decay of charged kaons generated in the beamline
downstream of the magnet mainly contributes the background.
In order to reduce the charged kaon background,
we installed a charged particle
veto detector ``Upstream Charged Veto'' (UCV)
before the data taking in 2021.
The UCV is a single layer of 0.5-mm square scintillating fibers,
and covers the beam region just upstream of the KOTO detector.
Charged kaons are detected with UCV, and the events are rejected.

\begin{figure}[h]
 \centering
 \includegraphics[bb=7 4 1045 845,clip,width=0.48 \textwidth]{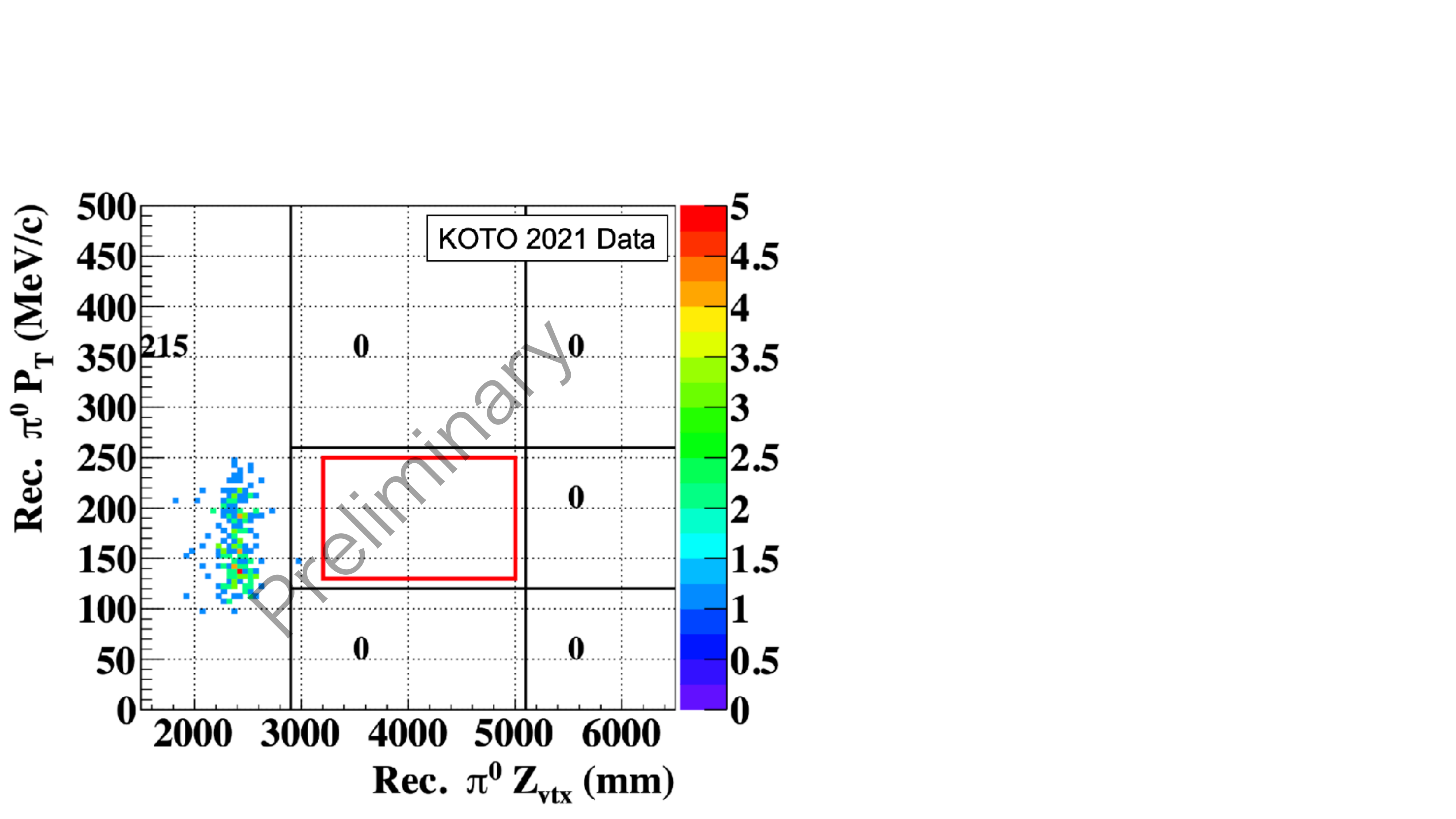}
 \caption{The final plot in $Z_\mathrm{vtx}$ and $P_\mathrm{T}$ plain
 in the analysis of data taken in 2021.\label{fig:ptz2021}} 
\end{figure}

For the data taking in 2021,
the beam power of the primary protons
reached 64.5 kW with a 2-s beam spill in a 6.2-s cycle.
In the analysis of the data,
we achieved the single event sensitivity of $9.26\times 10^{-10}$.
After unblinding the signal region 
with the expected number of the backgrounds 0.26, 
we found no event as shown in Fig.~\ref{fig:ptz2021}, and 
set a preliminary upper limit of $2.1\times 10^{-9}$ (90\% CL)~[6].

In order to suppress the charged kaon background further,
we upgraded the UCV to a thinner and more charged-particle
efficient version with a 0.2-mm thick plastic scintillator film
in the spring of 2023. 
We installed a permanent magnet in the downstream region of the beamline 
to reduce the charged kaons entering the detector
in the fall of 2023. 

Higher beam power of 80-100 kW is expected in the following years.
We upgraded the DAQ with high-speed data transfer
using optical fibers and a module to perform the event building
with an FPGA. 
We setup several PC nodes with GPUs for the data collection, 
the data compression, and the event selections.
In the KOTO experiment, we will take data with these upgrades
for the goal of the single event sensitivity below $10^{-10}$.

\begin{figure}[h]
 \centering
 \includegraphics[bb=24 2 1528 698,clip,width=0.48\textwidth]{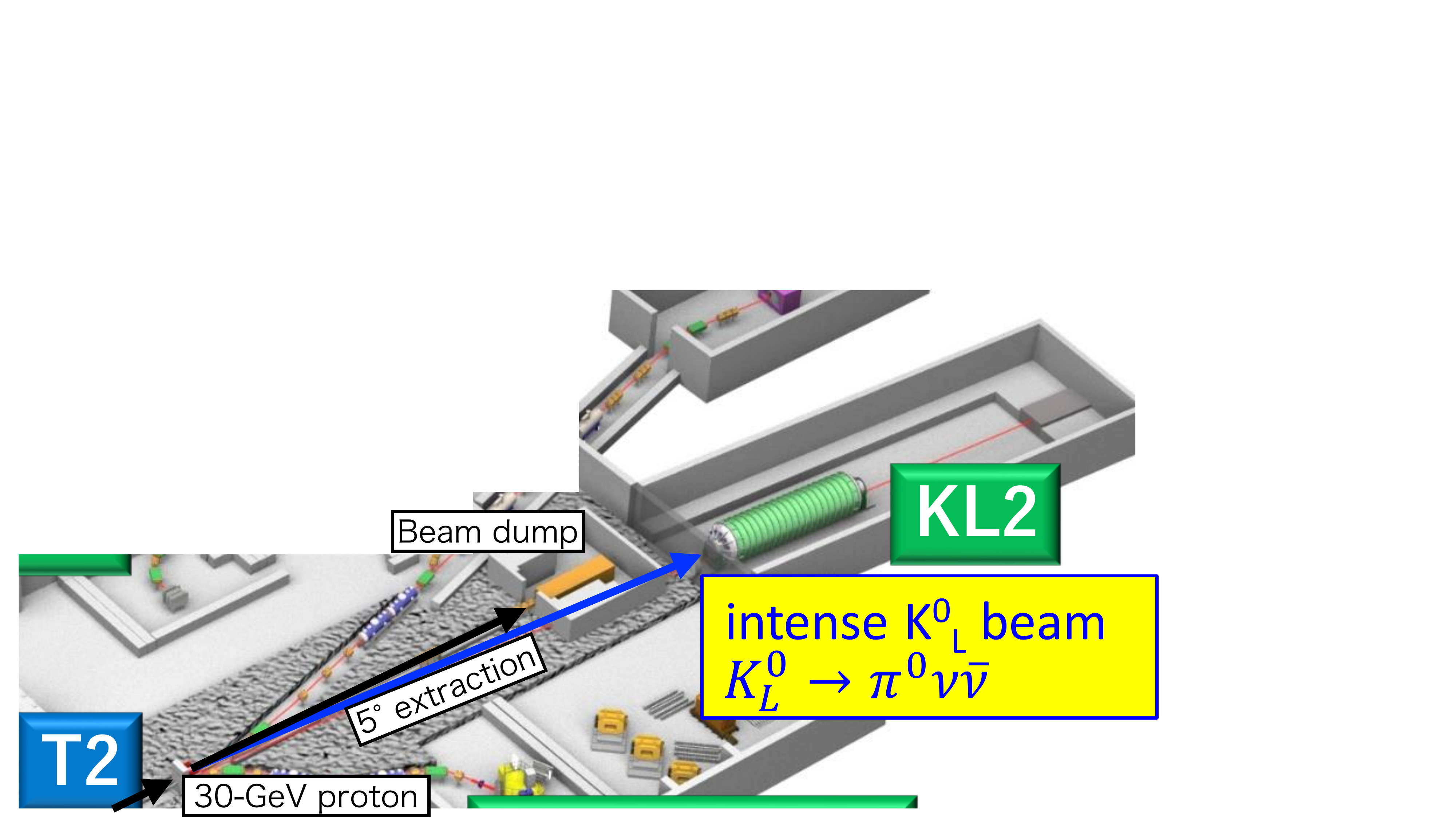}
 \hspace{0.01\textwidth}
 \includegraphics[bb=13 8 1495 891,clip,width=0.48\textwidth]{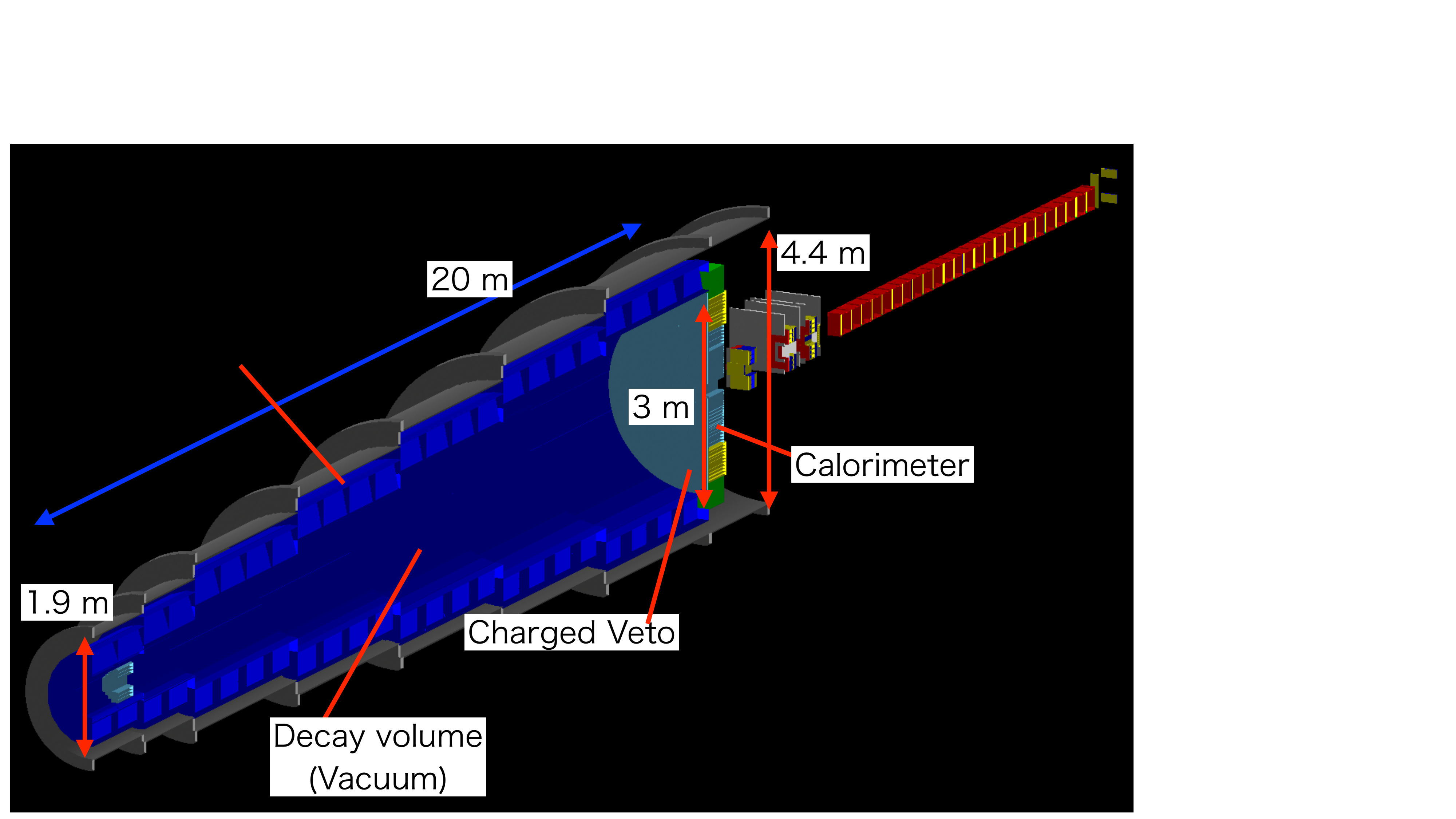}
 \caption{The KL2 beam line and the KOTO-II detector in the extended hadron experimental facility (left),
 and a cut-away view of the KOTO-II detector in the current design (right).\label{fig:KOTOII}}
\end{figure}

We are planning a next-generation experiment KOTO~II
at the extended hadron experimental facility of J-PARC as shown in Fig.~\ref{fig:KOTOII}
to measure the branching ratio of the $K_L\to\pi^0\nu\overline{\nu}$ decay~[7,8]. 
The production angle of $K_L$ is 5 degrees
with the detector behind the primary proton dump.
The peak of the $K_L$ momentum spectrum is 3~GeV/c,
which is larger than 1.4~GeV/c in KOTO.
The length of the beam line is 43~m, and
the detector starts at 44~m from the production target,
which is the origin of the coordinate
with the z axis along the $K_L$ beam direction.
A 3-m diameter calorimeter is at 20 m in z with a 20-cm square beam hole,
which gives the solid angle of the $K_L$ beam \SI{4.8}{\micro sr}.
The length of the decay region is 12 m from 3 to 15 m in z,
which is larger than 3 m for KOTO.
Owing to the larger momentum of $K_L$, the longer decay region
is effective with the larger calorimeter.
The counting rate of the detector 
was evaluated with simulations.
The contribution of the muon from the primary beam dump
was evaluated with a special experiment;
A vertical hole was dug into the underground behind the current beam dump,
and the muon flux was measured as a function of the angle from the primary-beam direction.
The contribution from the muon flux in the KOTO-II detector was negligible
with an additional 3-m thick iron shield.

The expected number of the signal is 35 for the value of the branching ratio
in the SM
with the 100-kW beam in the running time of $3\times 10^7~\mathrm{s}$,
while the expected number of the background is 40.
This could give the 5-$\sigma$ observation of $K_L\to\pi^0 \nu\overline{\nu}$.
If the measured branching ratio is different
from the value in the SM by 40\%,
it will be an indication of new physics at the 90\% CL.

-----------

\vfill  

\noindent{\bf References }
\begin{description}
\setlength\itemsep{-3pt}

\item {[1]} J. Comfort et al., Proposal for $K_L\to\pi^0\nu\bar{\nu}$ Experiment at J-Parc, \url{https://j-parc.jp/researcher/ Hadron/en/pac 0606/pdf/p14-Yamanaka.pdf}.
\item {[2]} T. Yamanaka, Prog. Theor. Exp. Phys. {\bf 2012}, 02B006 (2012).
 \item{[3]} A.~J.~Buras, Eur. Phys. J. {\bf C 83}, 66 (2023).
\item {[4]} J. Ahn et al., Phys. Rev. Lett. {\bf 122}, 021802 (2019).
\item {[5]} J. Ahn et al., Phys. Rev. Lett. {\bf 126}, 121801 (2021).
 \item{[6]} J.~Redeker, presentation at ICHEP 2024\,\url{https://indico.cern.ch/event/1291157/contributions/5896367/attachments/2898931/5083214/JRedeker_ICHEP_2024_KLpi0nunu.pdf}
 \item {[7]} K.~Aoki et al., arXiv:2110.04462
 \item {[8]} G.~Anzivino et al., Eur. Phys.J.C {\bf 84} (2024) 4, 37
\end{description}

\stepcounter{count}
\clearpage

\phantomsection
\addcontentsline{toc}{section}{
{\bf Spectroscopy of hyperons and singly heavy baryons} \\
E.~Santopinto$^*$ and H.~Garcia-Tecocoatzi}

\titl{Spectroscopy of hyperons and singly heavy baryons}

\name{
E. Santopinto$^{1}$ and H. Garcia-Tecocoatzi$^{1}$
}

\adr{
$^1$ INFN, Sezione di Genova, Via Dodecaneso 33, 16146 Genova, Italy
}

 The current status of the nonstrange and strange light baryons described as three quark system using the hypercentral quark model is briefly presented. We also discuss  the hyperons as quark-diquark systems. In addition, we discuss the channels where the strange baryons can be looked for in the forthcoming experiments as a part of the Extension Project for the J-PARC Hadron Experimental Facility. 
 Finally, we discuss  the singly charm baryons in a three quark and quark-diquark schemes. 

\vspace{7mm}
{\Large\bf 1~~Introduction}
\vspace{4mm}
\\
The program of the Extension Project for the J-PARC Hadron Experimental Facility is dedicated to hot topics, like i) Single strangeness hypernuclear physics, hyperon scattering experiment, and kaonic nucleus study, ii) Understanding baryon-baryon forces (nuclear forces) inside dense nucleus/hadronic matter to elucidate the mystery of the inner structure of the neutron stars. iii)  The study  of the hadron spectroscopy  with two or three strangeness and charmed baryons. Moreover, they plan to study the effective degree of freedom that describes hadrons, such as diquark correlation, and the properties of hadrons inside the nucleus.

Recent years showed significant improvement in our understanding of baryon resonances in the light quark
sector, predominantly from photo- and electroproduction experiments (from 22 3* and 4* N and $\Delta$ resonances listed in Particle Data Booklet PDG2000 to 34 in PDG2024 \cite{ParticleDataGroup:2020ssz}). However, the strange quark sector
showed stagnation - twenty years with few improvements( from 24 3* and 4* $\Lambda/\Sigma$  resonance in PDG2000 to 25 in PDG2024 \cite{ParticleDataGroup:2020ssz}, and still the same 7 $\Xi$ states).

In the other hand, recent advancements in the understanding of singly heavy baryons have been achieved through the application of non-relativistic quark models, QCD sum rules, and lattice QCD \cite{Chen:2022asf}. 
On the experimental front, observations from the LHCb and Belle experiments have consistently identified various charmed baryon states, including but not limited to $\Omega_{c}(3000)$, $\Omega_{c}(3050)$, $\Omega_{c}(3066)$, $\Omega_{c}(3090)$, $\Omega_{c}(3119)$ (observed only by LHCb), and $\Omega_{c}(3188)$ (reported in references \cite{LHCb:2017uwr,Belle:2017ext}).  The LHCb experiment has reported the observation of $\Xi^0_{c}(2923)$, $\Xi^0_{c}(2939)$, and $\Xi^0_{c}(2965)$ states \cite{LHCb:2020iby}. 

Very recently, two new excited states, $\Omega_c^0(3185)$ and $\Omega_c^0(3327)$ were observed in the $\Xi_c^+ K^-$ invariant-mass spectrum by the LHCb Collaboration~\cite{LHCb:2023sxp}.
The measured masses and widths of the two newly found states are
\begin{align}
	m[\Omega_{c}(3185)] & = 3185.1 \pm 1.7_{-0.9}^{+7.4} \pm 0.2 \mathrm{MeV},  \\
	\Gamma[\Omega_{c}(3185)] & = 50 \pm 7_{-20}^{+10} \mathrm{MeV},  \\
	m[\Omega_{c}(3327)] & = 3327.1 \pm 1.2_{-1.3}^{+0.1} \pm 0.2 \mathrm{MeV},   \\
	\Gamma[\Omega_{c}(3327)] & = 20 \pm 5_{-1}^{+13} \mathrm{MeV}.
\end{align}

In the present contribution,
  we discuss the nonstrange and strange light baryons described as three quark system using the hypercentral quark model. We also discuss  the hyperons as quark-diquark systems. In addition, we discuss the channels where the strange baryons can be looked for in the forthcoming experiments as a part of the Extension Project for the J-PARC Hadron Experimental Facility.   Finally, we discuss  the singly charm baryons in a three quark and quark-diquark schemes.
We use the model of Ref. \cite{Santopinto:2018ljf}, and its extension on D-wave singly heavy baryons \cite{Garcia-Tecocoatzi:2022zrf,Garcia-Tecocoatzi:2023btk}, with the aim of discussing the assignment of $\Omega_{c}(3327)$.

\begin{figure}
    \centering
    \includegraphics[width=0.8\linewidth]{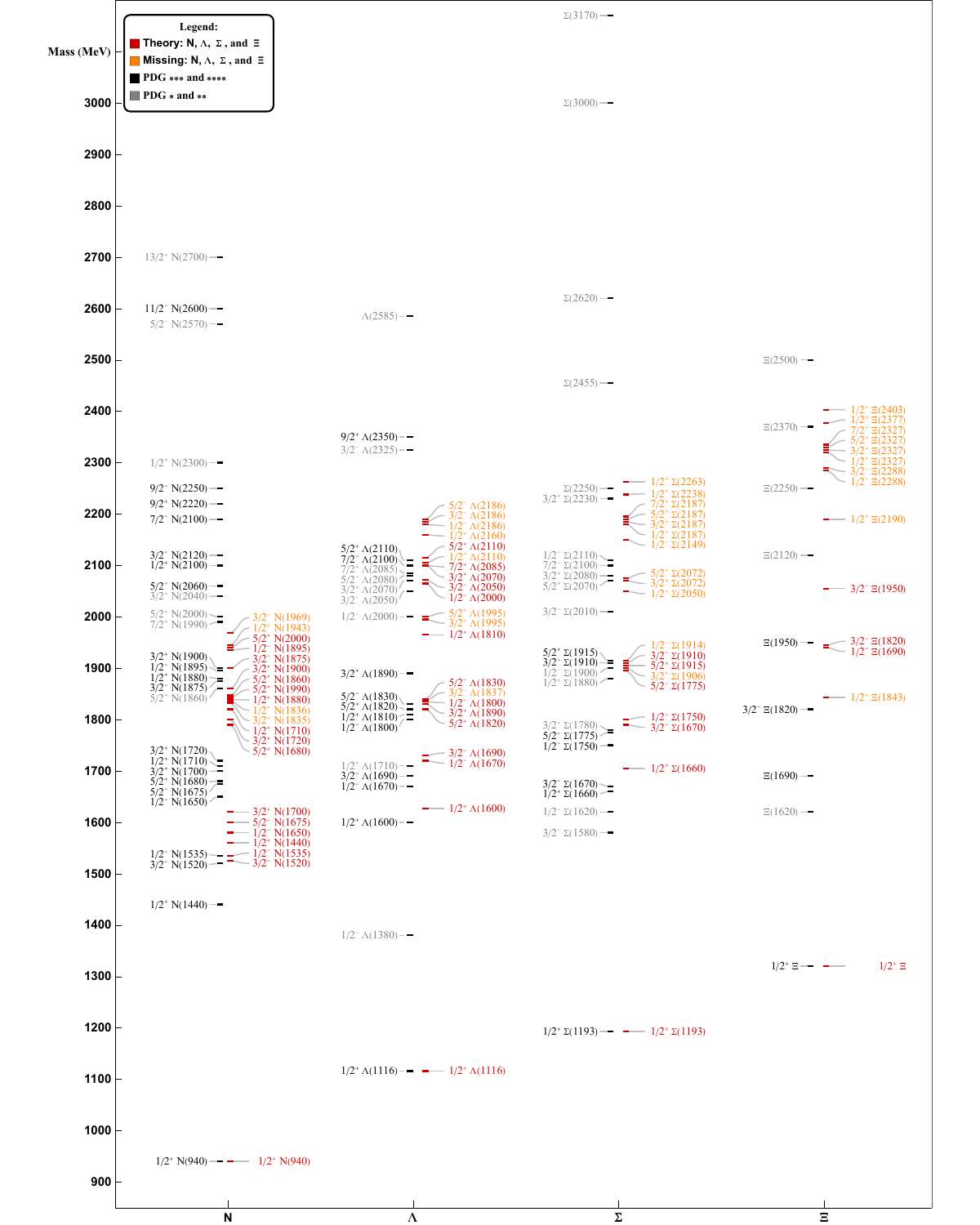}
    \caption{Theoretical octet baryon mass calculations up to D-wave within the hQM of Ref. \cite{Giannini:2005ks} (in red for  3 and 4 stars, in orange for missing states, and in violet for singlet states) compared with the experimental data as from PDG \cite{Workman:2022ynf} (in black).}
    \label{fig:hQM}
\end{figure}

\vspace{7mm}
{\Large\bf 2~~The model}
\vspace{4mm}
\\
{\large\bf 2.1~~Hypercentral quark model}
\vspace{4mm}
\\
 The hypercentral model (hQM) was introduced in Ref.~\cite{Ferraris:1995ui}, and it was extended  to the strangeness sector in Ref.~\cite{Giannini:2005ks}  where a G\"ursey Radicati inspired 
$SU(6)$ breaking interaction was  considered  to describe the splittings within each $SU(6)$ 
multiplet. The three quark Hamiltonian
\begin{equation}
  \label{eq:hcqmhamilt}
  H=H_0+H_{GR}
\end{equation}
with
$$H_0=3m+\frac{\mbox{\boldmath $p$}_{\lambda}^2}{2m}+\frac{
\mbox{\boldmath $p$}_\rho^2}{2m}+V(x)~,$$

and

$$H_{GR}=+A \div C_2[SU_{SF}(6)]+B \div C_2[SU_F(3)]+C \div C_2[SU_S(2)]$$
$$+DC_1[U_Y(1)]+E \div \left(C_2[SU_I(2)]-\frac{1}{4}(C_1[U_Y(1)])^2\right)~,$$
where  $
V(x)= -\frac{\tau}{x}~+~\alpha x$ is the hypercentral potential, i.e.  it depends on $x=\sqrt{\rho^2+\lambda^2}$,
and  m is the constituent quark mass.  The theoretical results as from Ref.~\cite{Giannini:2005ks}  for the octet baryons is shown in Fig. \ref{fig:hQM} and compared with the experimental data~\cite{Workman:2022ynf}. 
\\

\vspace{4mm}
{\large\bf 2.2~~Interacting quark-diquark model of baryons}
\vspace{4mm}
\\
The interacting quark-diquark model was introduced in Ref.~\cite{Santopinto:2004hw} for non-strange baryons. Later, the model was extended to describe the strange baryons in Ref. \cite{Santopinto:2014opa}. 
We consider a quark-diquark system, where $\vec{r}$ and $\vec{q}$ are the relative coordinate between the two constituents and its conjugate momentum, respectively. 
The baryon rest frame mass operator is 
\begin{equation}
	\begin{array}{rcl}
	M & = & E_0 + \sqrt{\vec q\hspace{0.08cm}^2 + m_1^2} + \sqrt{\vec q\hspace{0.08cm}^2 + m_2^2} 
	+ M_{\mbox{dir}}(r) 
	 +  M_{\mbox{ex}}(r)  
	\end{array}  \mbox{ },
	\label{eqn:H0}
\end{equation}
where $E_0$ is a constant, $M_{\mbox{dir}}(r)$ and $M_{\mbox{ex}}(r)$ respectively the direct and the exchange diquark-quark interaction, $m_1$ and $m_2$ stand for diquark and quark masses, where $m_1$ is either $m_{[q,q]}$ or $m_{\{q,q\}}$ according if the mass operator acts on a scalar or axial-vector diquark, with $[q,q]$ = $[n,n]$ or $[n,s]$ and $\{q,q\}$ = $\{n,n\}$, $\{n,s\}$ or $\{s,s\}$.

The direct term is, 
\begin{equation}
  \label{eq:Vdir}
  M_{\mbox{dir}}(r)=-\frac{\tau}{r} \left(1 - e^{-\mu r}\right)+ \beta r ~~,
\end{equation}
is the sum of a Coulomb-like interaction with a cut off and a linear confinement term.
 
\begin{figure}
    \centering
    \includegraphics[width=0.8\linewidth]{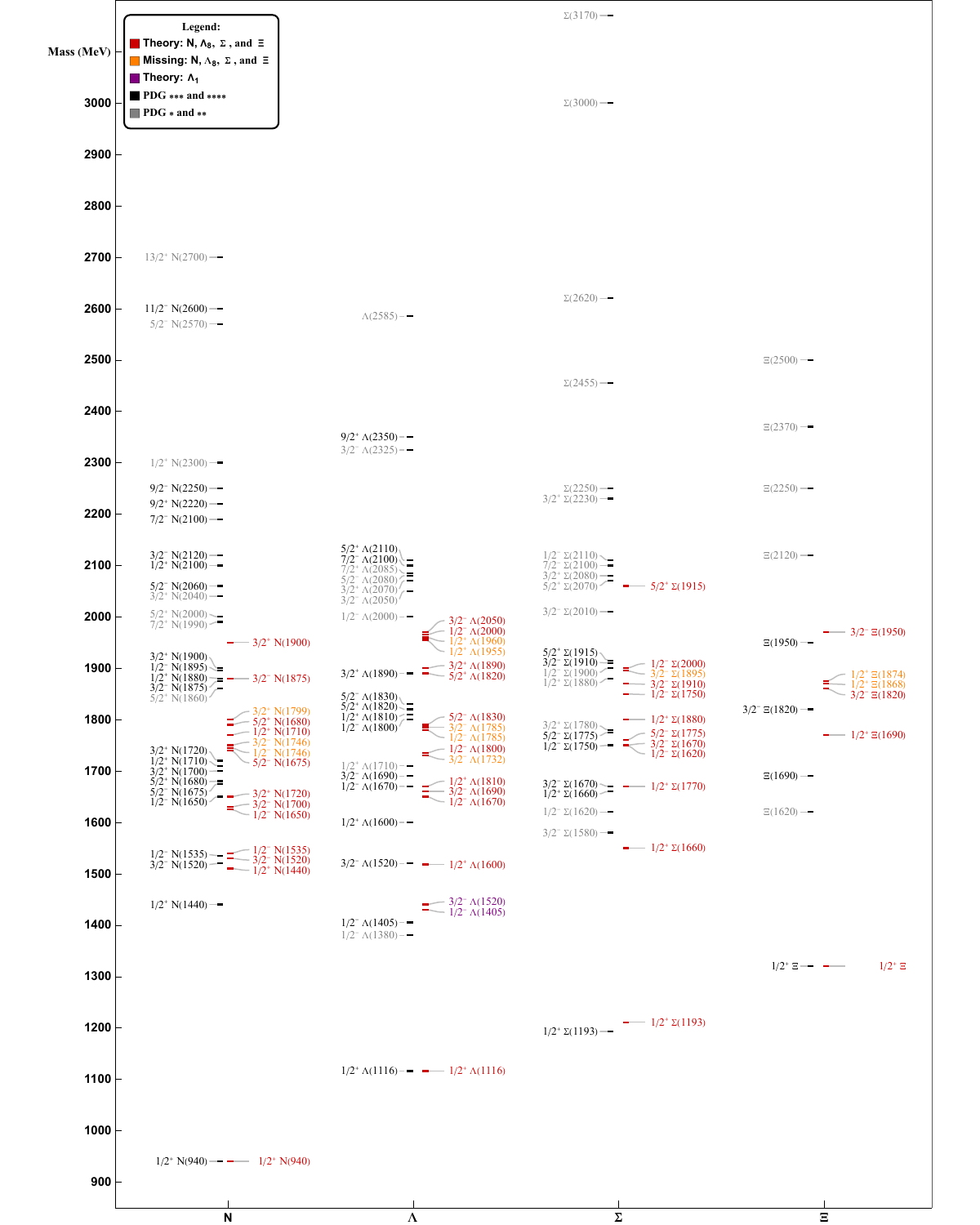}
    \caption{Theoretical octet baryon mass calculations up to 2 GeV within the interacting quark-diquark model of Ref. \cite{Santopinto:2014opa} (in red for  3 and 4 stars, in orange for missing states, and in violet for singlet states) compared with the experimental data as from PDG \cite{Workman:2022ynf} (in black). }
    \label{fig:diquark}
\end{figure}

It is considered an exchange interaction, since this is the crucial ingredient of a quark-diquark description of baryons \cite{Santopinto:2004hw,Lichtenberg:1981pp}. Thus, we consider the following G\"ursey-Radicati \cite{Gursey:1992dc} inspired interaction
\begin{equation}
	\begin{array}{rcl}
	M_{\mbox{ex}}(r) & = & \left(-1 \right)^{L + 1} \mbox{ } e^{-\sigma r} \left[ A_S \mbox{ } \vec{s}_1 
	\cdot \vec{s}_2  \right.+\left. A_F \mbox{ } \vec{\lambda}_1^f \cdot \vec{\lambda}_2^f \mbox{ } 
	+ A_I \mbox{ } \vec{t}_1 \cdot \vec{t}_2  \right]  
	\end{array}  \mbox{ },
	\label{eqn:Vexch-strange}
\end{equation}
where $\vec{s}$ and $\vec{t}$ are the spin and the isospin operators and $\vec{\lambda}^f$ the SU$_{\mbox{f}}$(3) Gell-Mann matrices. 
In the non-strange sector, we also have a contact interaction 
\begin{equation}
	\begin{array}{rcl}
	M_{\mbox{cont}} & = & \left(\frac{m_1 m_2}{E_1 E_2}\right)^{1/2+\epsilon} \frac{\eta^3 D}{\pi^{3/2}} 
	e^{-\eta^2 r^2} \mbox{ } \delta_{L,0} \delta_{s_1,1}  \left(\frac{m_1 m_2}{E_1 E_2}\right)^{1/2+\epsilon}
	\end{array}  \mbox{ },
\end{equation}
which was introduced in the mass operator of Ref. \cite{Ferretti:2011zz} to reproduce the $\Delta-N$ mass splitting.  
The theoretical results as from Ref.~\cite{Santopinto:2014opa}  are shown in Fig. \ref{fig:diquark} and compared with the experimental data~\cite{Workman:2022ynf}.

\vspace{7mm}
{\Large\bf 3~~Hyperon missing states}
\vspace{4mm}
\\
There still many hyperons that will be possible to find  thanks to the program of the Extension Project for the J-PARC Hadron Experimental Facility. The strange quark sector showed stagnation - twenty years few improvements (from 24 3* and 4* $\Lambda/\Sigma$  resonance in PDG2000 to 25 in PDG2024 \cite{ParticleDataGroup:2020ssz}, and still the same 7 $\Xi$ states).  In fact, in Figs. \ref{fig:hQM}  and \ref{fig:diquark}, we can see in orange the missing states predicted within the  hQM and the interacting  quark-diquark scheme, respectively. Moreover, in Tables \ref{tab:xi-missing} and \ref{tab:omega}, we present the $\Xi$ and $\Omega$ missing states, respectively, and their decay channels where these states can be studied in the forthcoming experiments.  More details on the strong decay calculations  reported in Tables \ref{tab:xi-missing} and \ref{tab:omega} can be found in Ref.~\cite{Bijker:2015gyk}.  In particular, it is important to notice that the main decay channels contain kaons, for example the $\Xi$ missing states  are strongly coupled to the  $\Sigma \overline{K}$, $\Lambda \overline{K}$, 
 $\Sigma^* \overline{K}$, $\Lambda \overline{K}^*$, or $\Sigma\overline{K}^*$ channels, See Table~\ref{tab:xi-missing}, while the $\Omega$  missing states can be found in the $\Xi \overline{K}$, $\Xi^* \overline{K}$, and  $\Xi\overline{K}^*$ channels, see Table~\ref{tab:omega}.  

\vspace{7mm}
{\Large\bf 4~~Three quark model for singly heavy baryons}
\vspace{4mm}
\\
The mass spectrum of the singly heavy baryons can be described by the mass formula introduced in Ref.~\cite{Santopinto:2018ljf}:

\begin{eqnarray}
M &=& \sum_{i=1}^3 m_i +\omega_\rho n_\rho+\omega_\lambda n_\lambda+AS(S+1)
+B\frac{1}{2}\left[J(J+1)-L(L+1)-S(S+1)\right]
\nonumber\\
&& +EI(I+1)+G\frac{1}{3}\left[p(p+3)+q(q+3)+pq\right],
\label{mass_formula}
\end{eqnarray}
where, $n_\rho$ and $n_\lambda$ represent the quanta in the $\rho$- and $\lambda$-oscillator, respectively.  The corresponding frequencies share the same spring constant $K_{b/c}$ but have different reduced masses:

\begin{eqnarray}
\omega_\rho &=& \sqrt{\frac{3K_{b/c}}{m_\rho}} = \sqrt{\frac{6K_{b/c}}{m_1+m_2}}, \qquad
\omega_\lambda = \sqrt{\frac{3K_{b/c}}{m_\lambda}} = \omega_\rho \sqrt{\frac{m_1+m_2+m_3}{3m_3}}.
\end{eqnarray}

\begin{figure}
\centering
\begin{minipage}{1\textwidth}
  \centering
  \includegraphics[width=0.75\linewidth]{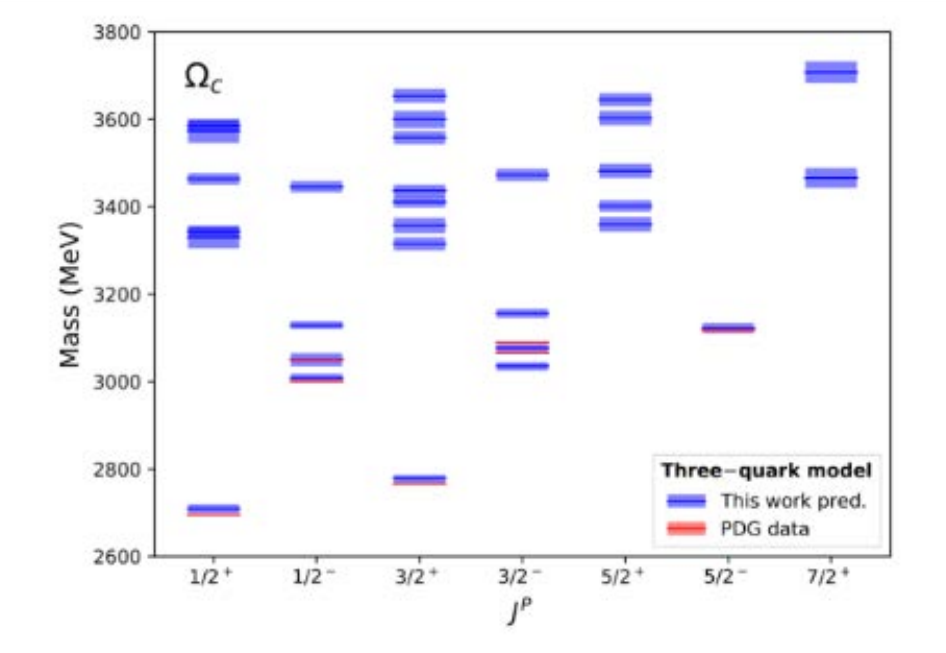}
  \caption{$\Omega_c$ mass spectra and tentative quantum number assignments based on the three-quark model Hamiltonian of Eq. \ref{mass_formula}. The theoretical predictions and their uncertainties (blue
lines and bands) are compared with the experimental results PDG\cite{Workman:2022ynf} (red
lines and bands). Figure taken from Ref. ~\cite{Garcia-Tecocoatzi:2022zrf} APS copyright.}
  \label{fig:ThreeQM}
\end{minipage}%

\begin{minipage}{1\textwidth}
  \centering
  \includegraphics[width=.75\linewidth]{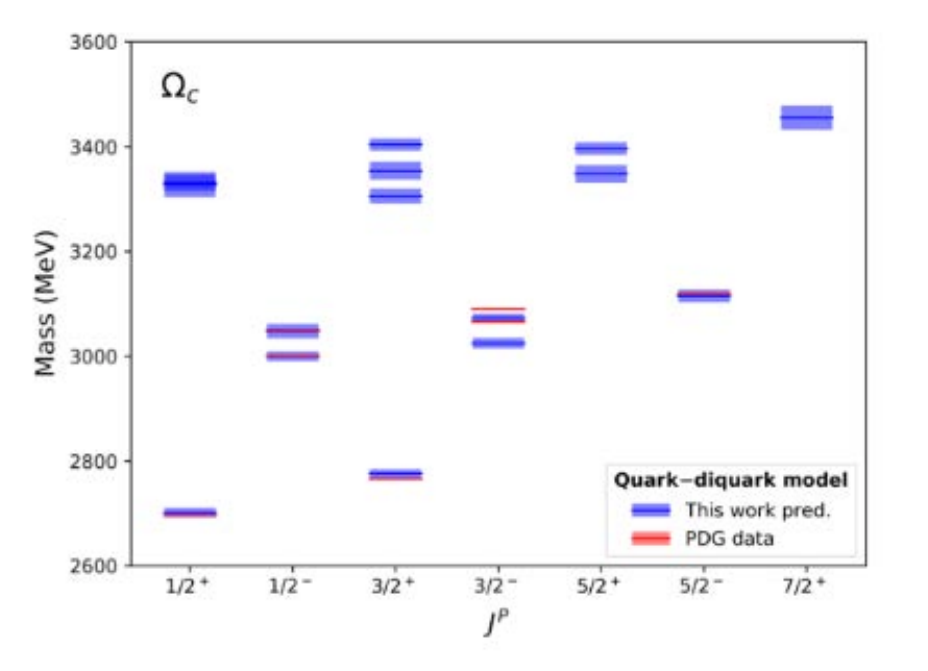}
  \caption{$\Omega_c$ mass spectra and tentative quantum number assignments based on the quark-diquark model Hamiltonian of Eq. \ref{MassFormulaqD}. the theoretical predictions and their uncertainties (blue
lines and bands) are compared with the experimental results PDG \cite{Workman:2022ynf} (red
lines and bands). Figure taken from Ref. ~\cite{Garcia-Tecocoatzi:2022zrf} APS copyright.}
  \label{fig:QDiquark}
\end{minipage}
\end{figure}

The labels ${L}$, ${S}$, ${J}$, and ${I}$ signify the orbital angular momentum, spin, total angular momentum, and isospin, respectively. The labels $(p,q)$ denote the $SU(3)$ flavor multiplets: the flavor sextet is labeled by $(2,0) \equiv \mathbf{6}$, the antitriplet by $(0,1) \equiv \mathbf{\bar{3}}$, the triplet by $(1,0) \equiv \mathbf{3}$, and the singlet by $(0,0) \equiv \mathbf{1}$. The parameters $A,B,E$ and $G$ were determined through a fit for single charmed baryons in Ref. ~\cite{Garcia-Tecocoatzi:2022zrf}, and  for the single bottom baryons in Ref. \cite{Garcia-Tecocoatzi:2023btk}. The theoretical results of Ref.~\cite{Garcia-Tecocoatzi:2022zrf} for the $\Omega_c$ are reported in Fig. \ref{fig:ThreeQM}.

\vspace{7mm}
{\large\bf 4.1~~Quark-diquark scheme for singly heavy baryons}
\vspace{4mm}
\\
The singly heavy baryons as quark-diquark systems can be described by the mass formula introduced in Ref.~\cite{Santopinto:2018ljf} 
\begin{eqnarray}
M^{qD}  &=& m_D + m_{b/c} + \omega_{r}\; n_{r} + a_{\rm S} \left[ S_{\rm tot}(S_{\rm tot}+1) \right]
 + a_{\rm SL} \frac{1}{2} \Big[ J_{}(J_{}+1) - L_{\rm tot}(L_{\rm tot}+1) \nonumber\\
&& - S_{\rm tot}(S_{\rm tot}+1) \Big] 
+a_{\rm I}\left[ I(I+1)  \right]+ a_{\rm F}\frac{1}{3} \left[ p(p+3)+q(q+3)+pq \right] ~.
\label{MassFormulaqD}
\end{eqnarray} 
where  $m_D$ and $m_{b/c}$  are the diquark  and heavy quark masses, respectively.  The theoretical results of Ref.~\cite{Garcia-Tecocoatzi:2022zrf} for the $\Omega_c$ within the quark-diquark scheme are reported in Fig. \ref{fig:QDiquark}.

\begin{table*}
\caption{Partial strong decay widths for missing $\Xi$  resonances using the hQM, as from Ref~\cite{Bijker:2015gyk}. APS copyright.} 
\label{tab:xi-missing} 
\begin{tabular}{cccccccccccc}
\hline
\hline
\noalign{\smallskip}
$\Xi$ & Mass & $\Sigma \overline{K}$ & $\Lambda \overline{K}$ & $\Xi \pi$ & $\Xi \eta$ 
& $\Sigma^* \overline{K}$ & $\Xi^* \pi$ & $\Lambda \overline{K}^*$&$\Sigma\overline{K}^*$& $\Xi\rho$&$\Xi\omega$ \\
\noalign{\smallskip}
\hline
\\
\multicolumn{12}{c}{$ $ hQM}\\
\multicolumn{12}{c}{$ $ \line(1,0){400}}\\
\noalign{\smallskip}
$^{2} 8_{1/2}[56,0_2^+]$ & 1843 & 125 & 6   & 5   & --    & --    & 15 & --& --& --& -- \\
$^{4} 8_{3/2}[70,1_1^-]$  & 2053 & 8 & 11 & 37 & { 0} & 223 & 154 & -- & --  & --& --\\
$^{2} 8_{1/2}[56,0_3^+]$ & 2190 & 0 & 0 & 0 & 0 & 0 & 0  & 0 & 0 & 0 & 0  \\
$^{2} 8_{1/2}[70,1_2^-]$  & 2288 & 0 & 0 & 0 & 0 & 0 & 0 & 0 & 0 & 0 & 0 \\
$^{2} 8_{3/2}[70,1_2^-]$  & 2288 & 0 & 0 & 0 & 0 & 0 & 0 & 0 & 0 & 0 & 0 \\
$^{4} 8_{1/2}[70,2_1^+]$ & 2327 & 3 & 1 & 8 & 1 & 6   & 5  & 8  & 10  & 40 & 1    \\
$^{4} 8_{3/2}[70,2_1^+]$ & 2327 & 2 & 1 & 4 & 0 & 35  & 32 & 16 & 16  & 62 & 1   \\
$^{4} 8_{5/2}[70,2_1^+]$ & 2327 & 6 & 7 & 24& 0 & 57  & 53 & 20 & 18  & 69 & 1   \\
$^{4} 8_{7/2}[70,2_1^+]$ & 2327 & 26& 33&108& 1 & 33  & 33 & 11 & 5   & 16 & 0   \\
$^{2} 8_{J  }[20,1_1^+]$  & 2377 & 0 & 0 & 0 & 0 & 0    & 0 & 0 & 0 &0 & 0   \\ 
$^{4} 8_{J}[70,1_2^-]$  & 2403  & 0 & 0 & 0 & 0 & 0 & 0 & 0 & 0 & 0 & 0   \\
\noalign{\smallskip}
\hline
\hline
\end{tabular}
\end{table*}

\begin{table}
\caption{As Table \ref{tab:xi-missing}, but for missing $\Omega$ resonances. APS copyright. } 
\label{tab:omega}
\begin{tabular}{cccccc}
\hline
\hline
\noalign{\smallskip}
$\Omega$ & Mass & $\Xi \overline{K}$ & $\Xi^* \overline{K}$&$\Omega \eta$& $\Xi\overline{K}^*$ \\
\noalign{\smallskip}
\hline
\\
\multicolumn{6}{c}{$ $ hQM}\\
\multicolumn{6}{c}{$ $ \line(1,0){290}}\\
\noalign{\smallskip}
$^{2}10_{1/2}[70,1_1^-]$ & 2142 & 26 & 48 & -- &--\\
$^{2}10_{3/2}[70,1_1^-]$ & 2142 & 68 & 403 & --&-- \\
$^{4}10_{3/2}[56,0_2^+]$ & 2162 & 68 & 102 & -- &-- \\
$^{4}10_{1/2}[56,2_2^+]$ & 2364 &109 & 34 & 27 &155\\
$^{4}10_{3/2}[56,2_2^+]$ & 2364 &  55 &137 & 88 &225\\
$^{4}10_{5/2}[56,2_2^+]$ & 2364 &  69 & 199 &117 &234\\
$^{4}10_{7/2}[56,2_2^+]$ & 2364 & 308 & 58 & 4 &23 \\
$^{2}10_{1/2}[70,1_2^-]$ & 2492 & 0 & 0 & 0 & 0 \\
$^{2}10_{3/2}[70,1_2^-]$ & 2492 & 0 & 1 & 0 & 0 \\
$^{4}10_{3/2}[56,0_3^+]$ & 2508 & 0 & 0 & 0 & 0 \\
\noalign{\smallskip}
\hline

\hline
\end{tabular}
\end{table}

\vspace{7mm}
{\Large\bf 5~~Discussion and conclusions}
\vspace{4mm}
\\
In Figs. \ref{fig:hQM}  and \ref{fig:diquark}  one can observe  in orange the hyperon missing states predicted within the hQM and the interacting  quark-diquark model, respectively. There are still many hyperons that can be studied  thanks to  the Extension Project for the J-PARC Hadron Experimental Facility.  Moreover, in Table \ref{tab:xi-missing} taken from Ref.\cite{Bijker:2015gyk}, we report the $\Xi$ missing states and their decay channels that can be studied in
 forthcoming experiments.  It is important to notice that the decay channels contain kaons, for example the $\Xi$ missing are strongly coupled to  the $\Sigma \overline{K}$, $\Lambda \overline{K}$, 
 $\Sigma^* \overline{K}$, $\Lambda \overline{K}^*$, or $\Sigma\overline{K}^*$ channels, See Table \ref{tab:xi-missing}, while the $\Omega$  missing states can be found in the $\Xi \overline{K}$, $\Xi^* \overline{K}$, and  $\Xi\overline{K}^*$ channels.

On the other hand, 
in Ref.\cite{Garcia-Tecocoatzi:2022zrf}, was provided the mass spectrum and strong decay width predictions for the $\Omega_c$ D$_\lambda$-wave states. By comparing Fig. \ref{fig:ThreeQM} with \ref{fig:QDiquark}, one can observes that in the quark-diquark scheme there are  much less missing singly charm bayrons. 

The model of Ref. \cite{Garcia-Tecocoatzi:2022zrf} predicts six $\Omega_c(ssc)$ D$_\lambda$-wave states, detailed in Table \ref{tab:All_mass_Omega}. Based on these results, one observes that the $\Omega_c^0(3327)$, observed by the LHCb Collaboration~\cite{LHCb:2023sxp}, can be assigned to either the $\vert l_{\lambda}=2, l_{\rho}=0 \rangle$, $^{2}D_{3/2}$ state with a mass of $3306^{+14}_{-14}$ MeV and a width of $10.6^{+5.3}_{-5.3}$ MeV, or the $\vert l_{\lambda}=2, l_{\rho}=0 \rangle$, $^{4}D_{1/2}$ state with a mass of $3330^{+25}_{-25}$ MeV and a width of $16.3^{+8.2}_{-8.0}$ MeV, both results are compatible with the experimental data, thus the measurement of the quantum numbers of   $\Omega_c$ D$_\lambda$-wave states is necessary for giving a unique assignment.  Additionally, this effort could lead to the discovery of the other five $\Omega_c$ D$_\lambda$-wave states.

\begin{table*}
\caption{ $\Omega_c(ssc)$ D$_\lambda$-wave  quantum number assignments (second column), predicted masses (third column) and strong decay widths (sixth column). The theoretical masses calculated within the diquark framework are reported(fourth column).  An $ssc$ state is characterized by the total angular momentum ${\bf J} = {\bf l}_{\rho}+{\bf l}_{\lambda} + {\bf S}_{\rm tot} $, where ${\bf S}_{\rm tot} = {\bf S}_{\rho}+\frac{1}{2}$, and the flavor multiple $\mathcal{F}$. Table taken from Ref. ~\cite{Garcia-Tecocoatzi:2022zrf} APS copyright.}
\begin{tabular}{c  c| c c c c c }\hline \hline
            &  & Three-quark &  Quark-diquark    &               &              &  \\ 
$\Omega_{c}(ssc)$&  & Predicted   &    Predicted   &  Experimental &  Predicted            & Experimental \\ 
 $\mathcal{F}={\bf {6}}_{\rm f}$  & $^{2S+1}L_{J}$ & Mass (MeV)  &   Mass (MeV)   &  Mass (MeV)   &  $\Gamma_{tot}$ (MeV) & $\Gamma$ (MeV) \\ \hline
\hline
$\vert l_{\lambda}=2, l_{\rho}\!\!=\!0 \rangle$ & $^{2}D_{3/2}$ & $3315^{+15}_{-14}$ & $3306^{+14}_{-14}$ & $\dagger$ & $10.6^{+5.3}_{-5.3}$ & $\dagger$ \\ 
$\vert l_{\lambda}=2, l_{\rho}\!\!=\!0 \rangle$ & $^{2}D_{5/2}$ & $3360^{+17}_{-16}$ & $3348^{+17}_{-17}$ & $\dagger$ & $24.4^{+12.0}_{-11.9}$ & $\dagger$ \\ 
$\vert l_{\lambda}=2, l_{\rho}\!\!=\!0 \rangle$ & $^{4}D_{1/2}$ & $3330^{+25}_{-25}$ & $3328^{+24}_{-23}$ & $\dagger$ & $16.3^{+8.2}_{-8.0}$ & $\dagger$ \\ 
$\vert l_{\lambda}=2, l_{\rho}\!\!=\!0\rangle$ & $^{4}D_{3/2}$ & $3357^{+18}_{-19}$ & $3354^{+17}_{-17}$ & $\dagger$ & $30.4^{+14.8}_{-14.9}$ & $\dagger$ \\ 
$\vert l_{\lambda}=2, l_{\rho}\!\!=\!0 \rangle$ & $^{4}D_{5/2}$ & $3402^{+13}_{-13}$ & $3396^{+12}_{-12}$ & $\dagger$ & $62.3^{+31.0}_{-31.1}$ & $\dagger$ \\ 
$\vert l_{\lambda}=2, l_{\rho}\!\!=\!0 \rangle$ & $^{4}D_{7/2}$ & $3466^{+23}_{-23}$ & $3455^{+23}_{-23}$ & $\dagger$ & $123.0^{+61.4}_{-62.1}$ & $\dagger$ \\ 

\hline \hline
\end{tabular}
\label{tab:All_mass_Omega}
\end{table*}


-----------



\stepcounter{count}
\clearpage

\phantomsection
\addcontentsline{toc}{section}{
{\bf Present and Future Kaonic Atoms Measurements with New Generation Radiation Detectors} \\
A.~Scordo}

\titl{Present and Future Kaonic Atoms Measurements with New Generation Radiation Detectors}

\name{A. Scordo 
$^{1}$ on behalf of the SIDDHARTA-2 Collaboration
}

\adr{
$^1$ Laboratori Nazionali di Frascati INFN, Via E. Fermi 54, Frascati, 00044, Italy
}

The study of kaonic atoms is crucial for understanding various aspects of nuclear physics and astrophysics [1]. Except for the most recent measurements at DA$\Phi$NE and JPARC on KHe [2-3] and KH [4-5], much of the knowledge on kaonic atoms dates back to the 1970s and 1980s [6]. These historical data serve as the experimental basis for developed theoretical models used to derive key parameters like KN and KNN interactions at threshold, nuclear density distributions, and neutron stars' equations of state (EOS).

Despite the historical data's significance, many of these measurements have large uncertainties on the strong interaction induced  shifts and widths; as a consequence, some measurements performed on the same transitions in different experiments are hardly compatible with each other [7-8], and quite often the relative and absolute yields of the main transition are de-facto unmeasured. 

Finally, the recent remeasurements of KHe transitions at DA$\Phi$NE and J-PARC have revealed inaccuracies in the older data, necessitating a new approach to improve the precision and reliability of the whole database of kaonic atom measurements, providing valuable information on nuclear absorption processes for $K^-$. 
Recent theoretical papers have revealed how, when using theoretical models of Kaon-nucleon(s) interaction to fit the existing kaonic atoms data, removing certain elements from the fits significantly improves the chi-squared values, suggesting that previous experimental points may have been incorrect [9].

Simultaneous measurements of different transitions in a single target may also offer insights into the very important atomic cascade models [10-11]. 

The X-rays emitted from kaonic atom $n_2\rightarrow n_1$ transitions range, depending on the Z of the element and on the $n_1$ and $n_2$ quantum numbers, from a few keV to hundreds of keV. Different detectors, such as crystal spectrometers, TES microcalorimeters, SDDs, Cd(Zn)Te, and HPGe, offer various advantages in terms of resolution, efficiency, and operational range. For instance, SDDs provide high efficiency in the 4-40 keV range, while Cd(Zn)Te detectors are suitable for the 20-300 keV range and operate at room temperature, making them ideal for intermediate mass kaonic atoms.

The DA$\Phi$NE collider delivers a nearly 4$\pi$ $K^-$ beam, and the SIDDHARTA-2 collaboration aims to exploit most of the solid angle and to maximize the potential of this unique facility utilizing a combination of SDDs for light kaonic atoms, HPGe for heavy kaonic atoms, and CdZnTe for intermediate kaonic atoms.

Initial tests with CdZnTe detectors have shown promising results, including good energy resolutions and linearity, both in the lab and at DA$\Phi$NE, with no observed radiation damage. Ongoing tests aim to improve the K/MIP identification, assess the real background at DA$\Phi$NE, and compare the performances of different detector sizes for future measurements [12-14].

More recently, the SIDDHARTA-2 collaboration managed to observe, in a very preliminary test run, kaonic atoms transitions from Cu, Al and Pb targets.

Ultra-fast timing CdZnTe detectors could be as well used for exotic atom spectroscopy both at J-PARC, where the high stopping power for K- in the target/degrader will ensure significantly higher statistics than those collectable at DA$\Phi$NE, enabling improved precision and expanding the kaonic atoms' database.

In conclusion, CdZnTe technology has proven to be ideal for intermediate mass kaonic atoms, demonstrating excellent in-beam performance during preliminary tests. The detectors' compactness and non-invasive electronics allow them to be used alongside existing experiments. 
The SIDDHARTA-2 collaboration is planning and designing new experimental setups for future measurements, with an increased detection surface area and improved spectroscopic performances, to be performed both on DA$\Phi$NE and on the K1.8BR beamline at J-PARC.

\vfill  

\noindent{\bf References }
\begin{description}
\setlength\itemsep{-3pt}
\item{[1]} C. Curceanu et al., Rev. Mod. Phys. {\bf 91} (2019) 025006.
\item{[2]} M. Bazzi et al., Phys.Lett.B {\bf 681} (2009) 310-314.
\item{[3]} S. Okada et al., Phys. Lett. B {\bf 653} (2007) 387.
\item{[4]} M. Bazzi et al., Phys.Lett.B {\bf 704} (2011) 113-117.
\item{[5]} M. Bazzi et al., Nucl.Phys.A {\bf 881} (2012) 88-97.
\item{[6]} E. Friedman et al. , Nucl. Phys. A {\bf 579} (1994) 518-538.
\item{[7]} C.J. Batty et al., Nucl. Phys. A {\bf 329} (1979) 407.
\item{[8]} P.D. Barnes et al., Nucl. Phys. A {\bf 231} (1974) 477.
\item{[9]} J. Obertova et al., Phys. Rev. C {\bf 106} (2022) 065201
\item{[10]} C. Batty et al., Sov. J. Part. Nucl, {\bf 13} (1982) 71.
\item{[11]} T. Koike et al., Genshikaku Kenkyu {\bf 49(6)} (2005) 159-164.   
\item{[12]} L. Abbene et al., Eur. Phys. J. ST {\bf 232(10)} (2023) 1487-1492.
\item{[13]} A. Scordo et al., Nucl.Instrum. Meth. A {\bf 1060} (2024) 169060.
\item{[14]} L. Abbene et al., Sensors {\bf 23(17)} (2023) 7328.
\end{description}

\stepcounter{count}
\clearpage

\phantomsection
\addcontentsline{toc}{section}{
{\bf Bound States of Two Decuplet Baryons with Total Spin $J = 3$ in the Constituent Quark Model} \\
T.~Sekihara}

\titl{\boldmath Bound States of Two Decuplet Baryons with Total Spin $J = 3$
  in the Constituent Quark Model}

\name{
Takayasu Sekihara$^{1}$
}

\adr{
$^1$ Graduate School of Life and Environmental Sciences,
  Kyoto Prefectural University, Sakyo-ku, Kyoto 606-8522, Japan
}


We evaluate the $S$-wave baryon-baryon potentials in the flavor
$SU(3)$ sector by solving the six-quark equation in terms of the
resonating group method (RGM) in the constituent quark model~[1].
While a similar analysis was conducted in Ref.~[2], our study
systematically explores the entire flavor $SU(3)$ sector, motivated by
recent experimental and lattice QCD advancements.

As a consequence, we find dibaryons in the flavor antidecuplet
($\bm{\overline{10}}$) states with total spin $J = 3$.  The properties
of these dibaryons are summarized in Table~\ref{tab:BS}.  While the
attraction in the $\Delta \Delta ( J = 3 , I = 0 )$ state was
previously calculated within the same model~[2], our systematic
exploration of the flavor $SU(3)$ sector enables us to further predict
the $\Delta \Sigma ^{\ast} ( J = 3 , I = 1/2 )$, $\Delta \Xi
^{\ast}$-$\Sigma ^{\ast} \Sigma ^{\ast} ( J = 3 , I = 1 )$, and
$\Delta \Omega$-$\Sigma ^{\ast} \Xi ^{\ast} ( J = 3 , I = 3/2 )$
dibaryons.  Although generated by constituent quark dynamics, the mean
squared distance between two baryons, $\sqrt{\langle r_{\rm D}^{2}
  \rangle }$, exceeds typical size of baryons ($\sim 1 \, \text{fm}$),
strongly suggesting that they are hadronic molecules rather than
compact hexaquark states.  The $\Delta \Xi ^{\ast}$-$\Sigma ^{\ast}
\Sigma ^{\ast}$ and $\Delta \Omega$-$\Sigma ^{\ast} \Xi ^{\ast}$ bound
states arise from coupled channels, with component fractions [$67 \%$
  ($33 \%$) for the $\Delta \Xi ^{\ast}$ ($\Sigma ^{\ast} \Sigma
  ^{\ast}$) component in the $\Delta \Xi ^{\ast}$-$\Sigma ^{\ast}
  \Sigma ^{\ast} (J = 3, I = 1)$ bound state, and $49 \%$ ($51 \%$)
  for the $\Delta \Omega$ ($\Sigma ^{\ast} \Xi ^{\ast}$) component in
  the $\Delta \Omega$-$\Sigma ^{\ast} \Xi ^{\ast} (J = 3 , I = 3/2)$
  bound state] consistent with the Clebsch--Gordan coefficient
predictions.

In Ref.~[1], we also present equivalent local potentials for
two-baryon systems in the flavor $SU(3)$ sector, along with their
strengths, within our model.

We anticipate that these flavor antidecuplet dibaryons may be
experimentally observed at J-PARC in the near future.

\begin{table}[!h]
  \caption{Properties of the dibaryon bound states~[1].
  }
  \label{tab:BS}
  \centering
  \begin{tabular}{lcl}
    \hline
    \hline
    System & Binding energy [MeV] & Note
    \\
    \hline
    $\Delta \Delta ( J = 3 , I = 0 )$ & 13.1 &
    $\sqrt{\langle r_{\rm D}^{2} \rangle } = 1.70 \, \text{fm}$
    \\
    $\Delta \Sigma ^{\ast} ( J = 3 , I = 1/2 )$ & 12.6 & 
    $\sqrt{\langle r_{\rm D}^{2} \rangle } = 1.68 \, \text{fm}$
    \\
    $\Delta \Xi ^{\ast}$-$\Sigma ^{\ast} \Sigma ^{\ast} ( J = 3 , I = 1 )$
    & 11.7
    & $\Delta \Xi ^{\ast}$ $67 \%$,
    $\Sigma ^{\ast} \Sigma ^{\ast}$ $33 \%$
    \\
    $\Delta \Omega$-$\Sigma ^{\ast} \Xi ^{\ast} ( J = 3 , I = 3/2 )$
    & 11.1
    & $\Delta \Omega$ $49 \%$,
    $\Sigma ^{\ast} \Xi ^{\ast}$ $51 \%$
    \\
    \hline
    \hline
  \end{tabular}
\end{table}

\vfill  

\noindent{\bf References }
\begin{description}
  \setlength\itemsep{-3pt}

  
\item{[1]} T.~Sekihara and T.~Hashiguchi,
Phys.\ Rev.\ C \textbf{108} (2023) 065202.

\item{[2]} M.~Oka and K.~Yazaki,
  Prog.\ Theor.\ Phys.\ \textbf{66} (1981) 556-571; 
  572-587.
\end{description}

\stepcounter{count}
\clearpage

\phantomsection
\addcontentsline{toc}{section}{
{\bf Fine-tuning of the $\bar{K}NN$ and $\bar{K}NNN$ calculations} \\
N.V.~Shevchenko}

\titl{Fine-tuning of the $\bar{K}NN$ and $\bar{K}NNN$ calculations}
\name{
N.V. Shevchenko$^{1}$
}

\adr{
$^1$ Nuclear Physics Institute, 25068 \v{R}e\v{z}, Czech Republic 
}


The lightest possible exotic system with strangeness $K^- pp$ was intensively studied theoreti-cally
using different methods and inputs. The predictions for the characteristics of the quasi-bound state
in the system are varying. The most recent E15 experiment at J-PARC [1] reported the first
clear signal of the $\bar{K}NN$ quasi-bound state. While the measured binding energy of the state is comparable
with some of the theoretical results, the width is much larger than the theoretical predictions.

We studied the quasi-bound state in the $K^- pp$ system [2] solving Faddeev-type dynamical-ly exact three-body 
Alt-Grassberger-Sandhas (AGS) equations [3] with coupled $\bar{K}NN - \pi \Sigma N$ channels using
different input. We demonstrated that the antikaon-nucleon interac-tion plays the main role in the description
of the $K^- pp$ system, while the dependence of the results on the nucleon-nucleon potential is weak.  

Recently, we studied what else can change a theoretical result [4]. For this sake, we calculated
binding energy and width of the quasi-bound state in the $K^- pp$ system with different potentials in the
lower $\pi \Sigma N$ channel. Namely, spin-dependent and spin-indepen-dent  versions of
the $\Sigma N - \Lambda N$ potential
were constructed with parameters fitted to experimental $YN$ cross-sections alone or together with $\Sigma N$
and $\Lambda N$ scattering lengths from "an advanced" recently proposed model of hyperon-nucleon interaction.
The depen-dence of
the $K^- pp$ quasi-bound state characteristics  on the $\pi N$ interaction was also studied by using different
versions of the pion-nucleon potential, fitted to pion-nucleon scattering lengths. It turned out that the dependence
of the $K^- pp$ result on the $YN$ interaction model can be quite strong, while $\pi N$ interaction changes
the predicted binding energy and width only slightly.

The next  step was constructing the antikaon-nucleon interaction models with three coupled channels: $\bar{K}N$,
$\pi \Sigma$, and $\pi \Lambda$.  While our chilly motivated antikaon-nucleon potential already coupled all
three channels, and its parameters were only refitted, the phenomenolo-gical $\bar{K}N - \pi \Sigma - \pi \Lambda$
potentials were newly constructed.
As before, we constructed two phenomenological models: one has a one-pole structure of the $\Lambda(1405)$
resonance, while the second one gives $\Lambda(1405)$ consisting of two poles. The new three potentials with
coupled $\bar{K}N - \pi \Sigma - \pi \Lambda$ channels equally well reproduce elastic and inelastic $K^- p$ cross-sections,
threshold branching ratios $\gamma$, $R_c$, and $R_n$  together with $1s$ level shift and width of kaonic hydrogen,
which is reproduced directly. The higher poles of all three potentials corresponding to the $\Lambda(1405)$ resonance
and situated very close one to the other. The lower poles of the chirally motivated and the two-pole
phenomenological antikaon-nucleon models are also not very different. The corresponding three-body Faddeev-type AGS
equations with coupled $\bar{K}NN - \pi \Sigma N - \pi \Lambda N$ channels were written down and solved. The resulting
$K^- pp$ binding energies and widths differ sufficiently from the previously calculated ones with the same remaining
two-body input.

Calculations with the new $YN$, $\pi N$ interaction models and the $\bar{K}N$ potentials with three coupled
channels  [4] lead to increasing the widths of the $K^- pp$ quasi-bound state calculated using two-pole
antikaon-nucleon potentials. However, the widths are
still smaller than the experimental value. In addition, the new result obtained with the two-pole phenomenological potential
reproduces the experimental binding energy [1] of the quasi-bound state in the $K^- pp$ system.

Finally, we performed new four-body calculations of the quasi-bound state in the $\bar{K}NNN$ system with the new
antikaon-nucleon interaction models coupling three $\bar{K}N - \pi \Sigma - \pi \Lambda$ channels. Four-body Faddeev-type
equations [5] were solved with separable forms of the three-body amplitudes and $2+2$ partition.
Since the four-body equations are one-channel ones, the corresponding antikaon-nucleon potentials were used there
in the exact optical form of the corresponding potential.  The four-body $\bar{K}NNN$ binding energies given by
the one- and two-pole phenomenological potentials and their widths become smaller than before and very close to each
other. On the opposite, the binding energy and width calculated with the new chirally motivated potential increased.
These preliminary results will be additionally checked.

\vfill  

\noindent{\bf References }
\begin{description}
\setlength\itemsep{-3pt}
\item{[1]} T. Yamaga,  EPJ Web Conf. {\bf 271} (2022 ) 07001. 
\item{[2]} N.V. Shevchenko, Few. Body. Syst. {\bf 58} (2017) 6. 
\item{[3]} E.O. Alt, P. Grassberger,W. Sandhas, Nucl. Phys. B {\bf 2} (1967) 167.
\item{[4]} N.V. Shevchenko, Few. Body. Syst. {\bf 65} (2024) 31. 
\item{[5]} P. Grassberger and W. Sandhas, Nucl. Phys. B {\bf 2} (1970) 181.
\end{description}

\stepcounter{count}
\clearpage

\phantomsection
\addcontentsline{toc}{section}{
{\bf The role of hyperons in cold and hot high density matter in astrophysics objects and in terrestrial collisions} \\
J.~R.~Stone}

\titl{The role of hyperons in cold and hot high density matter in astrophysics objects
  and in terrestrial collisions}

\name{Jirina R. Stone$^{1}$
}

\adr{
$^1$ University of Oxford, Department of Physics (Astrophysics),
Oxford, United Kingdom
}


The focus of this contribution is to compare matter created in heavy ion
 collisions (HIC) at low and medium beam energies with that found in 
 the cores of astrophysical (compact) objects where hyperons play a
 significant role.
 \vspace{0.2cm}
 
The quantities characterizing the two systems used in the comparison
include the relative population
of nucleons and hyperons, the maximum particle number density, 
 and neutron-proton asymmetry $\delta = (\rho_n - \rho_p)/(\rho_n+\rho_p)$.
Furthermore, temperature and pressure of matter in compact objects, the role of Coulomb effects in
HIC and the difference in microscopic properties of the two systems
are discussed.
\vspace{0.2cm}

Simulations were performed using the Boltzmann-Uehling-Uhlenbeck (BUU) transport theory [1,2] for HIC and the
quark-meson coupling model (QMC) [3,4] for cold neutron stars (NS) (T=0 MeV), hot
proto-neutron (PNS) stars (T=50 MeV)  and expected post-merger object of binary
NS collision (PMO) (T=100 MeV). 
\vspace{0.2cm}
\\
The main results are summarized as follows:

\begin{itemize}

\item{Matter in HIC consists of neutrons and protons and is governed
    solely by strong interactions with typical lifetimes too short
    for developing equilibrium states. No heavy baryons or leptons
    are present.
\vspace{0.2cm}
\\   
    In compact objects, both strong and weak interactions
    (with variable, temperature dependent lifetimes)
    act. Electrically-neutral equilibrium states, consisting of
    nucleons, heavy baryons, and leptons,
    develop, allowing determination of pressure as a
    function of density, temperature and composition, the equation of
    state. At T=0 MeV, hyperons
    appear at densities higher than a model-dependent threshold. At
    finite temperatures, hyperons are present at all densities.} 
    
  \item{The maximum total number density achievable in
      collisions of symmetric and asymmetric Ca, Sn and Pb ions,
      including the contribution of Coulomb interaction [5],
      does not exceed 2.5 $\rho_0$  ( $\rho_0$ = 0.16 fm
      $^{-3}$). This is in sharp contrast with the central density of 
      maximum mass NS, PNS and PMO which is expected to be in
      the range 5 -6 $\rho_0$ [3,4].}

  \item{The neutron-proton asymmetry in HIC does not  generally
      exceed the asymmetry in the initial state of the collision at all
      beam energies and tends to decrease during the reaction.
      The maximum asymmetry $\delta^{max}$ is 0.25 at $\rho$ = 0.4
      fm$^{-3}$ (2.5 $\rho_0$).
\vspace{0.2cm}

      In compact objects, maximum neutron-proton asymmetry $\delta^{max}$
      is strongly dependent on temperature, density and composition.
      At $\rho$ = 0.4 fm$^{-3}$, below the threshold density for
     appearance of hyperons at T=0 MeV, $\delta$ = 0.84.
     At finite temperature, without hyperons,
     $\delta$ is 0.67 and 0.37 at T=50 and 100 MeV, respectively.
     Including hyperons reduces the asymmetry further
     to $\delta$=0.59 (T=50 MeV), and $\delta$=0.15 (T=100 MeV).
     The large reduction of $\delta$ at high temperatures is
     likely related to the temperature dependence of 
     of nucleon-hyperon transmutations, e.g.
     $\Lambda \leftrightarrow p +\pi^-$ and
     $\Lambda \leftrightarrow n +\pi^0$,
     leading to a decrease in neutron and proton
     populations at different rates. }
     
 \item{A significant portion of the neutron-proton asymmetry in HIC
     originates from Coulomb forces, masking the pure
     nuclear contribution [5].  Since matter in compact objects
    is electrically neutral, any comparison with HIC matter must
    account for this effect. }

 \item{Appearance of hyperons at densities higher than a
     model-dependent
     threshold naturally softens the EoS of NS at T=0
     MeV as a new degree of freedom is added to the system, slowing down
     the rate of change in pressure with increasing density. At
     finite temperature, when hyperons are present at all densities,
     the pressure is lowered continuously with increasing density
     compared to matter with only neutrons and protons.
     However, with growing temperature, the thermal pressure counters the
     hyperon effect by up to 30\% at T=100 MeV.}
   \end{itemize}

     In summary, we contrasted the cold and hot matter in astrophysical
     objects with the matter created in HIC at low and medium beam
     energies and concluded that the two different forms of high
     density matter are incompatible and cannot provide meaningful
     mutual constraints.

\vfill  

\noindent{\bf References }
\begin{description}
\setlength\itemsep{-3pt}
\item{[1]} J.R.Stone, P.Danielewicz, and Y.Iwata,  Phys.Rev.C {\bf 96} (2017) 014612 
\item{[2]} J.R.Stone, P.Danielewicz, and Y.Iwata, Phys.Rev.C {\bf 109} (2024) 044603  
\item{[3]} J.R.Stone, V.Dexheimer et al.,  Mon.Not.R.Astron.Soc. {\bf 502} (2021) 3476 
\item{[4]} P.M.Guichon, J.R.Stone, and A.W.Thomas,  Phys.Rev.D {\bf 108} (2024) 083035 
\item{[5]} J.R.Stone, P.M.Guichon and A.W.Thomas,  Phys.Letts.B {\bf 826} (2022) 136915
\end{description}

\stepcounter{count}
\clearpage

\phantomsection
\addcontentsline{toc}{section}{
{\bf Plausibility of the LHCb $P_c(4312)^+$ in the GlueX $\gamma p \to J/\psi p$ total cross sections and J-PARC possibility for $\pi^-p \to J/\psi n$} \\
I.~Strakovsky}


\titl{Plausibility of the LHCb $P_c(4312)^+$ in the GlueX $\gamma p \to J/\psi p$ total cross sections and J-PARC possibility for $\pi^-p \to J/\psi n$}

\name{Igor~I.~Strakovsky$^{1}$
}

\adr{$^1$ Institute for Nuclear Studies, Department of Physics,
    The George Washington University, Washington, DC 20052, USA
}


\vspace{0.3cm}
I dedicated my talk to memory of my friend, Youngseok Oh. We lost him one year ago...
\vspace{0.2cm}

\begin{figure}[ht]
\vspace{-0.7cm}
\centering
{
    \includegraphics[width=0.4\textwidth,angle=90,keepaspectratio]{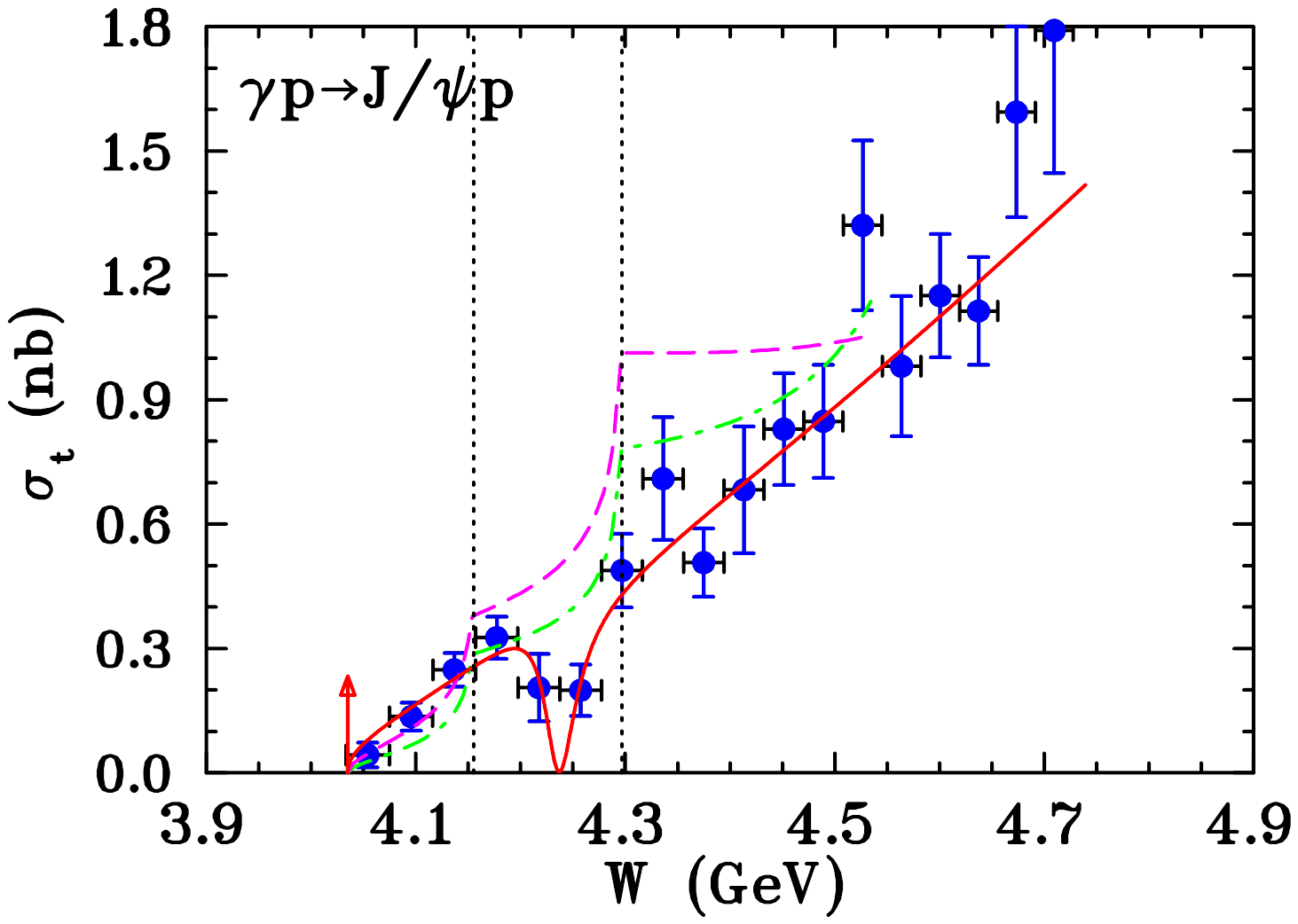}
}

\vspace{-0.5cm}
\centerline{\parbox{1\textwidth}{
    \caption[] {\protect\small 
    The GlueX total cross section for the $\gamma p\to J/\psi p$ photoproduction (blue filled circles)~[2]. The phenomenological best-fit result shown by red solid curve~[3]. The open charm model predictions~[4] shown by magenta dashed (green dash-dotted) curves with $q_{max}$ = 1~GeV/$c$ ($q_{max}$ = 1.2~GeV/$c$). 
    This model does not fit the GlueX data and has no normalization factor. 
    Vertical black dotted lines show $\Lambda_c\bar{D}^{(\ast)}$-thresholds. The red vertical arrow indicates the $J/\psi$ production threshold ($W = 4.035~\mathrm{GeV}$).} 
    \label{fig:fit1} } }
\end{figure}

New high-statistics total cross-section data for $\gamma p \to J/\psi p$ from the GlueX experiment~[1] is fitted in a 
search for the exotic $P_c(4312)^+$ state observed by the LHCb Collaboration in the reaction $\Lambda_b \to J/\psi pK^-$~[2]. There is no alternative confirmation of this LHCb observation. The fit of the GlueX data shows that destructive interference involving an S-wave resonance and associated non-resonance background produces a sharp dip structure about $77~\mathrm{MeV}$ below the LHCb mass, in the same location as a similar structure is seen in the GlueX data~[3]. The interference between open charm and gluon exchange may produce a dip~[4], but there is room for resonance~[3]. 
Limitations of the employed model the need for improved statistics (Fig.~\ref{fig:fit1}). High-p Experiment at J-PARC is capable to measure 
$\pi^-p \to J/\psi n$ cross sections, using $\pi 20$ beamline.
It allows us to understand dynamics of $c\bar{c}$ production at the threshold and to look for the effect of LHCb $P_c$. Additionally, it allows to determine $J/\psi N$ scattering length independently on the GlueX input which is dependent on Vector Meson Dominance model~[5].
More details about a future development of the J-PARC the Hadron Experimental Facility is given in Ref.~[6].
\vspace{0.5cm}
%
%
%
%
%
%

\vfill  

\noindent{\bf Acknowledgments: }
I thank Misha Ryskin, Takatsugu Ishikawa, and Atsushi Hosaka for useful remarks and continuous interest in this work. 
This work was supported in part by the U.~S. Department of Energy, Office of Science, Office of Nuclear Physics, under Award No.~DE--SC0016583.
\vspace{0.5cm}

\noindent{\bf References }
\begin{description}
\setlength\itemsep{-3pt}
\item{[1]} S.~Adhikari \textit{et al.} [GlueX Collaboration], Phys.\ Rev.\ C\ \textbf{108} (2023) 025201.
\item{[2]} R.~Aaij \textit{et al.} [LHCb Collaboration], Phys.\ Rev.\ Lett.\ \textbf{128} (2022) 062001.
\item{[3]} I.~Strakovsky, W.~J.~Briscoe, E.~Chudakov, I.~Larin, L.~Pentchev, A.~Schmidt, and R.~L.~Workman, Phys.\ Rev.\ C\ \textbf{108} (2023) 015202.
\item{[4]}  M.~L.~Du, V.~Baru, F.~K.~Guo, C.~Hanhart, U.~G.~Mei\ss{}ner, A.~Nefediev, and I.~Strakovsky, Eur.\ Phys.\ J.\ C\ \textbf{80} (2020) 1053.
\item{[5]} I.~Strakovsky, D.~Epifanov, and L.~Pentchev, Phys.\ Rev.\ C\ \textbf{101} (2020) 042201.
\item{[6]} \textit{ Task force on the extension of the Hadron Experimental Facility}, \href{https://doi.org/10.48550/arXiv.2110.04462}{arXiv:2110.04462 [nucl-ex], (2021).}
\end{description}

\stepcounter{count}
\clearpage

\phantomsection
\addcontentsline{toc}{section}{
{\bf $\bar{p}$ Physics Opportunities at J-PARC Hadron Experiment Facility} \\
K.~Suzuki}

\titl{$\bar{p}$ Physics Opportunities at J-PARC Hadron Experiment Facility}

\name{
Ken Suzuki$^{1}$
}

\adr{
$^1$ Research Center for Nuclear Physics (RCNP), University of Osaka, Ibaraki, 567-0047, Japan
}

One of the key new beam lines is the $\pi 20$ beam line, which will provide secondary particle beams with momenta up to 20 GeV/c and a momentum spread of 0.1\%. This beam line is capable of delivering high quality antiproton beam suitable for hadron physics studies. This will be the first to offer such high-quality antiproton beams in approximately 30 years. Fig.~1 shows the past hadron physics experiments with antiproton with their typical beam momenta range and their sensitivities indicating physics opportunities and the high potential that the $\pi 20$ beam line offers. The Inlets of Fig.~1 summarises the antiproton beam parameters of the beamline~[1], together with that of the $\overline{\rm P}$ANDA experiment at FAIR~[2].

The $\pi 20$ beam line at J-PARC represents a significant advancement for the hadron physics community. It will facilitate groundbreaking experiments in areas such as hyperon/anti-hyperon production, hadron spectroscopy, and studies involving antiprotons or other hadrons (such as hyperons, strange, or charm particles) within nuclei. Moreover, this facility will enable deeper exploration into hadron structure, offering unprecedented opportunities for scientific discovery.

\begin{figure}[h]
\begin{center}
\includegraphics[width=10cm]{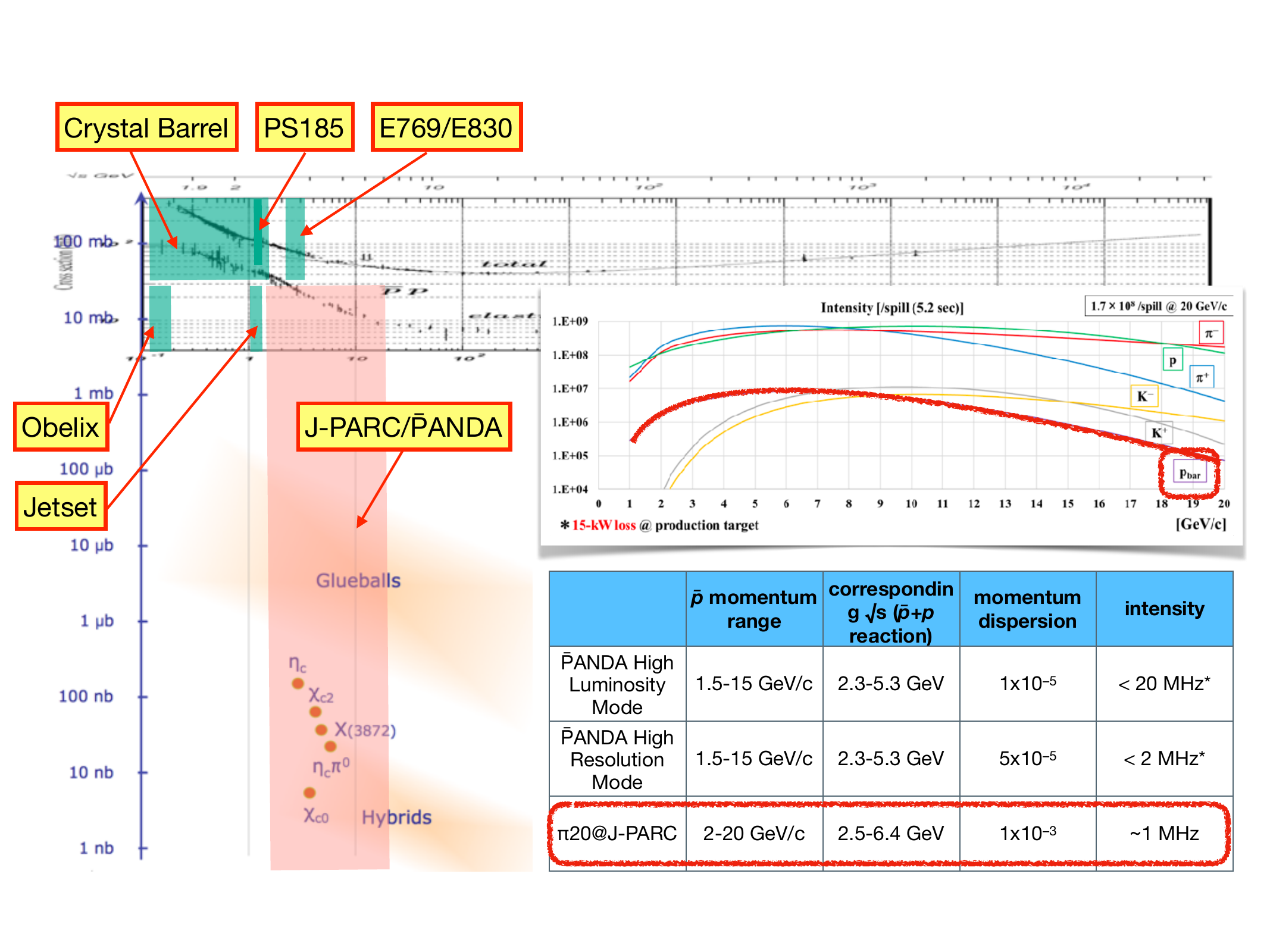}
\end{center}
\caption{$\bar{p}p$ reaction cross-sections with major experiments and their typical momenta and their sensitivities overlaid. The anticipated $\rm{\bar{p}}$ beam quality at the J-PARC hadron hall $\pi$20 beam line is also shown.}
\label{fig}
\end{figure}

-----------

\vfill  

\noindent{\bf References }
\begin{description}
\setlength\itemsep{-3pt}
\item{[1]} H. Takahashi, private communication (2024)
\item{[2]} Physics Performance Report $\overline{\rm P}$ANDA, \url{https://panda.gsi.de/oldwww/archive/public/panda_pb.pdf}
\end{description}

\stepcounter{count}
\clearpage

\phantomsection
\addcontentsline{toc}{section}{
{\bf Present and future of the Hadron Experimental Facility} \\
H.~Takahashi}

\titl{Present and future of the Hadron Experimental Facility}

\name{
Hitoshi Takahashi$^{1}$
}

\adr{
$^1$ Institute of Particle and Nuclear Studies, KEK, Tsukuba 305-0801, Japan
}

The J-PARC Hadron Experimental Facility has
three primary proton beam lines called as the A-line, B-line, and C-line.
In the A-line, 30-GeV primary protons are injected onto
a production target to produce various secondary beams.
A part of the primary protons is split from the A-line to the B-line,
and is used directly by experiments.
The C-line has recently been constructed, which provides
8-GeV primary protons for the COMET experiment.

SX tuning after the long shutdown started at
the beginning of February 2023.
The first beam arrived at the COMET area on February 9th, 2023.
The change of our license as an RI facility was
approved on March 15th after the facility inspection on March 14th.

The first 30-GeV run after the long shutdown was carried out in June 2023.
A new beam optics for the primary beam lines was applied
to reduce the micro spill structure in the B-line
by eliminating the vertical dispersion at the beam-splitting magnet.

After the beam time, a new permanent magnet was installed in the KL beam line
to sweep out charged $K$ mesons and to reduce the background.

The K1.8BR beam line is planned to be upgraded around 2026.
The beam line will be shortened by removing the last bending magnet
to increase $K^-$ yield.
The solenoid spectrometer will be replaced
by a larger superconducting solenoid
to cover a larger solid angle and to achieve higher neutron efficiency.

In order to increase the beam power,
a new rotating target is now being developed.
To achieve longer lifetime and higher rotation speed,
a He-gas lubricated bearing is planned, and some tests are in progress.
In addition to the basic rotation tests,
shake tests simulating a large earthquake and hard landing tests have been conducted.
It was found that seizure failure was suppressed
by using a bronze alloy for the radial bearing.
This was a great progress in the development of the rotary system.
Now, we are preparing a long-term rotation test,
and the detailed design of the new target could hopefully be determined in FY2024.

The HEF-ex project has made some progress over the past year.
The optical design of the HIHR beam line has been optimized
to reduce construction costs.
The optics were modified to focus simultaneously in the horizontal and
vertical directions at the mass slit,
resulting in the removal of 2 quadrupoles.
In addition, 6 quadrupoles just upstream of an experimental target
were removed for cost reduction.
A realistic design of the KL annex was performed
based on the radiation dose calculation,
including the design of a loading hatch and outbuildings
for equipment delivery using a rafter crane.
In order to evaluate the background level in the annex,
the radiation dose due to punch-through muons
downstream of the current primary beam dump was measured by the KL group.
The result was almost in agreement with the simulation.
It is also useful for planning additional shielding during the construction
of the extended hall.

\vfill  

\stepcounter{count}
\clearpage

\phantomsection
\addcontentsline{toc}{section}{
{\bf Recent results on heavy hadrons at Belle and Belle II} \\
M.~Takizawa on behalf of Belle and Belle II collaborations}

\titl{Recent results on heavy hadrons at Belle and Belle II}

\name{
Makoto Takizawa$^{1}$ on behalf of Belle and Belle II collaborations
}

\adr{
$^1$ Showa Pharmaceutical University, Machida, Tokyo, 194-8543, Japan
}

\hfill \break

After the briefly  introducing the  Belle and Belle II experiments, I presented two recent results concerning heavy hadrons from the Belle experiment and two from the Belle II experiment.

\hfill \break

The first result I presented evidence of a new excited charmed baryon decaying to \hfil \break
$\Sigma_c (2455)^{0, ++} \pi^\pm$~[1]. The Belle experiment studied $\bar B^0 \to \Sigma_c (2455)^{0, ++} \pi^\pm \bar p$ decays based on $772 \times 10^6$ $B \bar B$ events collected with the Belle detector at the KEKB asymmetric-energy $e^+ e^-$ collider. The  $\Sigma_c (2455)^{0, ++}$ candidates have been reconstructed via their decay to $\Lambda_c^+ \pi^\mp$
and $\Lambda_c^+$ decays to $p K^- \pi^+$, $p K^0_S$, and $\Lambda \pi^+$ final states.
A new structure has been found in the invariant mass spectra of 
$\Sigma_c (2455)^{0, ++} \pi^\pm$ with a significance of $4.2 \sigma$, tentatively named $\Lambda_c (2910)^+$.  It mass and width were measured
to be $(2913.8 \pm 5.6 \pm 3.8)$ ${\rm MeV}/c^2$ and $(51.8 \pm 20.0 \pm 18.8)$ ${\rm MeV}$, respectively. 

\hfill \break

The second result  involved studying   the lineshape of $X(3872)$ in the decay $B \to D^0 \bar D^{\ast 0} K$, using a same data sample as the first result~[2]. 
The peak near the threshold in the $D^0 \bar D^{\ast 0}$ invariant mass spectrum was fitted using a relativistic Breit-Wigner lineshape. The mass and width parameters has been determined to be $(3873.1^{+0.56}_{-0.50} \pm 0.10)$ ${\rm MeV}/c^2$ and $(5.2^{+2.2}_{-1.5}\pm 0.4)$ ${\rm MeV}$, respectively. The peak has been also studied using a Flatt\'e lineshape.

\hfill \break

The third result was the observation of $e^+ e^-  \to \omega \chi_{bJ}(1P)$ $(J = 0, 1, {\rm or}\,  2)$ using  samples at conter-of-mass energies $\sqrt{s} = 10.701, 10.745$, and 10.805 GeV, corresponding to integrated luminosities, 1.6, 9.8, and 4.7 fb${}^{-1}$, respectively~[3]. These data were collected with the Belle II detector during special operation of the SuperKEKB collider above the $\Upsilon(4S)$ resonance. A significant $\omega \chi_{b1}$ signal and evidence 
for the $e^+ e^- \to \omega \chi_{b2}$ process were foound at $\sqrt{s} = 10.745$ GeV. A strong enhancement of the Born cross section near 10.75 GeV was observed, with its energy dependence consistent with the $\Upsilon(10753)$ state.

\hfill \break

The last result was the first observation of  $\Upsilon(10753)$ decays to the $\pi^+ \pi^- \Upsilon(1S)$ and $\pi^+ \pi^- \Upsilon(2S)$ final states, 
using the same data sample as the third result~[4]. The mass and width of $\Upsilon(10753)$ have been measured to be 
$(10756.6 \pm 2.7 \pm 0.9)$ ${\rm MeV}/c^2$ and $(29.0 \pm 8.8 \pm 1.2)$ ${\rm MeV}$, respectively.  The significant difference 
between the slow production rate of $\omega \chi_{bJ}$ to $\pi^+ \pi^- \Upsilon(nS)$ at the $\Upsilon(10860)$ and the rate at $\Upsilon(10753)$ 
suggests these states may have different internal structures.

-----------

\vfill  

\noindent{\bf References }
\begin{description}
\setlength\itemsep{-3pt}
\item{[1]} \href{https://doi.org/10.1103/PhysRevLett.130.031901}{Y. B. Li et al., (Belle Collaboration), Phys. Rev. Lett. {\bf 130}, 031901 (2023).}
\item{[2]}  \href{https://doi.org/10.1103/PhysRevD.107.112011}{H. Hirata et al., (Belle Collaboration), Phys. Rev. D {\bf 107},112011 (2023).}
\item{[3]}  \href{https://doi.org/10.1103/PhysRevLett.130.091902}{I. Adachi et al., (Belle II Collaboration), Phys. Rev. Lett. {\bf 130}, 091902 (2023).}
\item{[4]} \href{https://doi.org/10.48550/arXiv.2401.12021}{I. Adachi et al., (Belle II Collaboration), 	arXiv:2401.12021 [hep-ex] (2024), accepted for publication in JHEP.}
\end{description}

\stepcounter{count}
\clearpage

\phantomsection
\addcontentsline{toc}{section}{
{\bf Measurement of Generalized Parton Distributions at J-PARC} \\
N.~Tomida}

\titl{Measurement of Generalized Parton Distributions at J-PARC}

\name{
Natsuki Tomida$^{1}$
}

\adr{
$^1$ Department of Physics, Kyoto University, Kyoto, 606-8502, Japan
}


Generalized Parton Distributions (GPDs) describe three-dimensional distributions of partons in nucleons.
There are four nucleon GPDs without parton helicity flip, $H, E, \tilde{H}$ and $\tilde{E}$.
They are described as a function of
the average momentum fraction carried by the struck parton $x$,
the longitudinal momentum fraction transferred to the struck parton $\xi$,
and the squared four momentum transfer $t$.
GPDs contain rich information on the internal structure of nucleons.
We can derive form factors and gravitational form factors from the moment of $x$ of GPDs.
GPDs have been measured with lepton and gamma induced reactions, namely Deeply Virtual Compton Scattering (DVCS), Deeply Virtual Meson Production (DVMP) and Time-like Compton Scattering (TCS) reactions.
In these measurements, GPDs are evaluated from Compton form factors and thus the $x$-dependence of GPDs is inaccessible.
In addition, the accessible kinematical region is limited to the DGLAP region ($-1<x<\xi$ or $\xi<x<1$).

At J-PARC, we can measure GPDs with unique reactions and can examine the $x$-dependence and the ERBL region ($-\xi<x<\xi$) of GPDs.
The 30 GeV proton beam at the high momentum beamline, the high momentum secondary beam at the future $\pi$20 beamline, and the MARQ spectrometer under construction give unique opportunity to measure GPDs using hadron beam.
Recently, J.-W. Qiu and Z. Yu proposed a new framework of processes that can access to GPDs.
The processes are called Single Diffractive Hard Exclusive Processes (SDHEPs)~[1].
They show that cross sections of 2$\to$3 processes, $B+N \to C+D+N'$ can be described with GPDs
when $C$ and $D$ have large opposite transverse momenta.
Here, $B$, $C$ and $D$ can be a lepton, gamma or hadron, and $N$ and $N'$ indicate a nucleon.
Among many SDHEPs, following processes are theoretically studied for measurements at J-PARC. 
(1) $p + p \to p + \pi + B$(where $B$ is a baryon)~[2],
(2) $\pi^- + p \to \gamma + \gamma + n$~[3],
and (3) $\pi^- + p \to \mu^+ + \mu^- + n$~[3,4].
The measurement of the process (1) can be carried out at the current high momentum beamline.
We can access the $x$-dependence and the ERBL region ($-\xi<x<\xi$) of GPDs from the measurement of differential cross sections of this reaction.
We can also examine the $x$-dependence of GPDs via the process (2).
We need an electromagnetic calorimeter for high energy photons up to 10 GeV to measure this reaction.
It is not included in the current MARQ spectrometer design.
The measurement of the process (3) can be carried out using the MARQ spectrometer.
Thanks to the high rate capability of the spectrometer, we can place a hadron absorber to identify muons at the downstream of the spectrometer.
We can identify exclusive events by analyzing momentum of di-muons at the upstream of the absorber.
Because the cross sections of the processes (2) and (3) are an order of 10 pb, we need a high intensity secondary beam to measure those reactions

There has been no GPDs measurement using hadron beam up to now. The measurement at J-PARC will open a new insight for studying the nucleon structure.

\vfill  

\noindent{\bf References }
\begin{description}
\setlength\itemsep{-3pt}
\item{[1]} J.-W. Qiu and Z. Yu, Phys. Rev. D {\bf 107} (2023) 014007.
\item{[2]} S. Kumano, M. Strikman and K. Sudoh, Phys. Rev. D {\bf 80} (2009) 074003.
\item{[3]} J.-W. Qiu and Z. Yu, Phys. Rev. D {\bf 109} (2024) 074023.
\item{[4]} E.R. Berger \textit{et. al.}, Phys. Lett. B {\bf 523} (2001) 265, 
S.V. Goloskokov \textit{et. al.}, Phys. Lett. B {\bf 748} (2015) 323,
T. Sawada \textit{et. al.}, Phys. Rev. D {\bf 93} (2016) 114034.
\end{description}

\stepcounter{count}
\clearpage

\phantomsection
\addcontentsline{toc}{section}{
{\bf Structures and production of medium-heavy hypernuclei in shell-model calculation} \\
A.~Umeya$^*$, T.~Motoba, and K.~Itonaga}

\titl{Structures and production of medium-heavy hypernuclei
in shell-model calculation}

\name{
Atsushi Umeya$^{1}$, Toshio Motoba$^{2,3}$, and Kazunori Itonaga$^{4}$
}

\adr{
$^1$ Liberal Arts and Sciences, Nippon Institute of Technology, 
345-8501 Saitama, Japan\\
$^2$ Research Center for Nuclear Physics, Osaka University, 
567-0047 Ibaraki, Osaka, Japan\\
$^3$ Yukawa Institute for Theoretical Physics, Kyoto University, 
606-8502 Kyoto, Japan\\
$^4$ Faculty of Medicine, Miyazaki University, 889-1692 Miyazaki, Japan\\
}

In addition to $p$-shell hypernuclear studies, 
the $\gamma$-ray measurement in ${}_{\;\Lambda}^{19}\mathrm{F}$ 
opened a new possibility of detailed spectroscopic analyses 
for $sd$-shell and heavier systems with strangeness $S=-1$~[1]. 
Correspondingly, a theoretical study of ${}_{\;\Lambda}^{19}\mathrm{F}$, 
as a gateway to $sd$-shell systems, 
has been performed to predict the detailed energy level structures 
and production cross sections~[2]. 
Recently, new projects of high-intensity and high-resolution 
$(K_{}^{-}, \pi_{}^{-}\gamma)$ and $(\pi_{}^{+}, K_{}^{+}\gamma)$ 
reaction experiments 
are being scheduled in the improvement of the J-PARC facility. 
Also, in the Jefferson Laboratory, 
a series of recent $(e, e'K_{}^{+})$ reaction experiments 
provide high-resolution data of the low-lying energy levels 
for $p$-shell hypernuclei~[3] 
and new experiments are also planned for heavier systems. 
\par
In anticipation of such new possibilities, 
here we take ${}_{\;\Lambda}^{27}\mathrm{Mg}$ 
as a typical $sd$-shell hypernucleus 
to be studied in the microscopic calculation, 
because the ${}_{}^{27}\mathrm{Al}$ nucleus is a unique target 
that provides a chance of combined analyses of different production reactions. 
Moreover, in the ${}_{\;\Lambda}^{27}\mathrm{Mg}$ structure, 
it is interesting to see the coupling feature 
between $\Lambda$ and the rotational bands 
of the even-even core nucleus ${}_{}^{26}\mathrm{Mg}$. 
The sufficiently large shell-model space is adopted in the calculations 
and partial results of our analyses 
for the ${}_{\;\Lambda}^{27}\mathrm{Mg}$ levels 
and ${}_{}^{26}\mathrm{Mg}$ band structures 
are briefly shown in the conference report~[4]. 
\par
Here we discuss the cross-section estimates of the different reactions 
leading to the ${}_{\;\Lambda}^{27}\mathrm{Mg}$ production. 
The left and right panels of Fig. 1 show the DWIA spectra of 
${}_{}^{27}\mathrm{Al}$ $(\gamma, K_{}^{+})$ ${}_{\;\Lambda}^{27}\mathrm{Mg}$ 
and 
${}_{}^{27}\mathrm{Al}$ $(\pi_{}^{-}, K_{}^{0})$ ${}_{\;\Lambda}^{27}\mathrm{Mg}$ 
reactions, respectively, 
that are calculated with the microscopic shell-model wave functions. 
The quasi-free contributions in the $E_{\Lambda}^{} > 0$ continuum region 
are omitted for simplicity. 
It is noted 
that the dominant contributions to the cross sections are different 
between the $(\gamma, K_{}^{+})$ and $(\pi_{}^{-}, K_{}^{0})$ reactions. 
In the doublet states which consist of a $J_{\mathrm{core}}^{}$ core 
and an $s_{1/2}^{\Lambda}$ hyperon, 
the $J_{\mathrm{core}}^{} + 1/2$ state has the larger cross section 
than the other in the $(\gamma, K_{}^{+})$ reaction, 
while the $J_{\mathrm{core}}^{} - 1/2$ state has the larger cross section 
in the $(\pi_{}^{-}, K_{}^{0})$ reaction. 
For example, 
in the first $(5/2_{}^{+}, 3/2_{}^{+})$ doublet 
of ${}_{\;\Lambda}^{27}\mathrm{Mg}$ 
obtained at $E_{\Lambda}^{} \simeq -15.1$ $\mathrm{MeV}$, 
we get the cross sections: 
$\frac{d\sigma}{d\Omega}(5/2_{}^{+}) = 14.356$ $\mathrm{nb/sr}$ vs. 
$\frac{d\sigma}{d\Omega}(3/2_{}^{+}) = 8.268$ $\mathrm{nb/sr}$ 
in the $(\gamma, K_{}^{+})$ reaction, 
while those are 
$\frac{d\sigma}{d\Omega}(5/2_{}^{+}) = 0.366$ $\mathrm{\upmu b/sr}$ vs. 
$\frac{d\sigma}{d\Omega}(3/2_{}^{+}) = 0.649$ $\mathrm{\upmu b/sr}$ 
in the $(\pi_{}^{-}, K_{}^{0})$ reaction. 
Another example is the $(7/2_{}^{+}, 9/2_{}^{+})$ doublet 
at $E_{\Lambda}^{} \simeq -12.6$ $\mathrm{MeV}$: 
both states are excited due to the spin-flip dominant nature 
in $(\gamma, K_{}^{+})$ reaction, 
while only $7/2_{}^{+}$ state is excited 
in the $(\pi_{}^{-}, K_{}^{0})$ reaction. 
The $(\gamma, K_{}^{+})$ reaction corresponds to the $(e, e'K_{}^{+})$ reaction 
to be done at the Jefferson Laboratory. 
On the other hand, 
the $(\pi_{}^{-}, K_{}^{0})$ reaction is one of the options for the experiment 
by using the high-intensity and high-resolution $\pi$-beam 
that can be performed at the J-PARC facility. 
Comparison of the different reaction processes 
including possible measurements of $\gamma$-rays 
can reveal the detailed energy level structure of hypernuclei 
not only in the $p$-shell but also in the $sd$-shell and heavier regions.

\begin{figure}[t]
\hfil\includegraphics[width=0.90\textwidth]{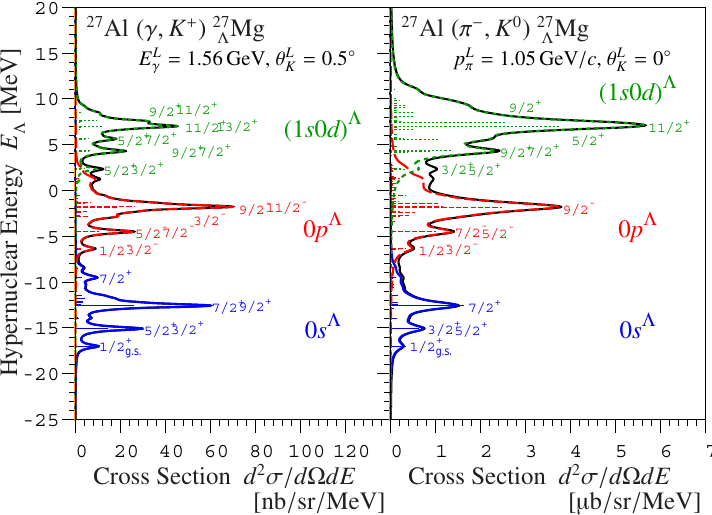}
\caption{The DWIA cross-section spectra of the 
${}_{}^{27}\mathrm{Al}$ $(\gamma, K_{}^{+})$ ${}_{\;\Lambda}^{27}\mathrm{Mg}$ reaction (left) 
and 
${}_{}^{27}\mathrm{Al}$ $(\pi_{}^{-}, K_{}^{0})$ ${}_{\;\Lambda}^{27}\mathrm{Mg}$ reaction (right).
They are preliminary results without the quasi-free contributions 
for the these reactions.}
\end{figure}

\vfill  

\newpage
\noindent{\textbf{References}}
\begin{description}
\setlength{\itemsep}{-3pt}
\item{[1]} S. B. Yang et al., Phys.\ Rev.\ Lett.\ \textbf{120}, 132505 (2018).
\item{[2]} A. Umeya and T. Motoba, Nucl.\ Phys.\ A \textbf{954}, 242 (2016). 
\item{[3]} T. Gogami et al., Phys.\ Rev.\ C \textbf{103}, L041301 (2021), and references therein.
\item{[4]} A. Umeya, Il Nuovo Cimento \textbf{47 C}, 234 (2024).
\end{description}

\stepcounter{count}
\clearpage

\phantomsection
\addcontentsline{toc}{section}{
{\bf Study of the three-body dynamics at short range via femtoscopy in nucleus-nucleus collisions} \\
O.~V\'azquez Doce}

\titl{Study of the three-body dynamics at short range via femtoscopy in nucleus-nucleus collisions}

\name{Ot\'on V\'azquez Doce$^{1}$}

\adr{$^1$ INFN, Laboratori Nazionali di Frascati, Frascati, Italy}


In this presentation, the current efforts to study the three-body dynamics and access in a direct way the effects of genuine three-body forces in hadron systems by using the femtoscopy technique have been shown.

In the recent years it has been demonstrated how by using femtoscopy, in particular in small systems [1], one can access information on hadronic interactions for stable and unstable particles produced in nucleus--nucleus collisions. The effect of the strong interaction among hadron pairs have been studied by the ALICE Collaboration at the LHC even in exotic systems with strangeness up to $S = - 3$, like the proton--$\Omega$ system [2], as well as in the charm sector. The femtoscopy technique is applied by measuring the correlation function of hadron pairs as a function of the relative momentum, that is compared with a theoretical expectation. To build the predicted correlation function two main ingredients are needed, namely the characteristics of the source of particles created after the collision, and the description of the final state interaction between the pair of hadrons. Provided that one has characterized \textit{a priori} the source, clear conclusions can be obtained regarding the interaction. In proton--proton (pp) collisions at the LHC, the size of the source is of about 1 fm, hence giving access to effects of the short-range strong interaction.

As a next step for femtoscopy, three-body systems are being accessed using two different methods, namely studies of hadron-deuteron correlations, and the extension of the femtoscopy method to the study of the three-particle correlation function.

\begin{figure}[h]
    \centering
    \includegraphics[width=0.53\textwidth]{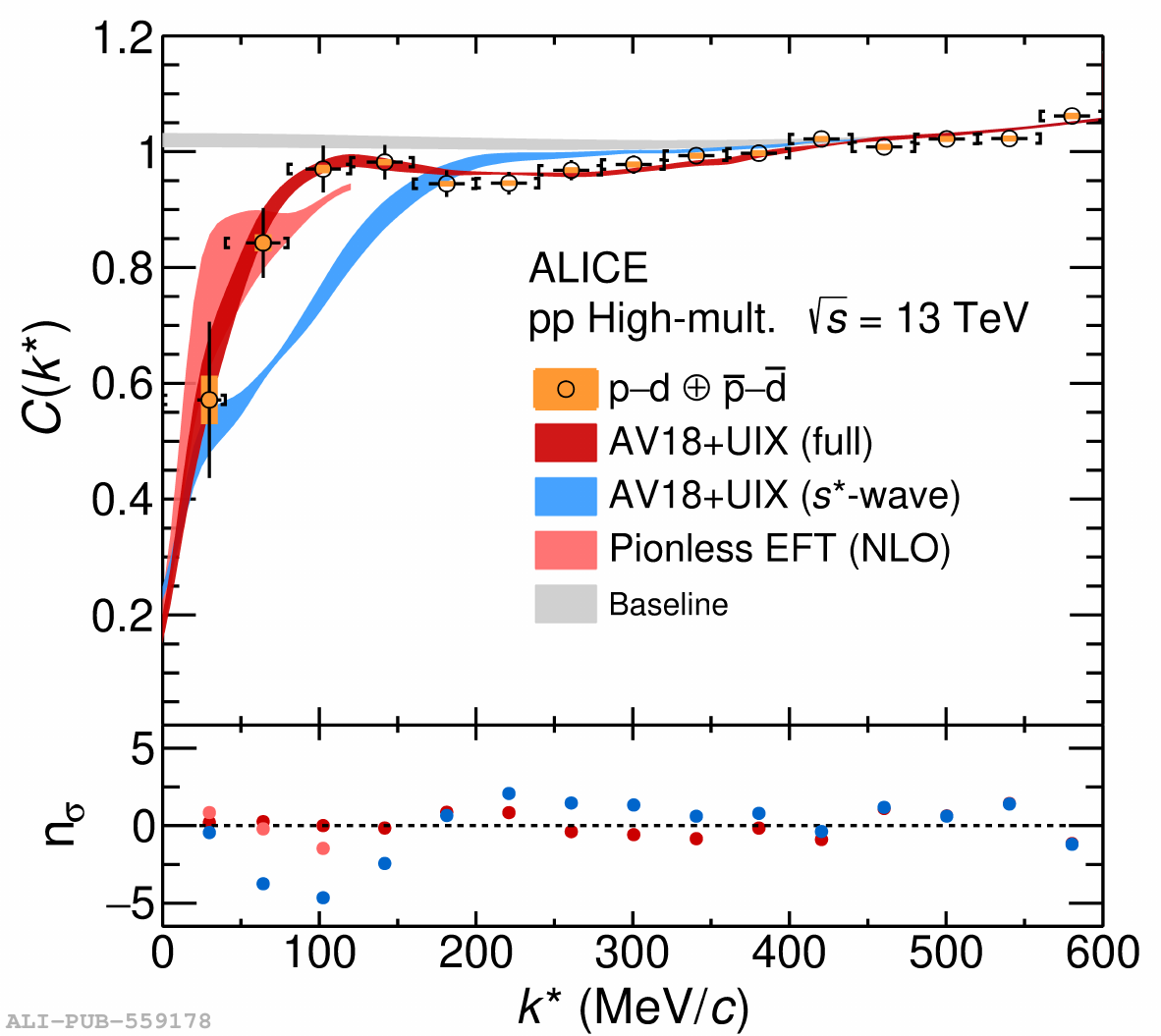}
    \includegraphics[width=0.46\textwidth]{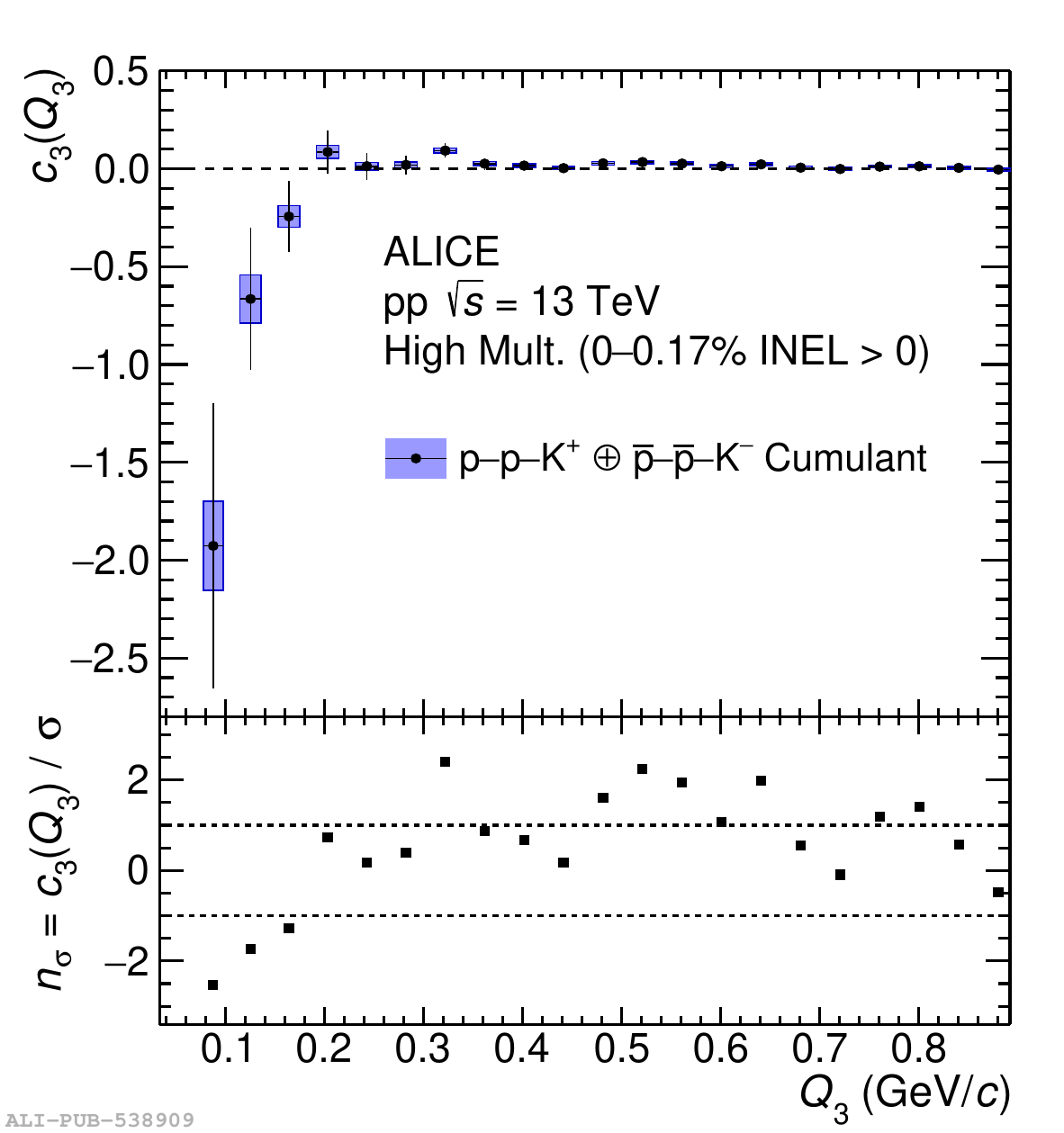}
    \caption{Right panel: proton--deuteron correlation function measured by ALICE in high-multiplicity pp collisions [3] compared with theoretical predictions. The red band corresponds to a full three-body calculation considering two- and three-body forces. Right panel: Cumulant for the p--p--K$^-$ system measured by ALICE in high-multiplicity pp collisions [4], see text for details.}
    \label{fig:femto}
\end{figure}

For the first approach, femtoscopic correlations among kaon--deuteron and proton--deuteron pairs as a function of the relative momentum $k^*$ have been measured by the ALICE Collaboration in pp collisions at $\sqrt{s}=13$~TeV [3]. Predictions using a model assuming two-particle scattering at the small distances that characterize the pp collisions describe very well the K$^{+}$--d data, demonstrating that the relative distances at which deuterons and proton/kaons are produced are around 1--2 fm. For the p--d case, a simple two-body calculation cannot describe the data, being necessary to develop a full three-body calculation that accounts for the internal structure of the deuteron, see the left panel of Fig. \ref{fig:femto}. The results show that femtoscopy with hadron-deuteron pairs in pp collisions is sensitive to the three-body dynamics at play, opening the possibility to extend this method to the strange sector by studying, for example, $\Lambda$--d pairs in the future.

The second approach gives an alternative to probe the dynamics of three-body systems by studying the final state interactions within triplet of particles emitted in the same pp collision. It has been demonstrated by ALICE that analysis of the three-body correlation function as a function of the hyper-momentum $Q_3$ gives access to possible genuine three-body effects, also in systems with strangeness like p--p--K$^+$ and p--p--K$^-$ [4]. The cumulant method allows one to subtract from the three-body correlations the effects originated by two-body interactions only, hence any deviation seen in the resulting cumulant can be attributed to genuine three-body effects, see the right panel of Fig. \ref{fig:femto}.

In summary, the femtoscopy technique is a promising tool to deliver new data on three-body systems at very small distances, also in the strangeness -- and even charm -- sectors, and constitutes a complementary approach to established methods like the study of hypernuclei.

\vfill  

\noindent{\bf References }
\begin{description}
\setlength\itemsep{-3pt}

\item{[1]} L. Fabbietti, V. Mantovani Sarti and O. V\'azquez Doce, Annu. Rev. Nucl. Part. Sci. {\bf 71} (2021) 377-402.

\item{[2]} ALICE Coll., Nature {\bf 588} (2020) 232-238.

\item{[3]} ALICE Coll., arXiv:2308.16120 [nucl-ex].
\item{[4]} ALICE Coll., Eur.Phys.J.A {\bf 59} (2023)  12, 298.

\end{description}

\stepcounter{count}
\clearpage

\phantomsection
\addcontentsline{toc}{section}{
{\bf Investigation of kaonic atom optical potential by the high-precision data} \\
J.~Yamagata-Sekihara}

\titl{Investigation of kaonic atom optical potential by the high-precision data}

\name{
Junko Yamagata-Sekihara$^{1}$
}

\adr{
$^1$ Department of Physics, Kyoto Sangyo University, Kyoto 603-8555, Japan
}


The ${\bar K}$-nucleus bound states, that is, kaonic atoms and kaonic nuclei, are promising for the studies of the kaon properties at finite nuclear density and the kaon-nucleon interaction in the nuclear medium.
The $2p$ states of the kaonic atoms in $^3$He and $^4$He have been observed very precisely by the J-PARC E62 experiment [1].

We theoretically investigate kaonic atom states in Ref. [2] based on the latest high precision data [1].
We consider the phenomenological form for the optical potential $U(r)$ proportional to the nuclear density distribution, defined as,
\begin{equation}
U(r)=(V_0+iW_0)\frac{\rho(r)}{\rho_0}
\end{equation}
where $\rho_0$ is the normal nuclear density $\rho_0=0.17~{\rm fm}^{-3}$, and $V_0$ and $W_0$ are parameters of the potential strength.
We assume the isoscalar form (IS) for the optical potential, which does not distinguish the proton and neutron densities, and extract the potential parameters consistent with the data of $^3$He and $^4$He atoms simultaneously.
By fitting to the data, we find two parameter sets, which can be categorized as weak-attraction strong-absorption potential ($V_0,W_0)=(-90,-120)$ MeV (IS-A) and strong-attraction weak-absorption potential ($V_0,W_0)=(-280,-70$) MeV (IS-B).

We apply these optical potentials to the global study of heavier kaonic atoms up to Sn.
We conclude the parameter IS-A~($V_0,W_0)=(-90,-120$) MeV is significantly better suited to describe the kaonic atom data across the wide range of the periodic table than IS-B.
 The potential with the parameter IS-B has a so large real part and provides the nuclear state with the same angular momenta as the atomic state.
 Consequently, the level repulsion between atomic and nuclear states makes the shift of the atomic states repulsive, while the potential with the parameter IS-A has the large imaginary part which makes the shift repulsive.
 The IS-A parameter set shares the same features with the potential studied in Ref. [3] and has the relatively weak real part and strong imaginary part.

We then consider the kaonic nuclear states in $^3$He and $^4$He by the same theoretical framework for the optical potential and investigate the effects of the nuclear states on the observables of the kaonic atoms. We find that the existence of the kaonic nuclear states with the same angular momentum as the atomic states affects the atomic state structure. Thus, we think that the determination of the quantum numbers of the kaonic nuclear states is important not only for the study of the kaonic nuclear states but also for the studies of the kaonic atoms and the optical potential.

\vfill  

\noindent{\bf References }
\begin{description}
\setlength\itemsep{-3pt}
\item{[1]} T.~Hashimoto \textit{et al.} [J-PARC E62], Phys. Rev. Lett. \textbf{128}, no.11, (2022) 112503.
\item{[2]} J. Yamagata-Sekihara, Y. Iizawa, D. Jido, N. Ikeno, T. Hashimoto, S. Okada, and S. Hirenzaki, to be submitted (2024).
\item{[3]} Y. Iizawa, D. Jido, N. Ikeno, J. Yamagata-Sekihara and S. Hirenzaki, [arXiv:1907.05626 [nucl-th]].
\end{description}

\stepcounter{count}
\clearpage

\phantomsection
\addcontentsline{toc}{section}{
{\bf Study of hypernuclei in J-PARC experiments} \\
T.O.~Yamamoto}

\titl{Study of hypernuclei in J-PARC experiments}

\name{
T.O.~Yamamoto$^{1}$
}

\adr{
$^1$ JAEA ASRC, Ibaraki, 319-1195, Japan
}

Recent experimental data on $S=-1$ and $-2$ systems promoted studies of baryon-baryon interactions.
In the J-PARC Hadron Experimental Facility (HEF), experimental progress was made on $YN$ scattering measurements~[1] and spectroscopic experiments of light hypernuclei to provide information on the $YN$ interaction.
In addition, the $YNN$ three-body interaction and baryon mixing effect in nuclear medium can be discussed by measurements on light to heavy hypernuclei.  
Furthermore, experimental data on heavy hypernuclei will provide information on the density dependence of the $YN$ interaction, which may be key to understanding the structure of neutron stars and has been discussed intensively.
Together with the present HFE programs, the future HEF-EX project will allow us to extend accessible physics related to hypernuclear study.

In the $S=-1$ sector, $\gamma$-ray spectroscopy of $\Lambda$ hypernuclei was performed successfully at HEF~[2], which promotes discussion on the charge symmetry breaking effect and the $\Lambda N-\Sigma N$ mixing effect in the $\Lambda N$ interaction.
$\gamma$-ray measurements will advance to heavier hypernuclear studies to investigate the density dependence of the $\Lambda N$ interaction.
Reaction spectroscopy with a sub-MeV resolution is also essential for future hypernuclear studies.
The HIHR beam line in HEF-EX is suitable for this purpose with the dispersion-matching method.
The beamline allows us to use pion beam intensity of 10$^8$ /spill with 0.4 MeV mass resolution, with which we can separate sub-major peaks of heavy $\Lambda$ hypernuclear states.
Precise binding energy measurements for a wide mass range of the target nuclei will be performed to study the density dependence of the $\Lambda N$ interaction.

In the $S=-2$ sector, missing mass spectroscopy via the $(K^-, K^+)$ reaction is one of HEF's flagship experiments.
We are aiming at the first observation of peak structures of $\Xi$ hypernuclear bound states with an improved mass resolution.
For this purpose, a high-resolution S-2S spectrometer was constructed at the K1.8 beam line in 2023.
The first data-taking will soon start with a carbon target followed by systematic measurements with other target nuclei to investigate central, spin- and isospin-dependent terms of the $\Xi N$ interaction.
In addition,  $\Xi$-nucleus bound systems are studied by HEF experiments using the emulsion technique and X-ray spectroscopy method, especially in terms of the $\Xi N-\Lambda\Lambda$ conversion strength.
The E07 group found a deeply bound state of $\Xi^--^{14}$N system in emulsion images~[3], suggesting weak conversion strength. 
We expect that future HEF experiments provide X-ray data on $\Xi$-atoms for a detailed discussion on the conversion strength.
Future physics programs at the present HEF and the HEF-EX may extend our knowledge of hypernuclei and the $YN$ interaction.

\vfill  

\noindent{\bf References }
\begin{description}
\setlength\itemsep{-3pt}
\item{[1]} K.~Miwa et al., PRL 128, 072501 (2022).
\item{[2]} T.O.~Yamamoto et al., Phys. Rev. Lett. 115, 222501 (2015).
\item{[3]} M. Yoshimoto et al., PTEP 2021, 7, 073D02 (2021).
\end{description}

\stepcounter{count}
\clearpage

\phantomsection
\addcontentsline{toc}{section}{
{\bf Interpretation of Compositeness through Energy Dependent Extension} \\
Z.~Yin$^*$ and D.~Jido}

\titl{Interpretation of Compositeness through Energy Dependent Extension}

\name{
Zanpeng Yin$^{1}$ and Daisuke Jido$^{1}$ 
}

\adr{
$^1$ Tokyo Institute of Technology
}

In this talk, we analyzed the problem with deuteron compositeness with a theoretical approach. We first give the formalism of compositeness starting by giving an energy-dependent version of formalism of compositeness.[1] By introducing surjective interpretation, we then reduced this formalism into one without energy dependency which is the origin of the calculation of compositeness based on weak-binding limit.[2] Based on this formalism, we continue on to state the problem as the conflict between the positivity of elementariness bounded by the structure of the Hilbert space and the negative elementariness calculated from phenomenology, which is highly unlikely to be avoided according to our numerical calculation. As a result, we concluded that the problem lies in surjective interpretation, and a possible solution through the introduction of a new quantity interactioness is given.[1]

-----------

\vfill  

\noindent{\bf References }
\begin{description}
\setlength\itemsep{-3pt}
\item{[1]} A Possible Solution to the Difficulty in the Interpretation of Deuteron Compositeness, \href{arXiv:2312.13582}{Z. Yin and D. Jido, arxiv:2312.13582}
\item{[2]} Comprehensive analysis of the wave function of a hadronic resonance and its compositeness, \href{https://doi.org/10.1093/ptep/ptv081}{T. Sekihara, T. Hyodo, and D. Jido, PTEP 63D04 (2015).}
\end{description}

\stepcounter{count}
\clearpage

\phantomsection
\addcontentsline{toc}{section}{
{\bf Exclusive diphoton mesoproduction at J-PARC for probing QCD tomography with enhanced sensitivity} \\
J.W.~Qiu, Z.~Yu$^*$}

\titl{Exclusive diphoton mesoproduction at J-PARC for probing QCD tomography with enhanced sensitivity}

\name{
Jian-Wei Qiu$^{1,2}$ and Zhite Yu$^1$
}

\adr{
$^1$ Theory Center, Jefferson Lab, Newport News, Virginia 23606, USA\\
$^2$ Department of Physics, William \& Mary, Williamsburg, Virginia 23187, USA
}

\newcommand{\fig}[1]{Fig.~\ref{#1}}
\newcommand{\eq}[1]{Eq.~\eqref{#1}}
\newcommand{\eqs}[2]{Eqs.~\eqref{#1} and~\eqref{#2}}
\newcommand{\refcite}[1]{Ref.~\cite{#1}}
\newcommand{\refs}[1]{Refs.~\cite{#1}}

\newcommand{\pp}[1]{\left(#1\right)}
\newcommand{\bb}[1]{\left[#1\right]}
\newcommand{\cc}[1]{\left\{#1\right\}}
\newcommand{\vv}[1]{\langle #1 \rangle}

\newcommand{\beq}[1][]{\begin{equation}\label{#1}}
\newcommand{\eeq}{\end{equation}}
\newcommand{\bse}[1][]{\begin{subequations}\label{#1}}
\newcommand{\ese}{\end{subequations}}
\newcommand{\nn}{\nonumber}

\renewcommand{\P}{\mathcal{P}}
\newcommand{\wt}[1]{\widetilde{#1}}
\newcommand{\sgn}[1]{{\rm sgn}\left[#1\right]}
\newcommand{\LQCD}{ \Lambda_{\rm QCD} }
\newcommand{\M}{\mathcal{M}}
\newcommand{\Mt}{\wt{\mathcal{M}}}
\renewcommand{\H}{\mathcal{H}}
\newcommand{\Ft}{\widetilde{\mathcal{F}}}
\newcommand{\Ct}{\widetilde{C}}
\newcommand{\ie}{{\it i.e.}}

\begin{figure}[htbp]
	\centering
		\includegraphics[scale=0.5]{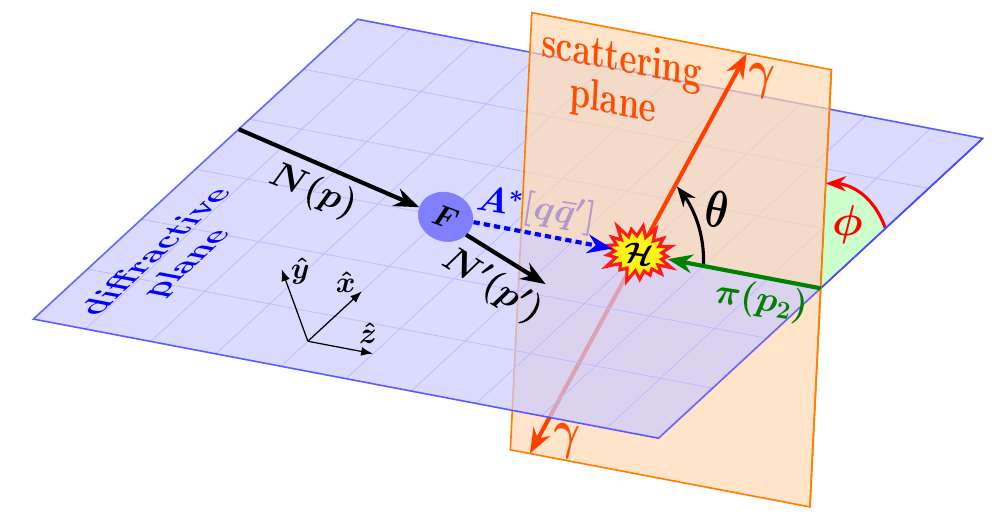}
	\caption{The frame to analyze the diphoton production process in \eq{eq:diphoton}.
		The $A^*$ represents the collinear quark and antiquark pair attaching the GPD to the hard part.
	}
	\label{fig:frame}
\end{figure}

The high-energy pion beam at J-PARC allows to probe hadron structures by extracting generalized parton distributions (GPDs)
through the hard exclusive diphoton production process~\cite{Qiu:2022bpq, Qiu:2024mny},
\beq[eq:diphoton]
	N(p) + \pi(p_2) \to N'(p') + \gamma(q_1) + \gamma(q_2),
\eeq
where the nucleon $N$ is diffracted to $N'$ in the near-forward direction.
This process is conveniently analyzed in the center-of-mass frame of the final-state photons, 
with the pion moving along the $-\hat{z}$ direction and the nucleons lying on the $\hat{x}$-$\hat{z}$ plane,
as shown in \fig{fig:frame} where the angles $\theta$ and $\phi$ of the photons are also defined.
In the kinematic region where the nucleon momentum transfer $t = (p - p')^2$ is much smaller than the transverse momenta of the final-state photons,
\beq
	q_T = |\bm{q}_{1T}| = |\bm{q}_{2T}| \gg \sqrt{|t|},
\eeq
the amplitude of \eq{eq:diphoton} can be factorized into nucleon transition GPDs $F^{[q\bar{q}']}_{NN'}(x,\xi,t)$ and $\wt{F}^{[q\bar{q}']}_{NN'}(x,\xi,t)$ 
and pion distribution amplitude $D^{[q'\bar{q}]}_{\pi}(z)$, 
convoluted with perturbatively calculable hard coefficients,
\begin{align}
	\mathcal{M}(\xi, t, \theta)
	&= \int_{-1}^{1} dx \int_0^1 dz \, D^{[q'\bar{q}]}_{\pi}(z) 
		\bb{ \wt{F}^{[q\bar{q}']}_{NN'}(x,\xi,t) \, C(x, \xi; z; \theta)				
			+ F^{[q\bar{q}']}_{NN'}(x,\xi,t) \, \wt{C}(x, \xi; z; \theta)
		}, 
\label{eq:factorize}
\end{align}
up to corrections suppressed by powers of $\sqrt{|t|} / q_T$.
Since only charged pion beams are accessible in experiments, the nucleon diffraction involves a flavor transition, 
with $(NN') = (pn)$ and $(qq') = (ud)$ for a $\pi^-$ beam and $(NN') = (np)$ and $(qq') = (du)$ for a $\pi^+$ beam.
These transition GPDs can be related to flavor-diagonal ones by isospin symmetry.

The importance of GPDs lies in the fact that they encode
tomographic parton images in slices of different parton momentum fraction $x$~\cite{Burkardt:2000za, Burkardt:2002hr, Ji:1994av, Ji:1995sv, Lorce:2017xzd, Metz:2020vxd, Ji:1996ek, Polyakov:2002yz, Polyakov:2018zvc, Burkert:2018bqq}. 
For this to be ever possible, we shall be able to precisely extract the $x$ dependence of GPDs.
This turns out to be rather difficult~\cite{Qiu:2022pla, Qiu:2023mrm}, 
because for most of the known processes, when factorizing their amplitudes similarly to \eq{eq:factorize}, 
the GPD convolutions at the leading order (LO) reduce to simple ``scaling'' forms like
\beq[eq:moment]
	F_0(\xi, t) = \int_{-1}^1 dx \frac{F(x, \xi, t) }{x - \xi \pm i \epsilon},
\eeq
which completely integrate over the $x$ dependence.
Inverting the full details of GPDs merely from such moments is an impossible task,
as one can construct an infinite family of analytic solutions $S(x, \xi, t)$ that give zero to \eq{eq:moment}.
These solutions, called shadow GPDs~\cite{Bertone:2021yyz, Moffat:2023svr}, would not be separable from the real GPDs 
if we only had moment-type sensitivity to GPDs like \eq{eq:moment}. 

This scaling property applies to the most well studied processes in the literature, 
including the deeply virtual Compton scattering~\cite{Ji:1996nm, Radyushkin:1997ki, Collins:1998be}, 
deeply virtual meson production~\cite{Brodsky:1994kf, Frankfurt:1995jw, Collins:1996fb}, 
timelike Compton scattering~\cite{Berger:2001xd}, and in particular, 
the exclusive Drell-Yan-type dilepton production~\cite{Berger:2001zn, Goloskokov:2015zsa, Sawada:2016mao},
\beq[eq:dilepton]
	N(p) + \pi(p_2) \to N'(p')+ \gamma^* [\to \ell^-(q_1) + \ell^+(q_2)],
\eeq
which could also be studied at J-PARC to constrain the polarized GPD $\wt{F}$.
In contrast, the diphoton process in \eq{eq:diphoton} overcomes this difficulty and provides enhanced sensitivity to the $x$ dependence of GPDs.
This is because the produced two photons can be radiated from two different parton lines 
that are necessarily linked by a virtual gluon, 
and then this internal gluon propagator carries a large $q_T$ flow to break the LO scaling.
In addition to simple moments like \eq{eq:moment}, the GPD convolutions in \eq{eq:factorize} also involve the special integrals,
\begin{align}
	I[\mathcal{F}^+]
	= \int_{-1}^{1} dx \frac{\mathcal{F}^+(x, \xi, t)}{x - x_p(\theta) + i \epsilon \, \sgn{\cos^2(\theta/2) - z} },
\label{eq:diphoton-special-int}
\end{align}
where $\mathcal{F}^+ = F^+$ or $\wt{F}^+$ is the charge-conjugation-even GPD combination, and the new pole $x_p$ depends on the 
polar angle $\theta$, which is equivalent to $q_T$ for fixed $t$ and $\xi$,
\beq[eq:x-pole]
	x_p(\theta)
		= \xi \bb{ \frac{1 - z + \tan^2(\theta / 2) \, z}{1 - z - \tan^2(\theta / 2) \, z} }.
\eeq
Varying $\theta$ shifts this pole around in the DGLAP region of the GPDs, 
causing an entanglement between the $x$ dependence of GPDs and the observable $\theta$ (or $q_T$) distribution.

\begin{figure}[htbp]
	\centering
		\includegraphics[trim={0 0 -2em 0}, clip, scale=0.5]{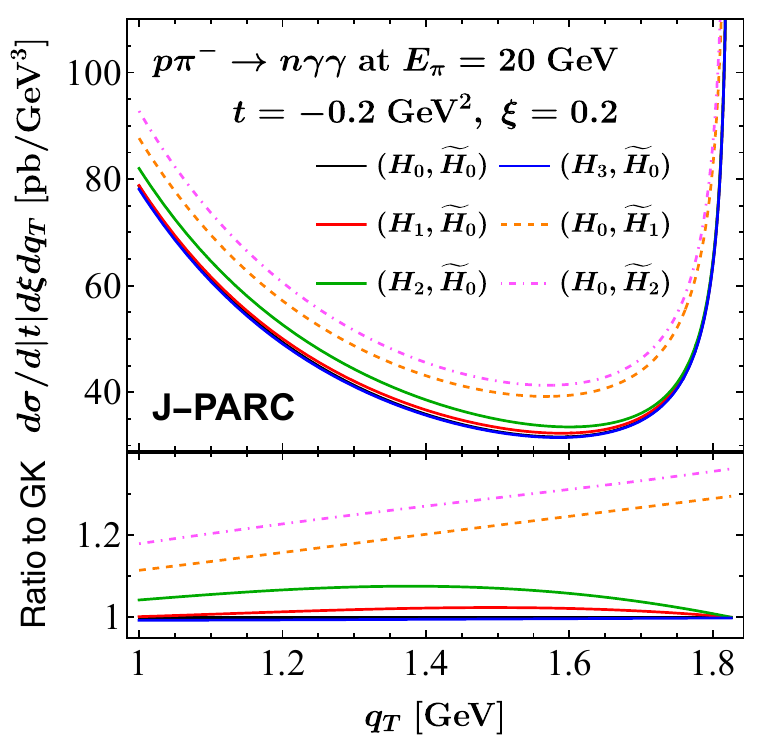}
		\includegraphics[trim={-2em -5em 0 0}, clip, scale=0.5]{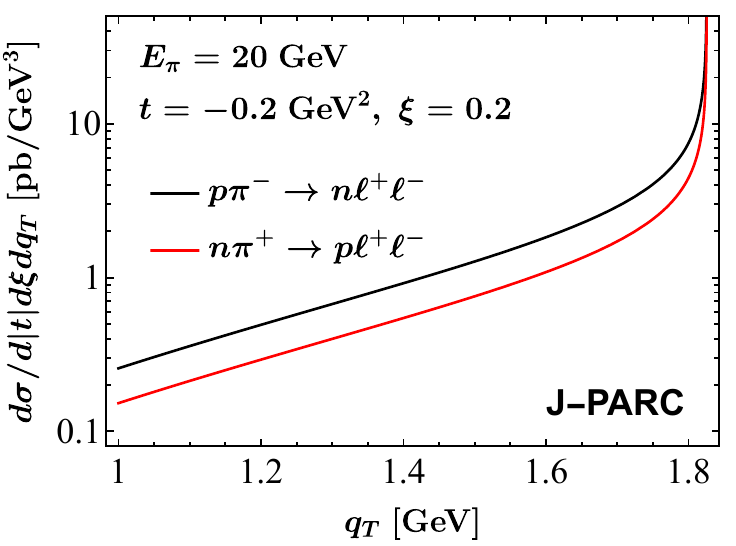}
	\caption{The $q_T$ distributions for the diphoton [Left; \eq{eq:diphoton}] and dilepton [Right; \eq{eq:dilepton}] production processes, 
		evaluated using different GPD models. For the dilepton process, different GPD models yield the same distribution.
	}
	\label{fig:qt-dist}
\end{figure}

Since the $\theta$ dependence cannot be separated from the GPD integrals in \eq{eq:diphoton-special-int},
it does not seem possible to construct a shadow GPD that gives zero to \eq{eq:diphoton-special-int} for all $\theta$ values.
The diphoton process can therefore distinguish the real and shadow GPDs (for the specific flavor transition), at least in principle.
To demonstrate the numerical effects, we take the Goloskokov-Kroll model~\cite{Goloskokov:2005sd, Goloskokov:2007nt, Goloskokov:2009ia, Kroll:2012sm} 
as the base GPDs $(H_0, \wt{H}_0)$,
and vary either $H_0$ or $\wt{H}_0$ by adding analytically constructed shadow GPDs to construct $(H_1, H_2, H_3)$ and $(\wt{H}_1, \wt{H}_2)$.
Taking these models, we calculate the differential cross sections for both processes in Eqs.~\eqref{eq:diphoton} and \eqref{eq:dilepton},
which are illustrated in \fig{fig:qt-dist} as distributions of $q_T$. 
Clearly, these different models yield different distributions for the diphoton process, in terms of both shape and magnitude.
The sensitivity to the polarized GPD $\wt{H}$ is higher than the unpolarized one $H$.
In contrast, since the shadow GPDs give zero contribution to \eq{eq:moment} by definition, these models are not separable by the dilepton process.
Furthermore, it is evident that the diphoton process also yields a higher rate than the dilepton one.

In conclusion, the exclusive diphoton production process offers J-PARC a novel and exceptional opportunity to explore the GPDs, especially their $x$ dependence. 
Realizing this potential could bring us significantly closer to obtaining the tomographic image of the proton.

\vfill  


\bibliographystyle{unsrt}
{\footnotesize
\bibliography{./Yu/reference}
}

\stepcounter{count}
\clearpage



\end{document}